\documentclass[twocolumn,floatfix]{aastex631}

\usepackage{graphicx}	
\usepackage{amsmath}	
\usepackage{amssymb}	
\usepackage{bm}		
\usepackage{float}
\usepackage{subcaption}

\newcommand{\ct}{\citealt}
\renewcommand{\thesubsubsection}{\thesubsection.\alph{subsubsection}}

\usepackage[T1]{fontenc}
\usepackage{ae,aecompl}

\usepackage{newtxtext,newtxmath}

\begin{document}
\title[Galactic Zoom]{Filamentary Hierarchies and Superbubbles:  Galactic Multiscale MHD Simulations of GMC to Star Cluster Formation.}

\correspondingauthor{Ralph E. Pudritz}
\email{pudritz@physics.mcmaster.ca}

\author{Bo Zhao}
\affiliation{Department of Physics and Astronomy, McMaster University, Hamilton, ON L8S 4M1}

\author[0000-0002-7605-2961]{Ralph E. Pudritz}
\affiliation{Department of Physics and Astronomy, McMaster University, Hamilton, ON L8S 4M1}
\affiliation{Origins Institute, McMaster University, Hamilton, ON L8S 4M1\\}
\affiliation{Universität Heidelberg, Zentrum für Astronomie, Institut für Theoretische Astrophysik, Albert-Ueberle-Str. 2, 69120 Heidelberg, Germany\\}
\affiliation{Max-Planck Institut für Astronomie, Königstuhl 17, D-69117 Heidelberg, Germany 
\\}

\author[0000-0002-3033-3426]{Rachel Pillsworth}
\affiliation{Department of Physics and Astronomy, McMaster University, Hamilton, ON L8S 4M1}

\author{Hector Robinson}
\affiliation{Department of Physics and Astronomy, McMaster University, Hamilton, ON L8S 4M1}

\author{James Wadsley}
\affiliation{Department of Physics and Astronomy, McMaster University, Hamilton, ON L8S 4M1}
\affiliation{Origins Institute, McMaster University, Hamilton, ON L8S 4M1\\}




\begin{abstract}
{ There is now abundant observational evidence that star formation is a highly dynamical process that connects filament hierarchies and supernova feedback from galaxy scale kpc filaments and superbubbles, to giant molecular clouds (GMCs) on 100 pc scales and star clusters (1 pc).  Here we present galactic multi-scale MHD simulations that track the formation of structure from galactic down to sub pc scales in a magnetized, Milky Way like galaxy undergoing supernova driven feedback processes.  We do this by adopting a 
novel zoom-in technique that follows the evolution of  typical 3-kpc 
sub regions without cutting out the surrounding galactic environment allowing 
us to reach 0.28 pc resolution in the individual zoom-in regions. 
We find  a wide range of morphologies and hierarchical structure, including superbubbles, turbulence, kpc atomic gas filaments hosting multiple GMC condensations that are often associated with superbubble compression; down to smaller scale filamentary GMCs and star cluster regions within them.  Gas accretion and compression ultimately drive filaments over a critical, scale - dependent line mass leading to gravitational instabilities that produce GMCs and clusters.  In quieter regions, galactic shear can produce filamentary GMCs within flattened, rotating disk-like structures on 100 pc scales.   Strikingly, our simulations demonstrate the formation of helical magnetic fields associated with the formation of these disk like structures.
 }
\end{abstract}

\keywords{magnetic fields -MHD- circumstellar matter - stars: formation}



\section{Introduction}
\label{Chap.Intro}

The advent of ALMA, JWST, and a host of  high resolution atomic hydrogen, molecular gas, and dust surveys make it clear that star formation is an intrinsically multiscale and highly dynamical process.  Hierarchies of filaments, beginning at galactic scales and extending down to sub pc scales, connect molecular clouds, star clusters and stars.  Going back up the hierarchy, stellar feedback from supernovae, in turn, impact the interstellar medium and sculpt these structures on as many scales.   Multiscale galactic simulations through many decades of physical scales including feedback processes are still highly challenging.  How mass and angular momentum flows as well as feedback processes affect the formation of structures within these hierarchies, therefore, are still not well spatially resolved nor physically understood.   In this paper we connect and track these processes by developing new simulation zoom-in methods from galactic down to sub pc scales.  

There are several recent observational breakthroughs that are important to address.  On galactic scales, the remarkable  JWST observations of the spiral galaxy NGC628 (M74) reveal that its atomic and molecular gas  is organized into plethora of interconnected filamentary  \citep{Thilker+2023} and superbubble structures \citep{Watkins+2022} associated with star formation, ranging over many decades in physical scales.  Nearly 1700 superbubbles ranging from 6 to 500 pc in radius fill its disk in nested structures, pressing gas into converging bubble walls in which some of the longest filaments are found.  Bubbles on kpc scales are also significantly stretched and distorted which is likely a consequence of galactic shear \citep{Watkins+2022}.    The early, deeply embedded phase of massive star formation within these  molecular clouds lasts just 5.1 Myr \citep{KimJWST+2022}, accounting for roughly 20\% of the overall cloud lifetime.  The stellar feedback arising from the supernova activity produced by massive OB associations in turn drive bubbles and superbubbles.  Kinematic studies of HI, CO, and HII gas across the most prominent bubble in NGC628 show that it is expanding at 15-50 km s$^{-1}$, being driven by ongoing SN activity in a coincident $10^5 \rm{M_{\odot}}  $ stellar association \citep{Barnes+2023}.   Superbubble lifetimes span a range of 7-42 Myr in this galaxy.

New long baseline HI surveys using the ATCA array of atomic gas between two kpc scale superbubbles in the LMC galaxy find that the atomic gas is highly filamentary with an average width of 21 (8-49) pc and containing a total of $ 8.5 \times 10^6 \rm{M_{\odot}} $ in the major HI ridge.  GMC mass clumps are distributed along the length of the filaments suggesting gravitational fragmentation as their origin \citep{Fujii+2021}. Dynamical studies have also turned up ubiquitous, periodic velocity fluctuations of molecular gas in the Milky Way and the galaxy NCG4321 over 4 decades in scale ($ 10^{-1} - 10^3 $ pc).  These are correlated with regularly spaced density fluctuations that likely arise from gravitational instabilities \citep{Henshaw+2020}. 

The PHANGS - ALMA survey of the mass distribution of molecular clouds in spiral galaxies such as NGC3627 is well modelled by a Schechter  function \citep{Rosolowsky+2021} - a power-law form with a high mass truncation whose highest mass is only 10 times greater than the peak.  Truncation of cloud populations is suggestive of the effects of stellar feedback (see \S 3.2) 

In the Milky Way, long atomic gas filaments extending over several kpc are among largest scale structures in our galaxy.   Good examples of these are the "Maggie" filament with a mass of $7.2 \times 10^5 M_{\odot}$ and line widths of $ 10$ km s$^{-1}$   in the THOR survey \citep{Syed+2022}, and a distinctly wavy molecular filament extending to 2-4 kpc \citep{Veena+2021}.   Giant molecular filaments extending over hundreds of pc have also been identified \citep{Ragan+2014, Goodman+2014, Abreu-Vicente+2016, Zucker+2018}.    Molecular cloud populations on these galactic scales are often associated with spiral arms and large scale filaments in the cold neutral medium (CNM) of atomic hydrogen.    
Tellingly, atomic filaments within 10 kpc of the galactic centre tend to be oriented perpendicular to or have no particular orientation with respect to the galactic plane while those beyond are parallel to the plane.  The difference in these orientations likely arises from dominant supernova feedback activity in the inner galaxy that blows material out of the galactic plane versus the dominance of galactic rotation and shear in the quieter outer regions ( \citet{Soler+2022}, also \S 3.1 ).  

On intermediate scales of the galactic solar neighbourhood,  GAIA stellar photometric surveys have produced 3D dust maps of the nearby galaxy, with resolutions down to 2pc \citep{Leike+2020}.   Such maps can then be used to deduce the full 3D structure of molecular clouds within 2 kpc.   They are found to be either filamentary (eg. Orion cloud) and/or sheet like (eg. California cloud) structures \citep{RezaeiKh+2020, Zucker+2021, RezaeiKh_Kainulainen2022}.    Moreover, molecular clouds are not isolated but are rather connected within long kpc-scale structures of widths of roughly 100 pc \citep{Zucker+2022}.   Many clouds show evidence of being affected by recent stellar or supernova feedback \citep{Zucker+2018}.   Surveys such as MIOP \citep{Beuther+2022} connect the larger scale filamentary clouds to star forming cores.    

On the smallest scales, the Herschel Space Observatory resolved galactic molecular clouds into networks of  filaments.  Of greatest importance is that individual star forming cores are strongly associated with gravitationally unstable filaments \citep{Andre+2010, Andre+2014, Henning+2010, Men'shchikov+2010, Arzoumanian+2011}. The critical condition is  that the line mass $m = M/ L$ ( usually measured in units of solar masses per pc) exceed a critical value for quiescent clouds forming low mass stars discussed below  \citep{Polychroni+2013, Andre+2014}.  Measurements of the Filament Line Mass Function (FLMF) on the pc scale find a power law relation beyond the critical line mass; $dN / dm \propto m^{-2.4}$ \citep{Andre+2019}, which suggests links with the core mass function of star formation \citep{Hennebelle_Chabrier2011,Lee_Hennebelle+2017}.  

It has also become clear that the formation of star clusters is driven in part by accretion flows from the larger scale environment onto such filaments and flow along these filaments towards overdense, cluster forming regions such as in the Serpens cloud cluster \citep{Kirk+2013}.    Massive clusters form at hubs at the intersection of several filaments  \citep{Myers2009, Kumar+2020} where they undergo particular large net filamentary accretion rates.  High resolution observations of regions of massive star formation on sub cluster scales suggest that they massive stars often form within fragmenting, disk like structures on scales of at least 1000 AU \citep{Beuther+2018, Beuther+2022, Ahmadi+2018, Ahmadi+2023}. 
   
Simulations of supersonic turbulence have long established that filamentary structure is a natural and highly dynamic consequence of intersecting shock waves and that overdensities in filaments can lead naturally to star formation \citep{Maclow_Klessen2004,Vazquez-Semadeni+2003, Krumholz_McKee2005, Hennebelle_Chabrier2011, Padoan_Nordlund2011,  Federrath_Klessen2012, Padoan+2014}.  In this picture, dense  cores form at the stagnation points in these flows and are able to collapse into stars only if the collapse time is faster than the time scale between shocks - which disperse the fluctuations. The statistical properties of the probability distribution of density fluctuations that exceed some critical density threshold for gravitational collapse in such turbulent media may then provide an explanation of the stellar IMF.  (see reviews   \citet{McKee_Ostriker2007, Offner+2014, Klessen_Glover2016}). 
\begin{figure*}
\includegraphics[width=\textwidth]{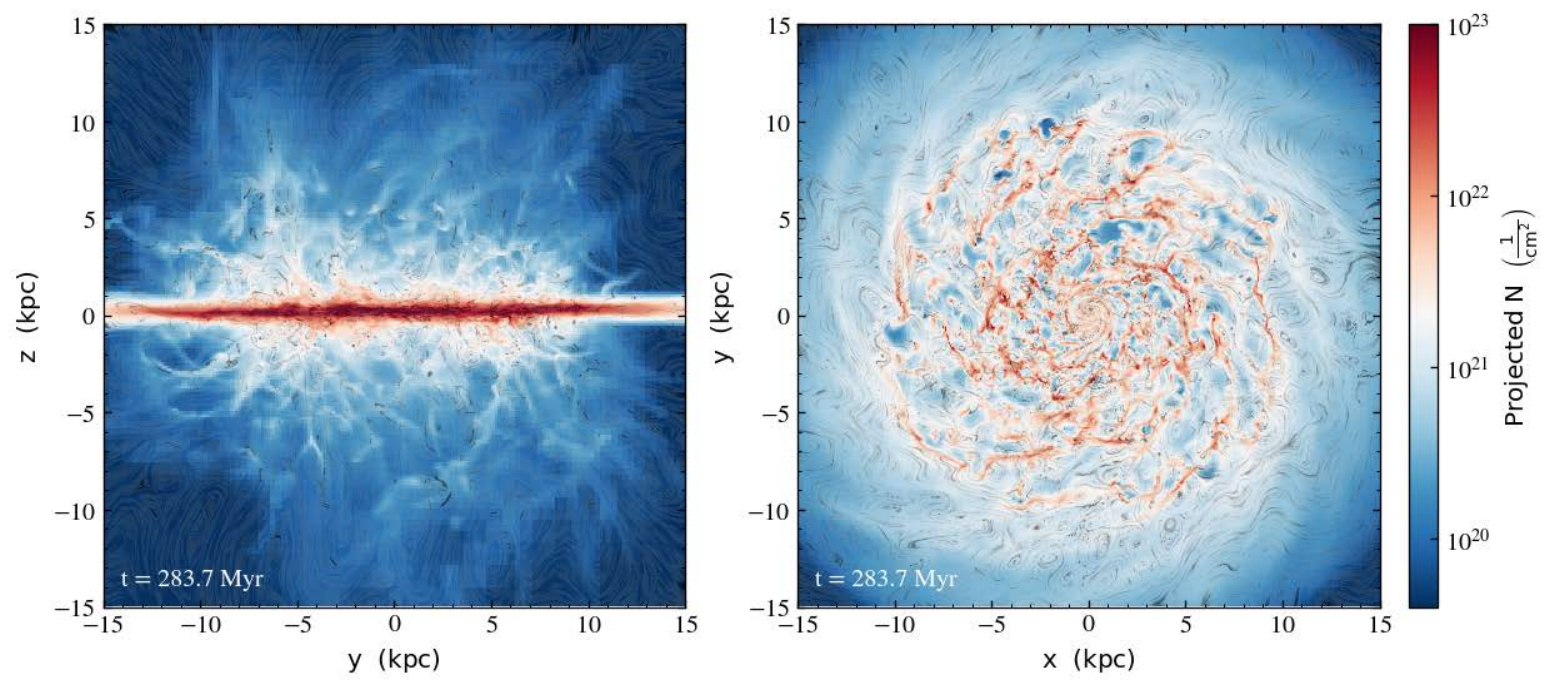}
\caption{Edge-on and face-on view of the logarithmic distribution of the column density of the galactic disk at t=283.7 Myr, overlaid with stream lines of magnetic field created via line integral convolution.}
\label{Fig:Proj}
\end{figure*}

This statistical gravoturblent picture, however, neglects the importance of the gravitational instability of filaments in creating structure. In particular, it is known both observationally and theoretically that gravitational instability of supercritical filaments gives rise to star forming cores.  Early theoretical work on self gravitating filamentary equilibria and their fragmentation \citep{Ostriker1964, Stodolkiewicz1963, Inutsuka+1997, Fiege_Pudritz2000a} showed that gravitational instability sets in when the line mass exceeds a critical value that depends only upon the sound speed;  $m_{crit} = 2c_s^2 / G  \simeq 16 M_{\odot}$ pc$^{-1}$ - the threshold for instability in thermal gas at $10$ K with sound speed;  $c_s \simeq 0.2$ km s$^{-1}$ where G is the gravitational constant.   This can be generalized to include gas pressure and magnetic fields, including helical field configurations  \citep{Fiege_Pudritz2000a, Fiege_Pudritz2000b}.  For unmagnetized, supersonic turbulence, the critical line mass becomes $m_{crit} = 2 \sigma^2 / G $ where $ \sigma^2 $ is the total velocity dispersion that includes a contribution from the non-thermal turbulence $\sigma^2 = c_s^2 + \sigma^2_{nt}$.  We shall demonstrate that this is important for GMC formation because the amplitude of supersonic turbulence on large kpc scales (5-10 km s$^{-1}$) results in critical line masses of a few  $10^4 M_{\odot}$ pc$^{-1}$ - sufficient for GMC formation within the most massive filaments.  When magnetic effects are included, a magnetic pre-factor occurs such that $m_{crit,B} = f_B .m_{crit}$. (We discuss these magnetic effects further in \S~\ref{Chap.ResultsII} and \S~\ref{Chap.Discuss}).  

Local cloud simulations show filaments are highly dynamical structures.  They are produced in shocks and will dissociate if they fail to approach their critical line mass \citep{Kirk+2015}.  Filaments which are gravitationally significant undergo continuous accretion from the surrounding medium. As fragmentation develops, filamentary flows along the filaments transport mass and local angular momentum onto the overdense fragments  \citep{Kirk+2015, Klassen+2017, Smith+2016, Smith+2020, Chira+2018}. Simulations of cluster formation in isolated GMCs that include radiative feedback, show that the most massive clusters accrete mass from filaments as well as smaller clusters out to 10s of pc across the cloud \citep{ Howard+2018}. The key result is that mass of the most massive clusters scales as nearly a linear power of the mass of the parent GMC.  Moreover, radiative feedback before supernovae limits, but does not cutoff their accretion of cold, dense filamentary streams into the forming massive cluster region \citep{Dale+2005, Dale+2012, Howard+2018}.  Thus  cluster masses are limited by the mass of their feeding reservoirs; more massive clouds build more massive star clusters \citep{Harris_Pudritz1994}. This is just one reason why galaxy scale processes are important to include. 

We propose, and will show, that the paradigm of gravitational fragmentation of filaments extends across the entire galactic hierarchical filamentary system.  This connects very well with the analysis of over 22,000 filaments gathered from over 40 individual observational studies; it show a clear scaling relation ranging over 8 decades in filament mass \citep{Hacar+2022} between filament mass and length;  $L \propto M^{0.5}$.   The origin of this relation is rooted in the size-linewidth relation for supersonic turbulence $ \sigma_{nt} $.  It also depends on the fragmentation line: the gravitational stability of filaments resulting when their line mass $m \ge  m_{crit} $ (see \S~\ref{Chap.ResultsI} ).

In this paper we feature "live" global galactic simulations with galactic disks in which supernova activity drives superbubble structures. We perform the first MHD simulations that adopt a novel 
zoom-in technique; we follow the evolution of two different 3-kpc 
regions in the galactic disk without cutting out the surrounding galactic environment.   We compute line masses within filaments constituting this hierarchy and find that gravitational instability drives the formation of key structures from kpc to sub pc scales:  giant molecular clouds and their star clusters.  Filaments form in converging  superbubbles in the inner disk or as a consequence of spiral wave action the quieter outer disk.  We also explore the role of galactic shear in shaping molecular clouds and their magnetic fields. We describe our methods and the initial conditions of our simulations in \S~\ref{Chap.Methods}, our results on global  ( in \S~\ref{Chap.ResultsI})  and then zoom-in subregions (in \S~\ref{Chap.ResultsII}), and then follow up with our discussion and conclusions in \S~\ref{Chap.Discuss}.  We defer to a second paper a detailed treatment of the statistical properties (distributions of mass, length, line mass, filament accretion rates, etc.) of the many hundreds filaments on the largest scales of these simulations (Pillsworth et al, 2024, in preparation).

\section{Methods}
\label{Chap.Methods}

\subsection{Previous Galactic Simulations}

In many global galaxy simulations \citep{ Kim+2002, Bournaud+2010,  Renaud+2013,  Kraljic+2014, Kortgen+2018, Kortgen+2019, Grisdale2021, Jeffreson+2020, Dobbs+2021, Tress+2020, Tress+2021}  spiral waves and spiral shocks are used as the seed for turbulence and the gas is compressed through the spiral shocks or various colliding flows. Thermal and gravitational instabilities fragment the gas into small molecular clouds.  At the cutting edge, spatial resolution in published simulations rarely reaches below 0.1 or 1 pc. However, we also incorporate MHD into global simulations because it can modulate feedback processes, filament confinement 
\citep{Fiege_Pudritz2000a, Schleicher_Stutz2018, Reissl+2018}, gravitational fragmentation \citep{Kortgen+2019}, as well as channel flows onto massive filaments  and forming clusters \citep{Klassen+2017, Pillai+2020}.

Simulators have tried several different approaches to bridge the large scale separation between galactic disks and cluster forming clumps in GMCs.  They generally rely on simplified, sub-grid star 
formation recipes based on stochastically spawning star cluster
particles of the same mass, or on the sink particle technique
where in this case sink particles represent a star cluster. \citet{Bournaud+2010} studied the properties of
the ISM substructure and turbulence in galaxy simulations with
resolutions up to 0.8 pc and $5\times 10^3~M_{\sun}$ with RAMSES. 
\citet{Renaud+2013} and \citet{Kraljic+2014} were able to capture the
transition from turbulence-supported to self-gravitating gas with
resolution up to 0.05 pc in simulations of Milky Way-like galaxies
using RAMSES. The results shed light on a handful of important questions 
on turbulence, clump mass, fragmentation. However, their inclusion of 
sink particle is sensitive to the resolution and causes artificial drift 
of particle positions.

\citet{Jeffreson+2020} used 3D AREPO simulations in a fixed gravitational potential to track the development and characteristics of thousands of clouds in three different galactic setups.  The highest resolution reached in some cells was 3 parsecs. They found the gravo-turbulence and star-forming 
properties of GMCs are decoupled from the galactic dynamics; however, 
the elongation, angular momentum, and velocity dispersions of GMCs 
can be affected by the the galactic rotation and gravitational instabilities.

\citet{Grisdale2021} focused on the evolution of the galactic disk with 
4.6~pc resolution using the RAMSES code. They addressed the impact of different star formation
models (molecular star formation versus turbulent star formation) on 
the star formation rate and GMC properties.  \citet{Dobbs+2021} carve out a few representative sub-regions from the 
galactic disk, focusing on the effect of photoionization on the formation 
of stellar clusters. However, they only turned on feedback on the zoom-in 
sub-regions, not on the galactic scale. The GMCs formed in such an 
environment may be less representative of the typical equilibrium state
of a galactic disk.   \citet{Tress+2020, Tress+2021} used the AREPO code \citep{Springel2010} to simulate molecular cloud formation 
and dynamics in an interacting M51-like galaxy.   They included detailed chemical reaction networks 
and ISM physics reaching sub-pc spatial resolution in some of the densest regions.  Spiral arms are naturally produced
but the resulting gas flows are of secondary importance in controlling the local
regions in which GMCs form; ISM physics coupled to stellar feedback take precedence.  Moreover, cloud populations in these
simulations show a large range in their virial parameters with a smooth transition from bound to unbound structures (see also \citet{Howard+2017})

Most of the above studies have ignored the magnetic field. A few studies 
have dedicated to the effect of magnetic field on galactic evolution 
and molecular cloud formation. \citet{Kim+2002} used a shearing box setup 
to investigate the nonlinear development of the Parker and magneto-Jeans 
instability (MJI). They find that MJI plays a more dominant role than 
Parker instability in forming bound cloud complexes in spiral galaxies. 
In comparison, \citet{Kortgen+2018,Kortgen+2019} found that the buoyancy of 
magnetic fields due to the Parker instability can efficiently guide 
the kilo-parsec galactic flows in forming filament and cloud structures. 
However, stellar feedback is ignored in both studies, which may constantly 
smooth out any of the magnetic related instabilities.


\subsection{Live Galactic Disk Set-Up} 

We carry out our MHD simulations using the adaptive-mesh-refinement (AMR) 
code \textsc{RAMSES}  \citep{Teyssier2002}, which is publicly available. As the first study, 
we employ standard \textsc{RAMSES} modules for MHD, star formation via star particles, stellar 
feedback, metal cooling and clumpfinder.  We implement gas cooling via Grackle's tabulated network for metal line cooling from Cloudy tables.  We enabled photoelectric heating in Grackle with a rate of $\zeta = 4 \times 10^{-26}$ erg cm$^{-3}$ s$^{-1}$, and no UV heating. We note that Grackle cooling has an effective temperature floor of 10 K below which the cooling routines in diffuse gas are no longer reliable.   We will leave radiative transfer and sink particle 
treatment in the subsequent zoom-in studies at individual cluster and 
stellar scales. 

We initialize the simulation using the isolated AGORA disk model 
\citep{Kim+2016} for Milky Way like galaxies at redshift $\sim$1. 
The simulation box is set to 300~kpc, which is half of the 
size of the original AGORA setup but is large enough to enclose 
the majority ($\sim$90\%) of AGORA's dark matter halo mass. The 
resulting circular velocity $\varv_{150}=160$~km~s$^{-1}$. 
The density distribution of the dark halo follows the NFW profile 
\citep{Navarro+1997} with concentration parameter c=10 and spin 
parameter $\lambda$=0.04. We follow the standard RAMSES treatments for gravity and softening as used in the original AGORA runs \cite{Kim+2016}. We also include a non-thermal Jeans pressure floor, which prevents artifical fragmentation due to unresolved pressure gradients \citep{Truelove+1997}. This is a  minimum allowed pressure in each cell of the form 
\begin{equation}
    P_{min} = \frac{1}{\gamma \pi} N_{Jeans}^2 G \rho^2 \Delta x^2
\end{equation}
where $\gamma$ is the adiabatic index and $\rho$ is the gas density. We note that $N_{Jeans}$=4 is the number of cells the Jeans length must be resolved by (the  Truelove criterion) in our simulation, where $\Delta x$ is the size of the cell. The pressure floor is implemented via temperature a polytrope, similar to the AGORA project \citep{Kim+2016} but with a lower Jeans temperature of 70 K due to our higher resolution.Only a very small portion of our gas is affected by this.  The adiabatic index for the hydrodynamics calculations is set at $\gamma = 5/3$; this index is always used during the hydro steps.  Grackle heating and cooling is done separately from the hydrodynamics.

The galactic disk is located at the box center. Its total mass amounts to 
$M_{\rm d}=4.3\times10^{10}~M_{\sun}$, with 80\% in stars and 20\% 
in gas. The gaseous disk is initialized with the following density profile:
\begin{equation}
\rho_{\rm d,gas}(r,z) = \rho_0 e^{-r/r_{\rm d}} e^{-|z|/z_{\rm d}}~,
\end{equation}
where the scale length $r_{\rm d}$=3.432~kpc, the scale height 
$z_{\rm d}$=0.1~$r{\rm _d}$, and 
$\rho_0=M_{\rm d, gas}/(4\pi r_{\rm d}^2 z_{\rm d})$. The gas is initialized with a temperature of $10^4$ K and Solar metallicity.
The stellar bulge component has a total mass of 
$M_{\rm b}=4.297\times10^9$~$M_{\sun}$, which follows the Hernquist 
profile with $M_{\rm b}/M_{\rm d}$=0.1. 
Note that the initial dark matter halo, stellar disk and bulge 
are initialized as background star particles, which are distinguished 
from the new stars by their formation time and age.

In addition, we have initialized a toroidal magnetic field in the AGORA 
disk, where the field strength scales with the gas density as; 
$B = B_o (\rho / \rho_o)^{2/3}$, with an initially rather weak magnetic field of $B_o = 0.1 \mu {\rm G} $ in a diffuse gas density of $\rho_o = 4.18 \times 10^{-25} $  g cm$^{-3}$ (or $n_o = 0.25 $ cm$^{-3} $) (see also \citet{Kortgen+2019} ).   This field is far below equipartition values initially.  Numerical simulations with this same galaxy model, and with the same weak initial field MHD setup as ours \citep{Robinson_Wadsley2023}, show that a turbulent dynamo quickly grows this magnetic field to equilibriate at an average 
magnetic field strength of $ 5 \mu $ G in the midplane at around 8kpc. 
The dynamo processes in the galaxy  amplify the magnetic field 
strength, transferring mechanical energy into magnetic energy.

Star formation is done stochastically in any cell above a threshold number density of 100 cm$^{-3}$, such that star particles are formed at a rate given by a Schmidt law,

\begin{equation}
    \frac{d\rho_*}{dt} = \frac{\epsilon_*\rho}{t_{\textrm{ff}}},
\end{equation}
where $\rho$ is the gas density, the free-fall time is $t_{\textrm{ff}}=\sqrt{3\,\pi/32\,G\,\rho}$ and $\epsilon_*$ is the efficiency per free-fall time, set to 0.1.

For every 100~$M_{\sun}$ of stars formed in a population, one 
star will undergo a supernova event that deposits $\sim$10$^{51}$~ergs 
of energy into the surrounding gas 
\citep{Agertz+2011,Krumholz+2014}. We use the \citet{Agertz+2011} approach, built in to RAMSES, where star particles deposit supernova energy in an adiabatic form without radiative cooling after a 5 Myr delay.  This energy is converted to normal, cooling thermal energy with a e-folding time of 5 Myr.   Feedback bubbles thus evolve adiabatically at first and radiative cooling becomes important later.  This feedback is effective at producing bubble features and driving turbulence.
In this way, a realistically low star formation rate can be reproduced 
even with a star formation efficiency parameter of $\epsilon_* = 0.1$ \citep{Robinson_Wadsley2023,Agertz+2015,Semenov+2018}. 

Our simulations are divided into two stages. We first evolve the 
galactic disk for 200--300~Myr at 16 levels of refinement, to give a resolution of 4.58~pc at the highest level.  By 300 Myr the 
galactic disk has reached a relatively quasi-equilibrium state with 
star formation rates of about $\sim$5~M$_{\sun}$~yr$^{-1}$ 
regulated by stellar feedback. At this time, the galactic structure 
is comprised of spiraling filamentary atomic structures, between which supernova driven
feedback bubbles expand. Such a simulation is quite similar to those 
carried out in \ct{Grisdale2021} and \citet{Robinson_Wadsley2023}.


\subsection{Zoom-in Technique}

To evolve the zoom-in region together with the galactic disk, 
we restart the galactic simulation and adjust the adaptive mesh refinement 
within a sub-box enclosing the region of interest, whereas the mesh in 
the rest of the simulation box is allowed to derefine.  
Note that the \textsc{RAMSES} factor of two maximum difference between neighbouring blocks means that the transition to low resolution occurs very gradually away from the sub-box. 
Because the physical structure in the zoom-in region revolves with 
the galaxy,
we impose on the sub-box of refinement an angular speed that is the same as the 
local galactic shear. 

The original mass refinement is $ 8562 \rm{M_{\odot}} $ per cell.  The zoom-in is performed via a restart to initiate the new 3 kpc zoom-in box, which has a mass refinement of $ 93 \rm{M_{\odot}} $.  We de-refine the galactic disk outside of the box down to the 10th level, which provides a spatial resolution of 300 parsecs. The maximum  
reached in the 3 kpc box is $20$ levels of refinement, at which point the usage of Hilbert space-filling curve decomposition in \textsc{RAMSES} becomes necessary \citep{Teyssier2002}.   This refinement 
strategy allows us to follow the zoom-in region with a grid resolution 
of 0.286~pc at a reasonable computational cost. The smallest time steps 
are on the order of a few 10$^1$ yr in physical units.


In this study, we focus on the early formation and evolution of GMCs and their cluster forming clumps; 
hence new star formation and stellar feedback are turned off for the whole simulation upon restart of 
the simulation, where we only run the zoom-in simulation for 10 Myr.
We are performing numerical experiments in these subregions to examine the short-term evolution of turbulent ISM material
prior to star formation.    There is of order $10^{8-9}$ solar masses of pre-existing stars belonging to the disk in these regions that, together with 
the dark matter and gas itself, provide a background potential for a short period (10 Myr) of subsequent gas evolution.

We do not invoke the idea that star formation is negligible or irrelevant on even this short timescale.
Instead, we chose regions where star formation was not currently occurring and evolved them for a short time
at higher resolution to examine their behavior in more detail. These zoom-ins are expensive, and we end them
before too long (10 Myr).  The larger-scale turbulent ISM/disk environment should provide realistic boundary conditions for these
regions on this timescale without excessive dissipation of the turbulence, as we will show.

\begin{figure}
\includegraphics[width=\columnwidth]{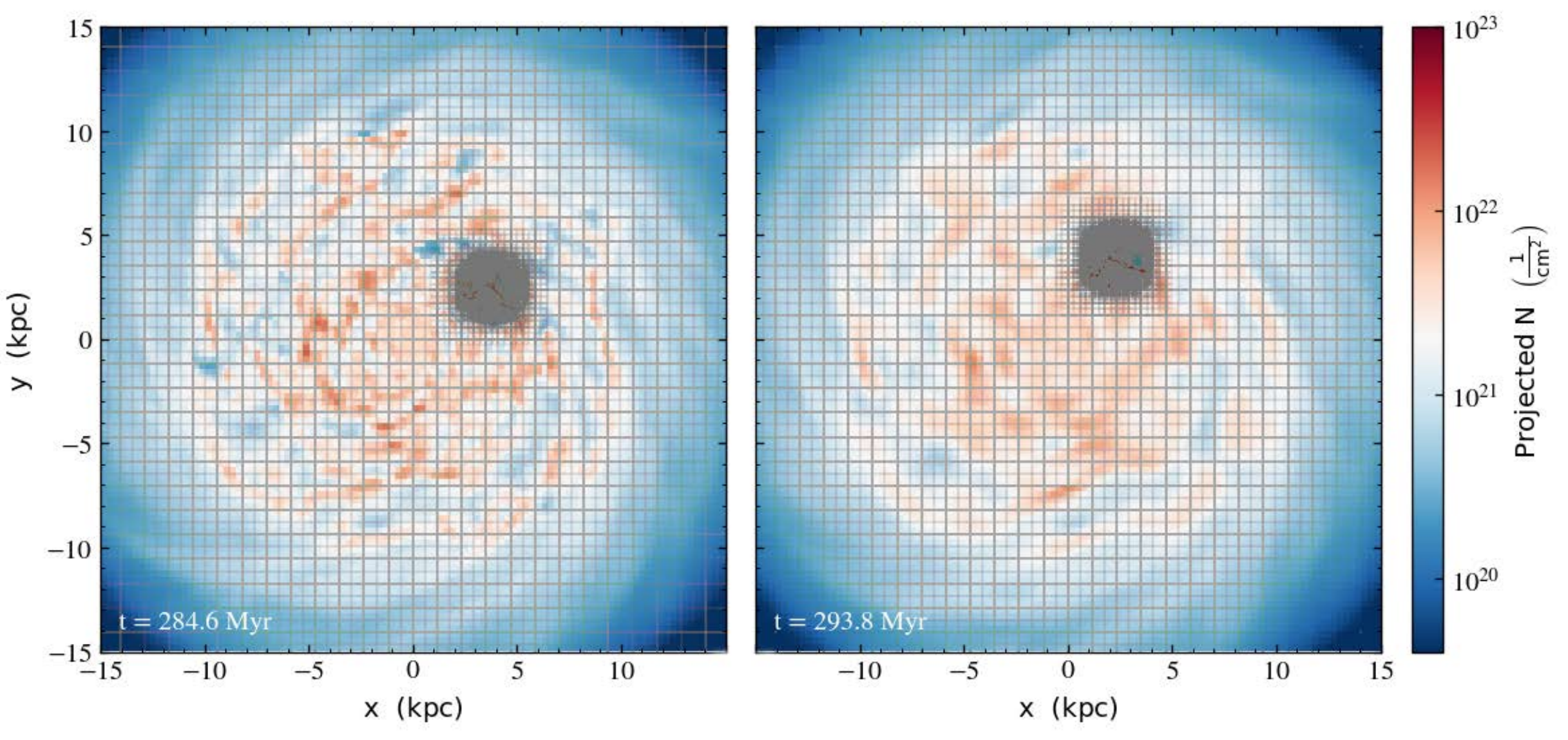}
\caption{The grid mesh layout of the zoom-in region at two different times. The grid mesh of the rest of the simulation box is fixed at lower levels, which also smooth out the structures over time.}
\label{Fig:gridEvol}
\end{figure}

\section{Results:  Galactic Scale Structure}
\label{Chap.ResultsI}

Before delving into the individual zoom-in regions, we first examine 
the overall galactic disk evolution and compare the high density structures 
with recent observations.  

\begin{figure*}
\includegraphics[width=\textwidth]{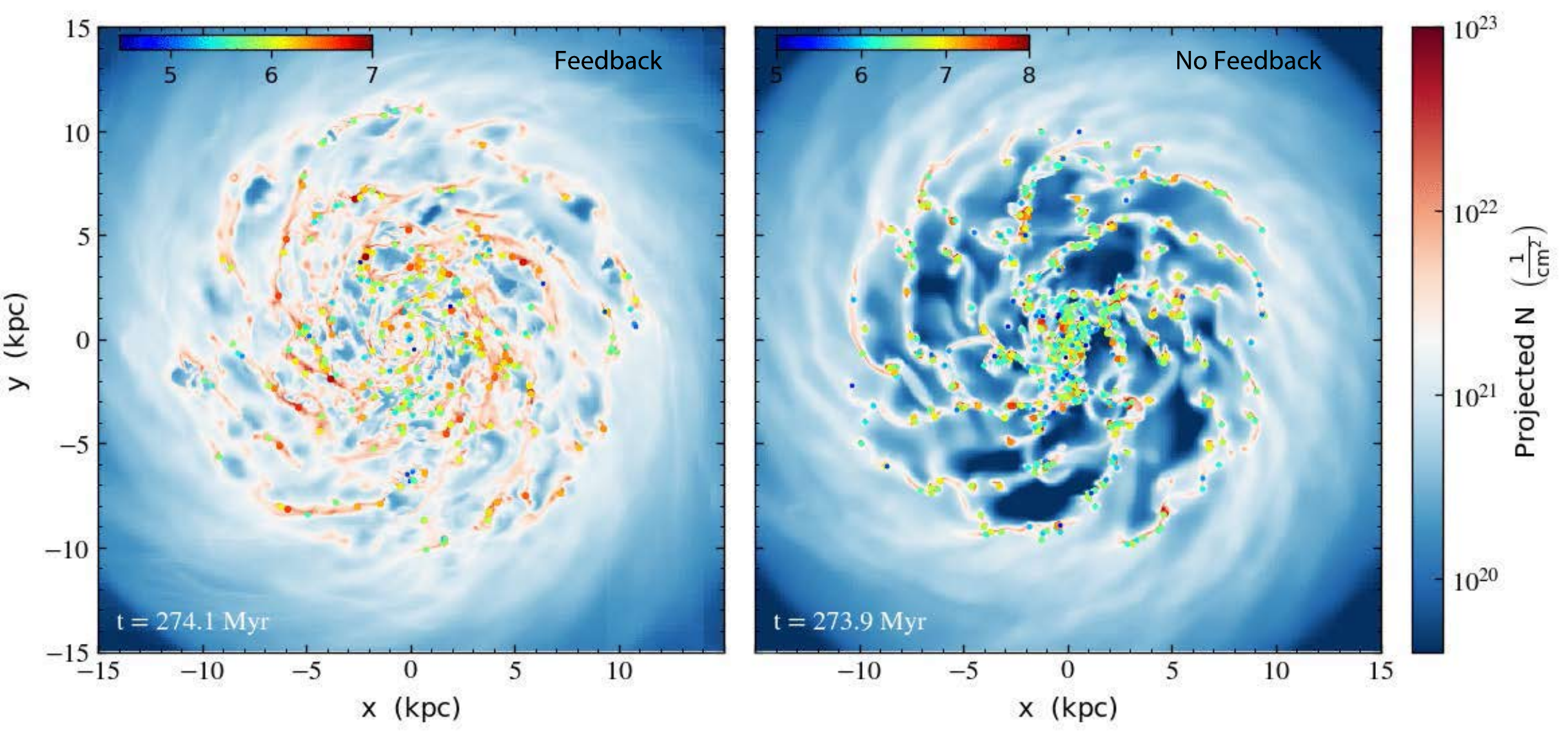}
\caption{Clumps identified within the galactic disk with markers whose colors represent a logarithmic scale of clumps masses ranging from $10^5 - 10^7 $  M$_{\sun}$, indicated in the top color bar. The column density map of the galaxy is shown in the background, with a logarithmic color scale indicated on the sidebar. Note the effectiveness by which gas is swept up and gathered into dense structures in the no feedback case shown in the right panel.  Superbubbles on kpc scales are clearly seen in the feedback simulation in the left panel. }
\label{Fig:clumps}
\end{figure*}

\subsection{Overall Galactic Structure}
In Fig.~\ref{Fig:Proj} we show edge-on and face-on views of the logarithmic distributions of the column density of our simulated galaxy at t=283.7 Myr after the run has started.  The stream lines of the magnetic field are overlaid.  The edge-on view shows the effects of supernova feedback in driving material into the galactic halo which extends out 10-15 kpc from the plane.  Systems of filaments oriented perpendicular to the disk dominate this outflow driven structure inside 10 kpc scales while outside of this disk radius, the gas is mostly quiescent and filaments remain restricted to the plane.  Magnetic field lines are also blown out into filamentary morphologies that track the filaments in many cases.  One also notes a hierarchy of bubble like structures that are the extension of superbubbles in the galactic plane that have expanded out of the galactic disk.  This transition from predominantly vertical to horizontal organization of HI filaments with respect to the galactic plane as one moves radially outward is clearly linked to the extent of SN feedback activity in the disk. These results strongly resemble the overall HI filament structure of the THOR HI survey \cite{Soler+2020, Soler+2022} discussed in the introduction.

The face on view of the galaxy reveals a number of spiral arms but lacks the obvious large scale, 2 arm spiral pattern seen in NGC628 as an example.  There is a range of dense filamentary structures, the longest of which are at least 5 kpc long and may arise in portions of spiral arms.  Others are clearly located at the intersections of large superbubbles which have created a plethora of feedback driven cavities ranging up to a few kpc in size.  This organization of bubble and filament hierarchies is strikingly reminiscent of the recent JWST observations of NGC628 \citep{Watkins+2022}.

The in-plane patterns of the magnetic field lines are also quite varied.  On the outskirts of the disk one sees remnants of the toroidal field pattern in the initial conditions.  However, the field is swept up into bubble walls in the inner 10 kpc of the galaxy where feedback processes are highly active. In some regions, the field is perpendicular to the more massive filaments, but this is not uniformly the case.  Note the variety of orientations of magnetic fields lines with respect to the 
filament axes of the clouds - perpendicular to the dense filaments in many instances but not exclusively so. 

The face on view also shows that there is a wide variety of different local environments in the galaxy.   In the inner regions  we see superbubbles converging on one another with large filaments in the shells that are being actively pressed together.  As we move to the quieter, outer regions of the galaxy, superbubbles are less prominent.  Instead we see segmented spiral arm structure,  more dominated by filaments.  In order to explore the similarities and differences in structure formation in these different regions, we will apply our zoom in mesh at two different locations in galactic radius, chosen to represent an active inner, and a more quiescent outer regions of the galaxy. 

 An illustration of the grid structure is shown in Fig.~\ref{Fig:gridEvol} where we show two snapshots of a co-rotating 3kpc highly refined mesh within the less  refined, larger scale galactic background.

\subsubsection{Behaviour of the ISM}
In Fig.~\ref{Fig:gas_phases} of Appendix~\ref{App.A}, we present the pressure-temperature, and temperature-density phase diagrams for the full ISM of the galaxy at the time t = 283.7 Myr (roughly the time of our restarts), when supernova feedback is fully included. Our temperature-density phase diagram is very similar in structure to the corresponding diagram for the stellar feedback, global RAMSES simulations of the AGORA galaxy model \cite{Kim+2016} (see their Fig. 17) as well as in EMP-Pathfinder cosmological zoom-in  simulations of the Milky Way \citep{Reina-Campos+2022}. As in these latter simulations, we see a good deal of low density, hot gas that is ejected from the disk due to the supernova feedback.

Fig.~\ref{Fig:magfield_density} in Appendix A shows the relation between the mass weighted magnetic field and gas density in the galaxy as a whole.  This Figure shows that the final field strength obtained in these simulations is about $ 10 \mu \rm{G}$ in the bulk of the ISM - clearly indicating that dynamo action has strongly amplified the initially weak field.   We overlay $B \propto \rho^{1/2}$ and $B \propto \rho^{2/3}$ (red and black lines, respectively) on these plots as a guide to the numerical trends.   We see that the relation is in rough agreement with the 2/3 power law for bulk of the magnetic field data above a density of $10^{-23}$ g cm$^{-3}$, as in the observational $ B - \rho$ relation of \citet{Crutcher2019} at higher densities. Note that while the initial conditions of our model had a very weak initial magnetic field with this scaling, this would have been quickly overwritten by the turbulent dynamo which has amplified the initial field by 2 orders of magnitude.  However, as lower densities, we find that there is a down turn to a shallower power law as compared to the $B \simeq const$ result given in \citet{Crutcher2019}.  We note that the observations have very large error bars below this density (see also \citet{Robinson_Wadsley2023}). We investigate this in greater detail in another forthcoming paper.

\subsection{GMCs in the Galactic Disk}
\label{S.Clumps}

We use the \textsc{Clumpfind} method in \textsc{RAMSES} that identifies
local 3D structures with number densities above 10~cm$^{-3}$. The method 
is based on \textsc{PHEW} (Parallel HiErarchical Watershed) algorithm 
implemented by \citet{Bleuler+2015}, which can be broken down into 
four main steps: (1) Watershed segmentation (from image processing 
applications), (2) Saddle point search, (3) Noise removal, and 
(4) Substructure merging. The merging process results in a tree-like 
representation of substructures similar to the dendrogram method
\citep[e.g.,][]{Rosolowsky+2008}. We refer readers to \citet{Bleuler+2015} 
for more details on the algorithm.

In Fig.~\ref{Fig:clumps} we show the spatial distributions of the clumps whose positions are marked with filled circles whose colors code for the total mass of the clump.
The most prominent difference between the models with and without 
feedback is that the density contrast in the galactic structure is 
larger in the model without feedback, with most of the mass concentrated 
on the the spiral and filamentary structures.  In the model with 
feedback on the other hand, the galactic structure shows more even distribution of materials - feedback
prevents the large concentration of mass into the arms and filaments that would otherwise occur.
As a result, the clumps generally have more mass in the model without 
feedback.  

The feedback snapshot in the left panel shows that some of the spiral arm features towards the periphery are populated with 
GMCs that have a fairly regular spacing - as an example the nearly 6 kpc spiral arm segment in the lower left at ( x: - 7 to -1 kpc ; y: -8 kpc) which 
has 5 GMC mass clouds spaced by 1 kpc.  The interior regions ( inside 5 kpc) of the galaxy are relatively more impacted with many superbubble structures
although there still are some more quiescent regions.  These are clearly signatures of gravitational instability which we investigate thoroughly in our zoom-ins.

\begin{figure*}
\includegraphics[width=\textwidth]{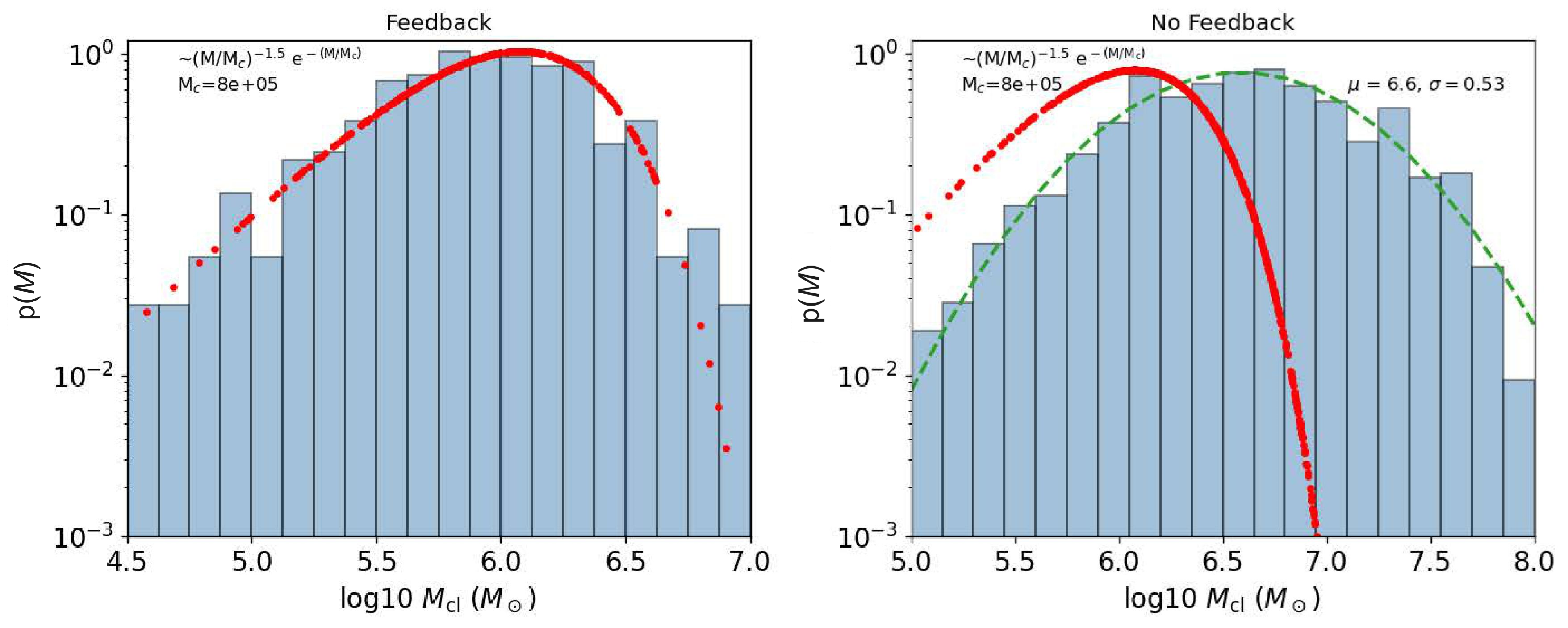}
\caption{Probability distribution functions of clump mass for the models with feedback (left) 
and without feedback (right). The same Schechter function is shown in red curve
in both panels, while the log normal fit is shown in the right panel.}
\label{Fig:clumps_fit}
\end{figure*}
  
Recent observations of cloud populations in nearby galaxies\citep{Rosolowsky+2021} have been fit by an empirical Schechter function 
\citep{Schechter1976} which has the functional form of a power-law with an exponential cut-off above some mass;  

\begin{equation}
p(M) = C*\frac{M}{M_c}^{\beta} exp(-\frac{M}{M_c})~,
\end{equation}
where $C$ is a normalization coefficient.

This probability distribution function was used to  fit the GMC populations consisting  of a total of nearly 5000 GMCs in ten different 
nearby galaxies in the PHANGS survey \citep{Rosolowsky+2021}.  These observational fits have cut off masses of galactic GMC distributions in the range $ M_c: 10^6 - 10^7 M_{\sun} $, with power law indices in the range $ \beta: (- 0.6)-  (- 3.4) $.  These parameters vary for GMC populations from galaxy to galaxy, as well as for GMCs situated in different regions of a galaxy. 

In Fig.~\ref{Fig:clumps_fit} we plot the probability density distribution $p(M)$ of clumps from the whole galaxy, comparing two hydrodynamic models with and 
without stellar feedback.
In the right panel (without stellar feedback),  the distribution of clump mass 
can be fit by a log-normal distribution with mean of 10$^{6.6}$~M$_{\sun}$, 
and standard deviation of 0.53 dex.  Lognormal distributions are always expected in multiplicative processes 
such as occur in supersonic turbulence where a parcel of gas is repeatedly compressed by 
passing shock waves.  Many simulations show that lognormal distributions of the gas density arise in supersonic shocked media \citep{Vazquez-Semadeni1994,Nordlund_Padoan1999, Kritsuk+2007,Pudritz_Kevlahan2013}.  In the high resolution simulations these fits are excellent over 8 decades in density \citep{Kritsuk+2007}.
 
 In contrast, we find that the presence of feedback 
skews  GMC mass distribution away from a log-normal. Thus in the left panel of Fig.~\ref{Fig:clumps_fit}, we a Schecther function fits our data well, with a cutoff mass of $M_c = 0.8 \times 10^6 M_{\sun} $, and a power-law index of $\beta = - 1.5 $. 
Our simulation has a mass cutoff near the bottom of the range observed in the PHANGS survey, and a power law index that is near the average for the PHANGS survey.  

The key result here is that the removal of massive GMCs from the mass spectrum by supernova feedback is responsible for the cut off 
in the Schechter function.  Without this feedback, cloud populations grow to over an order of magnitude greater in characteristic mass, as indicated in our lognormal results.  The feedback process is more extreme for the most massive clouds as they form more massive clusters with large OB star  populations \citep{Howard+2018}.  We note that we don't have UV or wind feedback associated with ours.  We suggest that the origin of the cutoffs seen in the PHANGS GMCs distributions has a similar origin. 


\section{Results: Zoom-ins of Cloud Forming Regions}
\label{Chap.ResultsII}

Having validated the galactic structures in our simulations against observed galactic structures, we now present the main 
results of the zoom-in regions. As already noted,  the formation and evolution 
of GMCs inside a galactic disk may be affected by the different environments in which they form.  This motivates us to choose two distinct regions 
to demonstrate how the galactic environment may impact the morphologies 
and properties of GMCs.  

\begin{figure*}
\includegraphics[width=\columnwidth]{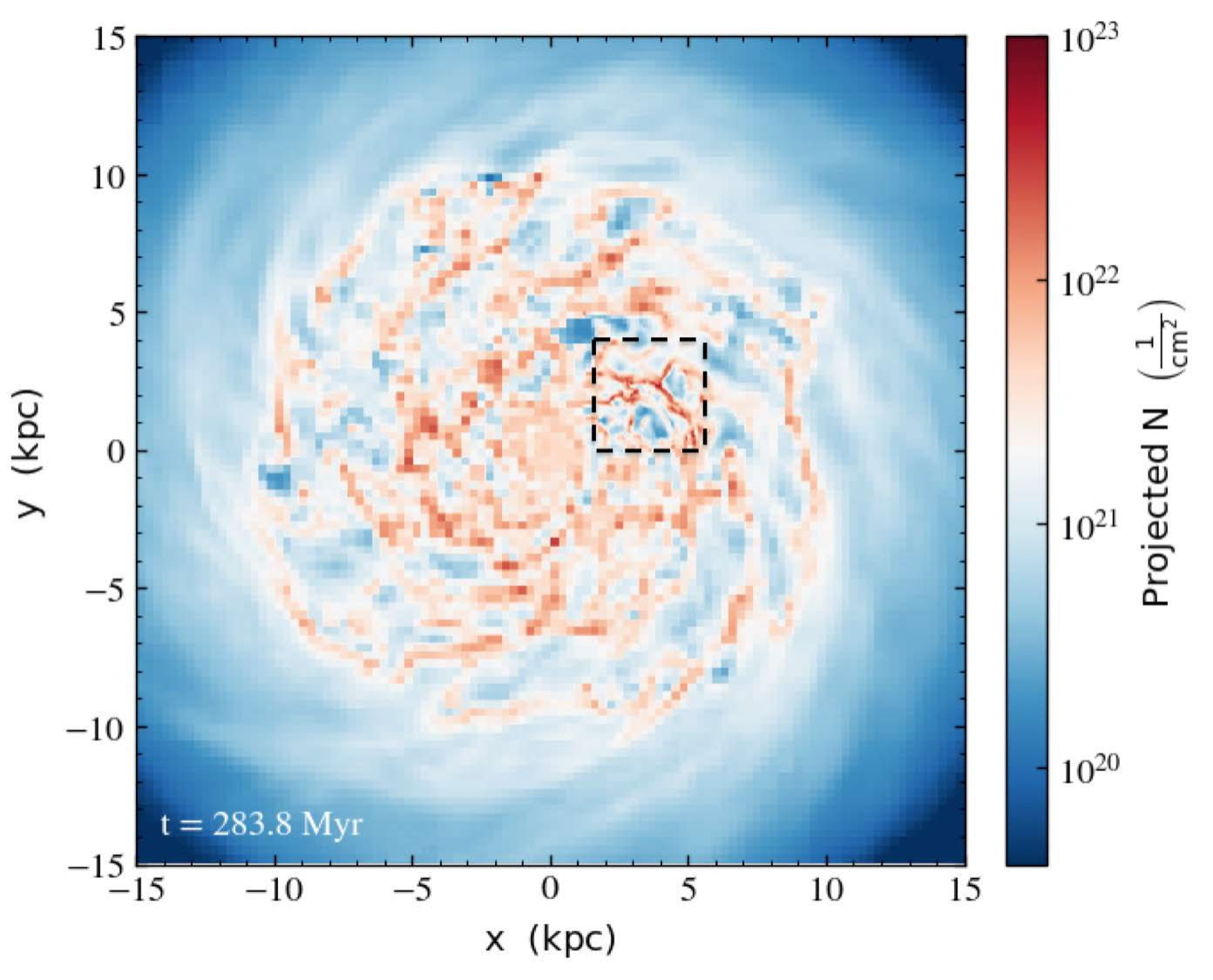}
\includegraphics[width=\columnwidth]{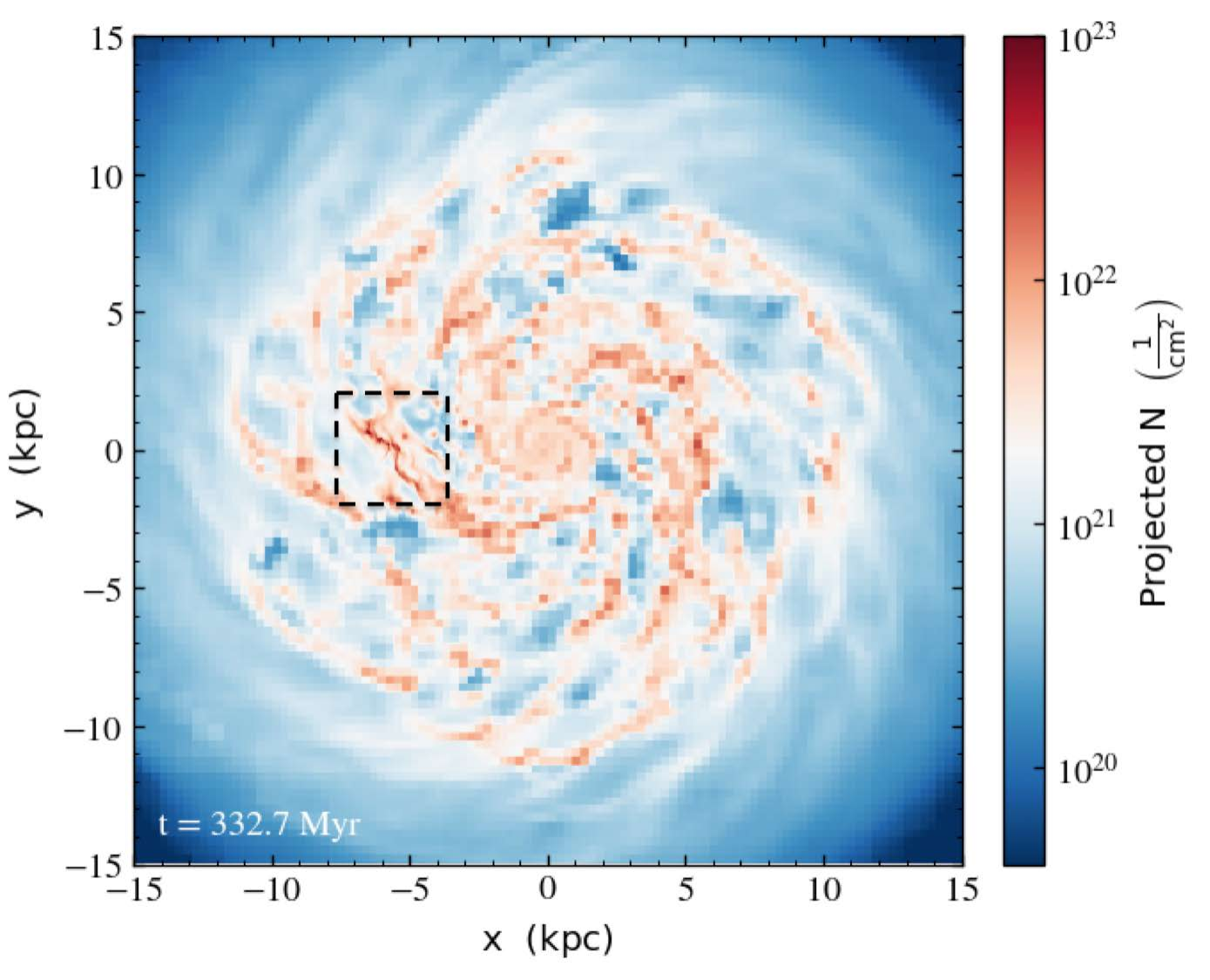}
\caption{Zoom in regions superimposed on the galaxy with feedback, as boxes.  The left panel shows the active region (more affected by feedback) centered at a galactic radius of 4 kpc, and the right panel shows the quiescent region centered at 6 kpc.  Note also the different times. }
\label{Fig:zoom_regions}
\end{figure*}

Fig.~\ref{Fig:zoom_regions} presents two, co-rotating, 3 kpc regions.  The left panel shows a zoom in region whose center is approximately 4 kpc from the galactic center. We chose our 10 Myr evolution to start at 286 Myr.  This region is surrounded by several
active feedback ``bubbles'' which are very common in the inner galaxy.  

The right panel of the Figure shows our 3 kpc frame located further out, at 6 kpc from the galactic center.  Superbubbles are still present but this is a more quiescent region, less affected by feedback structures.  The major filament here is more a consequence of spiral wave action.  We start our 10 Myr zoom in this region a bit later, at 332 Myr noting that structure further out in the galaxy takes somewhat longer to establish itself. 

\subsection{Connection with large scales: Mass Reservoir and Turbulence 
Properties}
\label{S.MassRes}

As compared to simulations in isolated boxes, where artificial boundary 
conditions are imposed to mitigate flows at the box boundaries, our 
set-up connects with the galactic scale and allows a natural flow of 
materials in and out of the zoom-in regions. 

\begin{figure}

\includegraphics[width=\columnwidth]{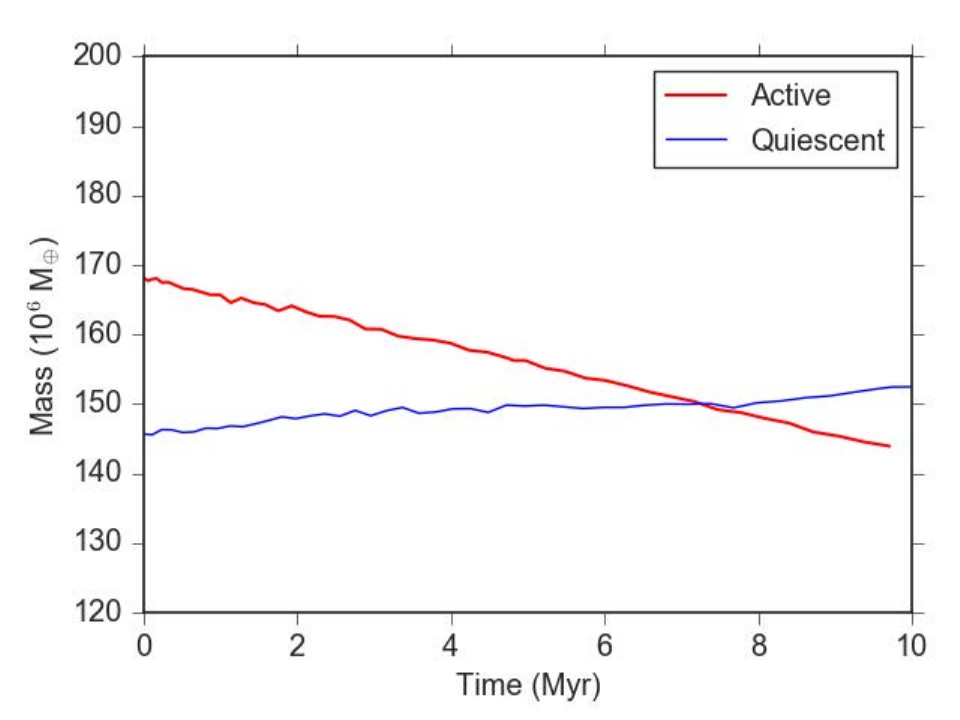}
\caption{Evolution of mass reservoir in the 3~kpc zoomed-in regions.}
\label{Fig:mass_evol}
\end{figure}

In Fig.~\ref{Fig:mass_evol},
we examine the time evolution of the mass reservoir in both the active and 
the quiescent regions. The active region is losing mass at a rate of 
$\sim$2.5~M$_{\sun}$~yr$^{-1}$ due to the expansion of the feedback bubbles. 
On the other hand, the quiescent region is gaining mass from the galactic 
environment at a rate of $\sim$0.6--0.7~M$_{\sun}$~yr$^{-1}$. Over the 
course of 10~Myr, the active region has lost approximately 
2.5$\times$10$^7$~M$_{\sun}$ 
in total, whereas the quiescent region has gained a sum of about 
7--8$\times$10$^6$~M$_{\sun}$. 
The significant loss of the mass reservoir of the 
active region is a result of the expansion of the feedback bubble in that 
region. 

In Appendix~\ref{App.B} (see Fig.~\ref{Fig:power_spec})  we calculate and present the energy $E(k)$ and power spectra $P(k)$ of supersonic turbulence in our two regions. 
These spectra are related as  $E(k)=4 \pi k^2 P(k)$.   We first transform the 3D velocity components to  
Fourier space and analyze the power spectrum for each region\footnote{The Fourier analysis requires 
a uniform grid, onto which we have interpolated the AMR data with a smallest 
cell of $ \simeq 25 \rm{pc} $}.   We find that at the 3~kpc scale, the energy spectrum of the three velocity components 
largely follow the Burgers power law: $ E(k) \propto k^{-2}$, down to 30~pc.  Its spectrum is a consequence of the spatial jump in velocity (step function) that characterizes shocks.  Specifically; 
$  \sigma_{nt}^2(k)= \int E(k) dk \propto k^{-1} $ in Fourier space, which then gives a size-line width relation  $ \sigma_{nt}(L)\propto L^{0.5}$.  This is a different index than that of \cite{Larson1981} for which  $ \sigma_{nt}(L)\propto L^{0.37}$ on scales 0.1 - 1 kpc.   The Burgers spectrum, which is slightly steeper than the Kolmogorov profile ($ E(k) \propto k^{-5/3}$), is especially apparent at the 10 Myr time point where the turbulence has had a chance to settle into a more 
steady state. 

High resolution simulations (with no gravity) show that the velocity power spectrum found in the case of highly compressive turbulence is close to the Burgers scaling while mild, transonic turbulence is closer to Kolomogorov \citep{Kritsuk+2007}.  
Our turbulence results are consistent with the very strong shocks that are occurring in our simulations, primarily due to supernova feedback.

It is difficult to disentangle the relative contributions of feedback and/or gravity 
to turbulence generation in the simulation.  The comparison of these two regions shows 
that stronger feedback in the active region drives turbulence amplitudes to higher values.   Our simulations without any feedback also produce power law
spectra which might be more prevalent in the quiescent region. Thus gravitationally driven motions, such as the shocks produced by our broken 
spiral patterns, also contribute.     

\subsection{Active Region: Expansion of Feedback Bubbles}
\label{S.Active}

\begin{figure}
\includegraphics[width=\columnwidth]{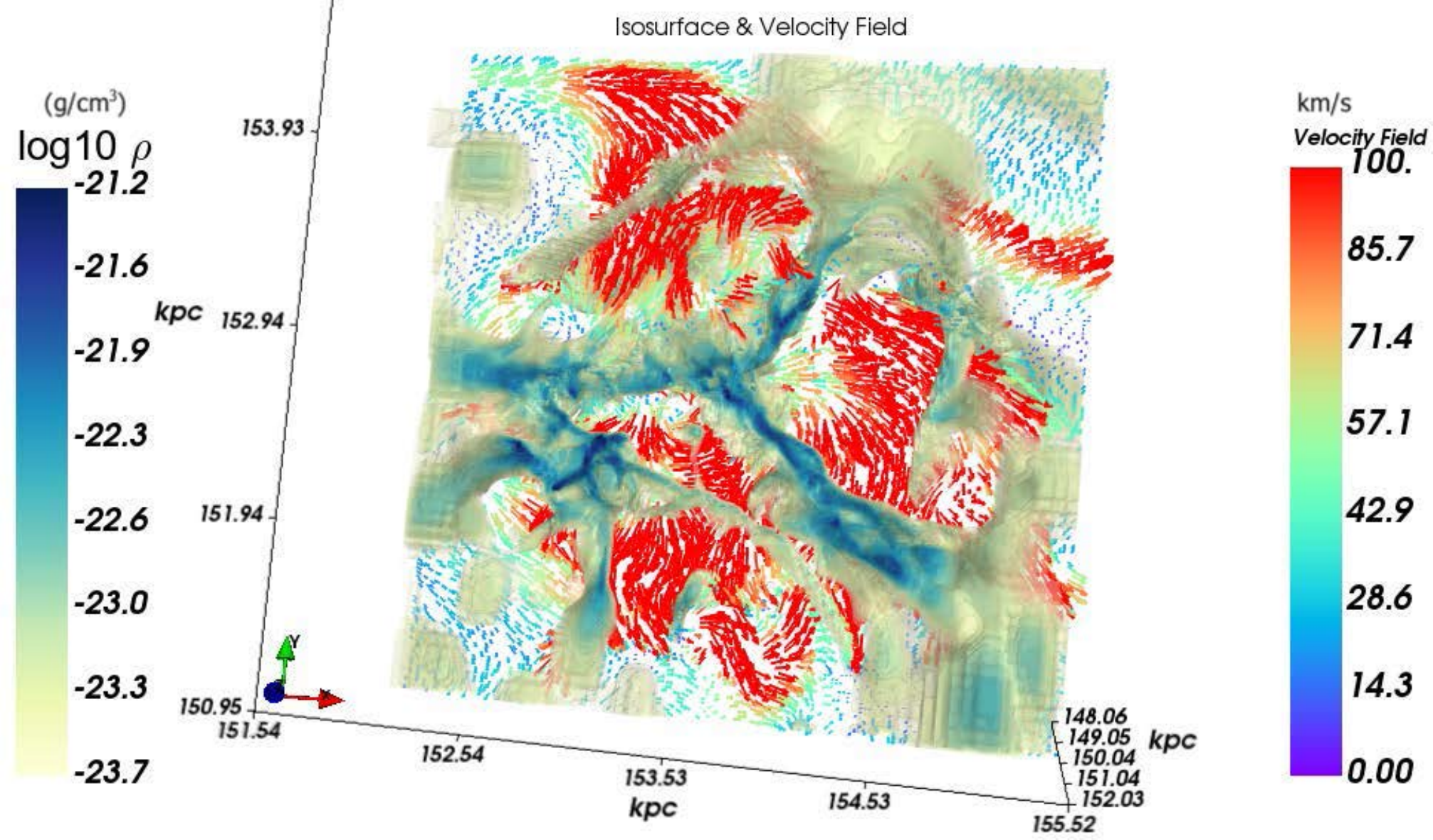}
\caption{Density isosurface and velocity field across the atomic filamentary 
structures in the large 3 kpc active region. The velocity vectors 
are drawn along x-y, parallel to the galactic plane.}
\label{Fig:active_0083}
\end{figure}

 Fig.~\ref{Fig:active_0083} shows the density and 
velocity distributions at 3.2~Myr in the active region after the restart of the zoom-in region. 
The active zoom-in region is composed of three  large atomic filamentary structures, each a
few kpc in length.  They are joined together at the interface of three previous feedback bubbles - expanding and squeezing 
all three of the filaments.   Along the three main filaments, the gas density is around 10~cm$^{-3}$, 
with only the few densest sites reaching molecular density. The velocity 
fields are pointing towards the top and bottom right filaments. We also note that the longest of these three filaments has a distinct, wave-like oscillation in the gas density and accompanying molecular cloud spatial distribution with a wavelength that is about 1 kpc.  This is reminiscent of a Radcliffe wave identified in the Milky Way \citep{Konietzka+2024}.

In Fig. 8 we show both the magnetic structure (upper panel) as well as details of the column density of these filaments (lower panel). The magnetic field line structure with respect to the filaments shows a number of interesting features: in   places field lines are parallel to the filaments -  an indication of fields that have been pushed into the walls of the superbubbles and the filaments that form there.  There are also other regions along the filament with field perpendicular to the major filament. Yet a third feature is where field enters perpendicular to a filament but then it is bent parallel to the filament axis in the interior.  This is evidence of filament aligned flow that drags the field lines in the dense filament along into the cluster forming clump within it \citep{Klassen+2017}  

\begin{figure}
\centering
\begin{subfigure}{\columnwidth}
    \includegraphics[width=\columnwidth]{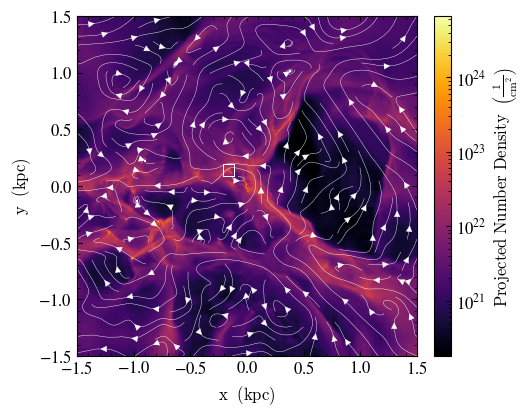}
    \caption{}
    \label{Fig:active_projection}
\end{subfigure}
\hfill
\begin{subfigure}{\columnwidth}
    \includegraphics[width=\columnwidth]{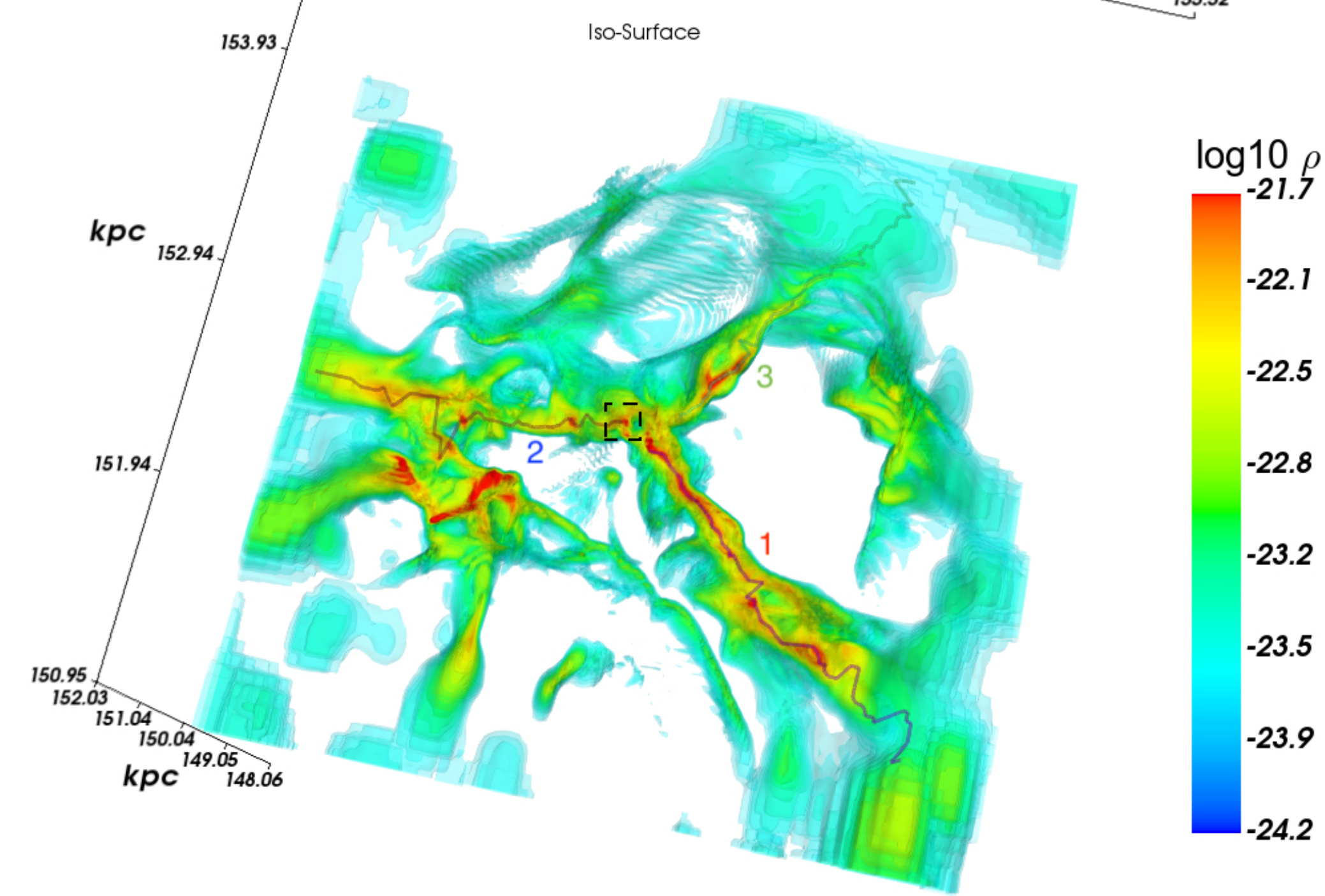}
    \caption{}
    \label{Fig:traceFine_large}
\end{subfigure}
\caption{a) Density projection of the large 3~kpc active region at $t=3.2$~Myr after the restart of the simulation. White streamlines trace magnetic field, while the white box indicates the region of interest for further zooms. b) Filaments traced in the same large 3~kpc active region. Refer to text for discussion of filaments labelled 1-3.}
\label{Fig:projection_trace}
\end{figure}

The
filaments have significant density variations along them that while not strictly periodic, are regularly 
spaced. In particular, the density peaks along Filament 1 (as labelled in Fig.~\ref{Fig:traceFine_large}) have a variable spacing of $ \lambda_{frag} \simeq 200 pc $.  We turn to investigate this spacing in \S5.  

To address the stability of these atomic filaments, we have developed 
a method to trace the dense structures along the filaments in 3D space 
(see Appendix~\ref{App.B} for details). As shown in 
Fig.~\ref{Fig:traceFine_large}, each atomic filament has a length of 
about $1-2~kpc$, and as already noted, the three filaments connect at a central 
``hub''-like location. Because these filaments traced here are linked 
by a series of discrete line segments, we can draw circular cross sections 
at each point along the filament and thus examine the density distribution 
in each cross section. It is also straightforward to derive the mass per 
unit length, which can be compared with the general stability condition:
${2 \sigma^2 / G}$, where $\sigma$ is the velocity dispersion 
 for filaments supported by turbulent motions \ct{Andre+2014, Fiege_Pudritz2000a}.

In the left panel of  Fig.~\ref{Fig:lineMass_large}, we plot the measured line mass (solid lines) for each of these three filaments, and 
compare them to the critical lines masses for gravitational instability (shown 
in three different colours in the dashed lines).   The latter quantities are computed from the 
measured velocity dispersions $\sigma$  in the filaments.  
The radial density structure of the filaments are shown in the three right panels of the figure.  

The bulk segments of filament 1 and 2 
have reached or exceeded their respective stability criteria which 
is consistent with the onset of multiple molecular condensations along 
these two filaments.   The peak values of the unstable line masses exceed 
$m_{crit} \simeq 10^4 M_{\odot}$ pc$^{-1}$ over regions 10s of  pc,  so that these fragments are
in the mass range, size, and density of molecular clouds.   Other regions lie
below their critical thresholds at this point in time and so would be 
expected to be loosely bound.  The low line mass for the segment of Filament 3 
corresponds to the low density end shown in Fig.~\ref{Fig:traceFine_large};
only the inner 0.5~kpc segment of Filament 3 close to the central ``hub'' 
reaches a quasi-stable state. 

One further question arises as to whether supernova expansion is driving the evolution of filaments past criticality, or whether pure radial gravitational collapse is occurring.  The radial collapse of filaments has been studied by a number of authors; \citet{Tilley_Pudritz2003} analyzed the self-similar collapse of infinite, idealized filaments, while \citet{Clarke_Whitworth2015} combined analytical and numerical simulations of finite systems with aspect ratios $A_o$.  In the former case, the self-similar time scale is $t_{col}=(\pi G \rho_c)^{-1/2} \simeq 12.0 \rm{Myr}$ for our massive atomic gas filament with densities approaching 10 cm$^{-3}$.  For the latter more physically realistic case of a finite length cylinder, $t_{col}= (0.49 + 0.26 A_o)(G \rho_c)^{-1/2}$ which for long filaments, $A_o = 10$, has a collapse time scale that is 5.48 times longer. These are both longer time scales than we see in the dynamics of our filament; it is superbubble expansion that ultimately is driving filaments into the critical condition. 

\subsubsection{Radial density profiles of filaments}

\begin{figure*}
\includegraphics[width=\textwidth]{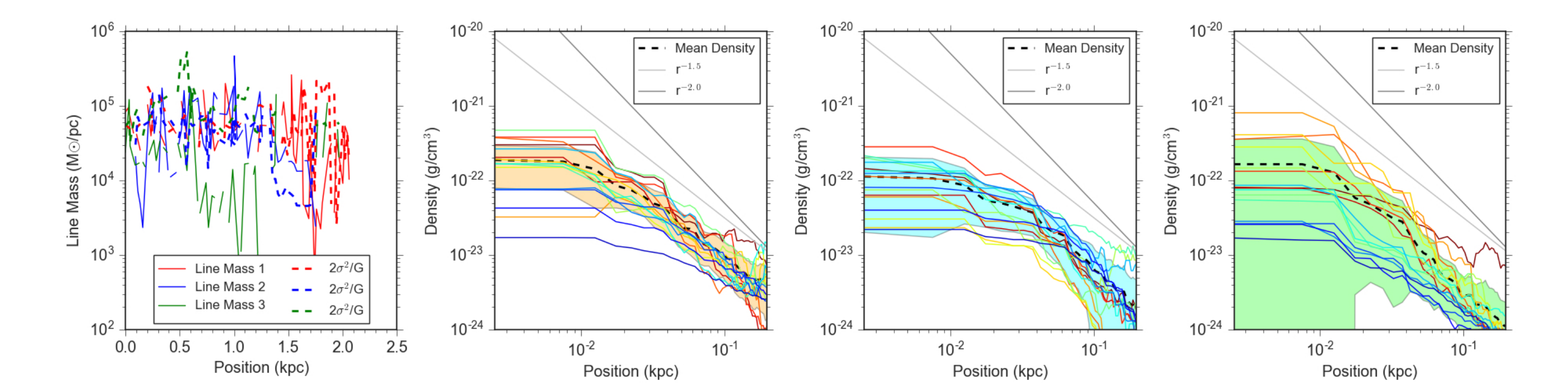}
\caption{Line mass along the three filaments (first panel) in the large 3 kpc active region
Fig.~\ref{Fig:traceFine_large} and the 
radial density profiles at the sampling locations (multi-colored lines) 
for filament 1 (second panel), 2 (third panel), and 3 (fourth panel), for 
the time frame at 3.2~Myr after the zoom-in restart. 
The shaded region in the density profile represents the standard deviation 
of the density spread.}
\label{Fig:lineMass_large}
\end{figure*}

In the right 3 panels of Fig.~\ref{Fig:lineMass_large}, we plot a number of the radial density profiles along the three filaments in our active region.  These are then averaged for each filament.   The radial profiles of filaments depends to some degree on the equation of state of the gas, magnetic fields, and dynamical processes.  For reference, the \citet{Ostriker1964, Stodolkiewicz1963} solution for an equilibrium, self gravitating cylinder of isothermal gas provides an excellent benchmark against which to compare the effects of non-thermal, dynamic, or other processes:

\begin{equation}
    \rho(r) = \rho_o \frac{1}{\left ( 1 + (r^2/8r_o^2 \right )^{p/2}}
\end{equation}
where $ p = 4$ is a power law distribution of gas in the envelope of the filament and $r_o^2 = c_s^2 / 4 \pi G \rho_o$ relates the natural core radius $r_o$ of a self gravitating system too its core density of $\rho_o$.  
Other
equilibrium models have shallower profiles for different physical effects: magnetic fields \citep{Fiege_Pudritz2000a, Kirk+2015, Tomisaka2014, Kashiwagi_Tomisaka2021}, external pressure \citep{Fiege_Pudritz2000a, Fischera_Martin2012}, polytropic equations of state \citep{Gehman+1996}, and rotation \citep{Recchi+2014}.

Herschel observations of thermal filaments in nearby molecular clouds show that the related Plummer profiles, with the definition 

\begin{equation}
R_{flat}^2 = 8 r_o^2 = 2 c_s^2 / \pi G \rho_o, 
\end{equation}
fit observations best for 
power law indices 
$ p \simeq 1.5 - 2.5$ \citep{Arzoumanian+2011, Palmeirim+2013, Hacar+2022}.  

As can be seen in Fig.~\ref{Fig:lineMass_large}, the highly dynamical
filaments forming in our active region have profiles lie in the range, with $ p \simeq 1.5 - 2.0$.   These overlap with observations of the Maggie filament reasonably well.  For the kpc atomic gas filaments profiled in Fig.~\ref{Fig:lineMass_large}, we can read off the flat part of the radial density plot for the kpc atomic gas filaments, giving $R_{flat,a} \simeq 10 - 15 pc $.  

Additionally, there appears to be a correlation between the stability of each filament and 
the density distribution across it.  For the densest filament 1, a profile of 
$r^{-1.5}$ best fits the density drop off in its cross section. For filament 
3, the densities of the cross-section at most sampling points are relatively 
low, with the outer radii matching a $r^{-2.0}$ profile. Thus, the density 
spread in the cross-sections of the filaments is the smallest for filament 1, 
and the largest for filament 3. For comparison, 
the density profiles in the cross-sections of filament 2 lies somewhat 
in between $r^{-1.5}$ and $r^{-2.0}$.  We note however, that 
the plateau in the inner radius is an upper limit - the result of the limited image 
resolution ($\delta x$$\lesssim$25~pc) after extracting the AMR grid 
onto a uniform grid.

\subsubsection{Filaments at 200pc scales}

\begin{figure}
\centering
\begin{subfigure}{\columnwidth}
    \includegraphics[width=\columnwidth]{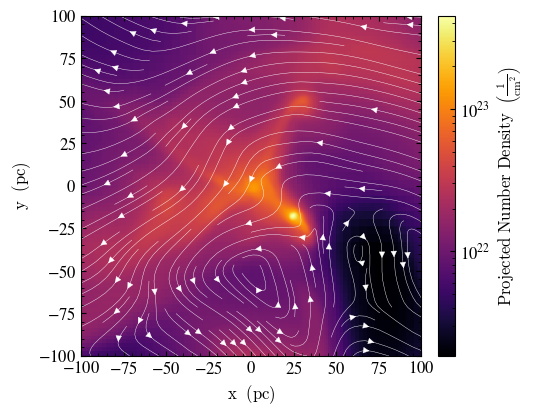}
    \caption{}
    \label{Fig:zoom_projection}
\end{subfigure}
\hfill
\begin{subfigure}{\columnwidth}
    \includegraphics[width=\columnwidth]{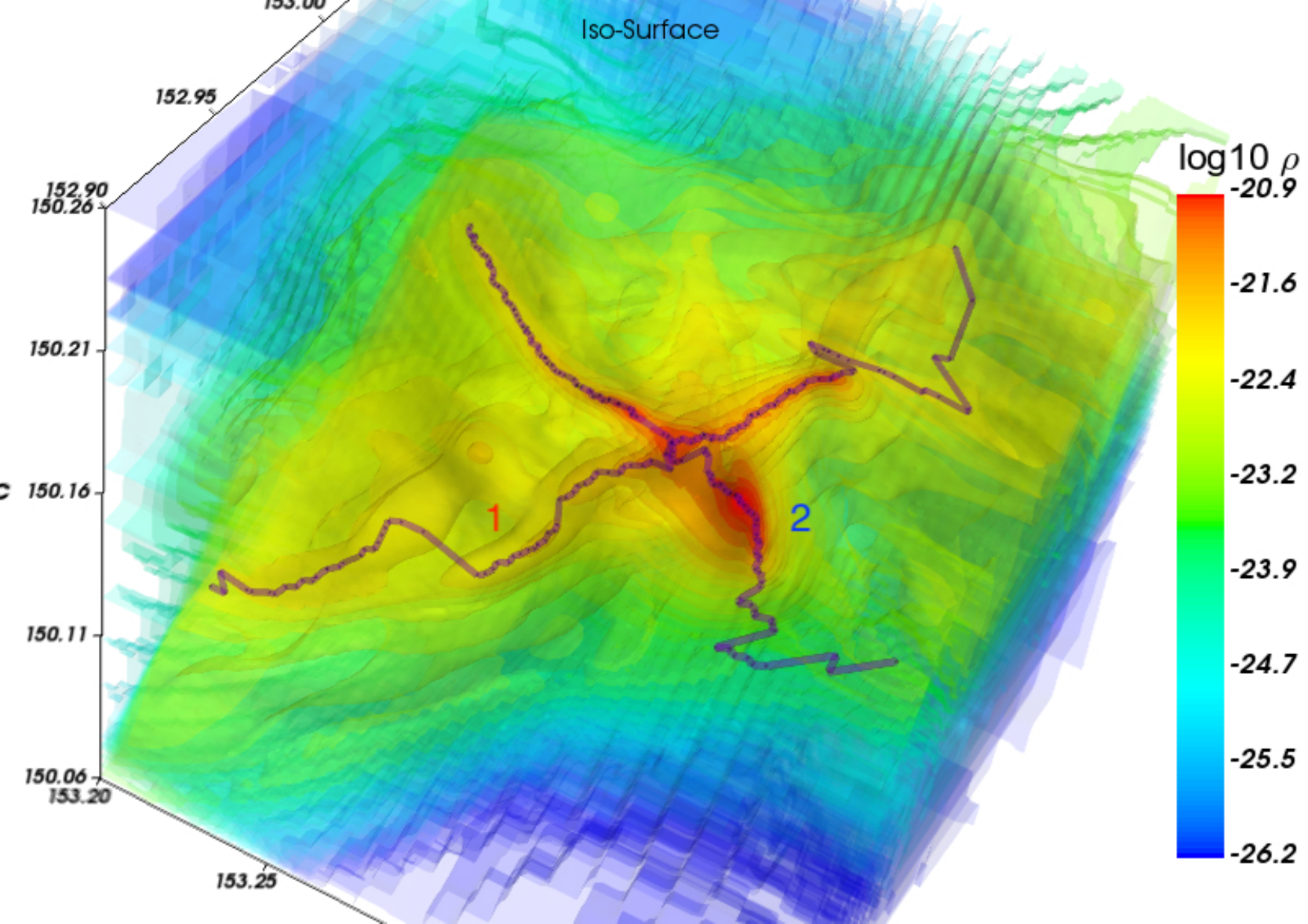}
    \caption{}
    \label{Fig:traceFine_small}
\end{subfigure}
\caption{a) Density projection within a 200~pc active subregion at $t=3.2$~Myr after restart of the simulation. White streamlines show magnetic field morphology. b) Filament trace with the same region.}
\label{Fig:zoomprojection_trace}
\end{figure}

As our numerical resolution is much higher ( $ 0.286  $ pc ), in Fig.~\ref{Fig:traceFine_small} we delve into even 
smaller scales by enlarging the 200~pc region at the tip of filament 2, near the 
hub or junction of the three filaments
(black squared region in Fig.~\ref{Fig:traceFine_large}) and applying 
the same method of filaments tracing. This particular region is still 
in an early stage of evolution, with only the central elongated structures 
reaching a density above 100~cm$^{-3}$. Using our filament tracing method, 
we have identified two filamentary structures in this region. 

\begin{figure*}

\includegraphics[width=\textwidth]{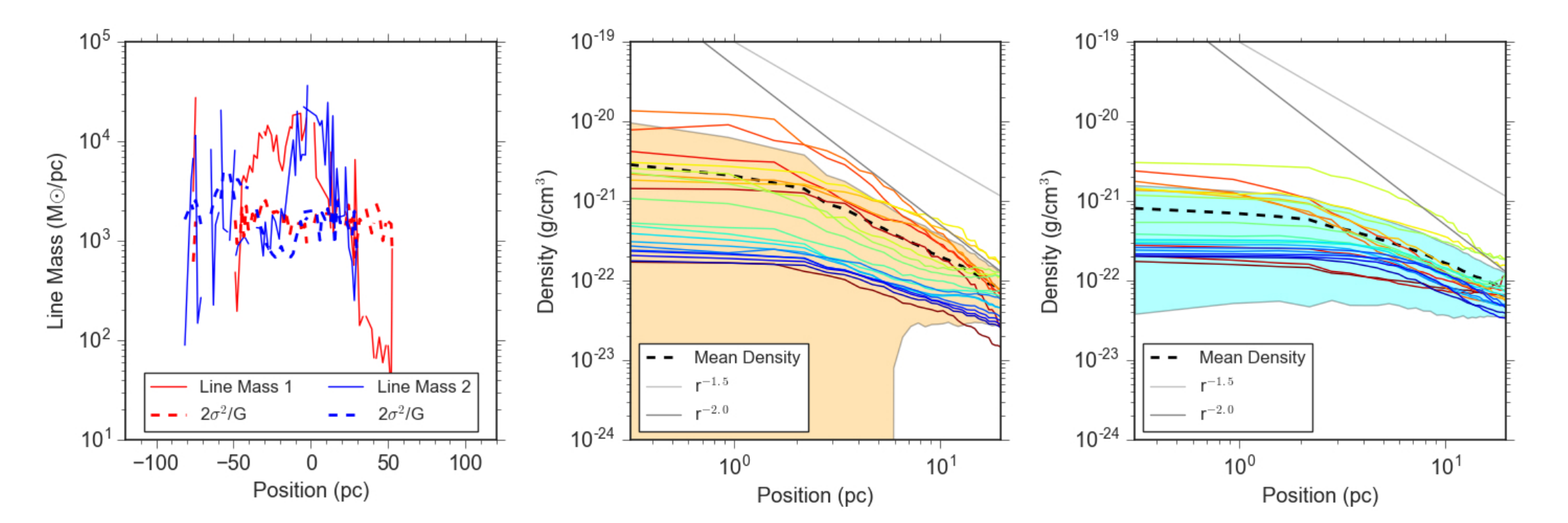}
\caption{Line mass along the two filaments (first panel) within the 200 pc active subregion
Fig.~\ref{Fig:traceFine_small}; and the 
radial density profiles at the sampling locations (multi-colored lines) 
for filament 1 (second panel) and 2 (third panel), for 
the time frame at 3.2~Myr after the zoom-in restart. 
The shaded region in the density profile represents the standard deviation 
of the density spread.}
\label{Fig:lineMass_small}
\end{figure*}

In Fig.~\ref{Fig:lineMass_small} we examine   
individual filaments that have the properties of a GMC, 50-100 pc in length.  In the panel on the left, we see that Filament 2 is mostly unstable 
by the stability criterion ${2 \sigma^2 / G}$, while Filament 1 is less so except for the densest segments of the filaments.  The peak
values of line masses are of the $m_{peak} \simeq 10^4 M_{\odot}$ pc$^{-1}$ with critical
values being an order of magnitude less at  $m_{crit} \simeq 10^3 M_{\odot}$ pc$^{-1}$. Given the large difference between these values, we surmise that these clumps have entered a nonlinear regime fed by filamentary flow. It is important to note that in our highly dynamical simulations, the initial regular fragmentation predicted by the equilibrium theory of smooth, infinite cylinders is at best a rough guide and deviations from strict periodic spacing are expected in this regime.

The density 
profiles of the two filaments are shown in the right hand panels of Fig.~\ref{Fig:lineMass_small}.  Here, the power-law fits to the envelopes of each filament are more consistent 
with the r$^{-1.5}$ power law. Because Filament 2 contains more 
higher density segments than Filament 1, 
the spread in the density profile is also narrower within it. 
We can read off the flattening radius for theses GMC filaments from  Fig.~\ref{Fig:lineMass_small} and find  $ R_{flat, GMC} \simeq 1~pc $.  We note that our numerical resolution starts to flatten the density profile as well, so that this should be taken as an upper limit to the core radius at the GMC scale.


Comparing the flattening radii of kpc filaments with those of the 100 pc GMC filaments, we note that they are quite different in value.  However, they scale with the filament length;  $ L_{a} / R_{flat, a} \sim  L_{GMC} / R_{flat, GMC} \simeq 100 $.  The implied linear scaling $R_{flat} \propto L$ has been observed for real filaments \citep{Hacar+2022}, however, we caution that as with the observations, higher resolution simulations studies are needed.     

\subsubsection{Time evolution of active filaments in 3kpc region}

\begin{figure}
\includegraphics[width=\columnwidth]{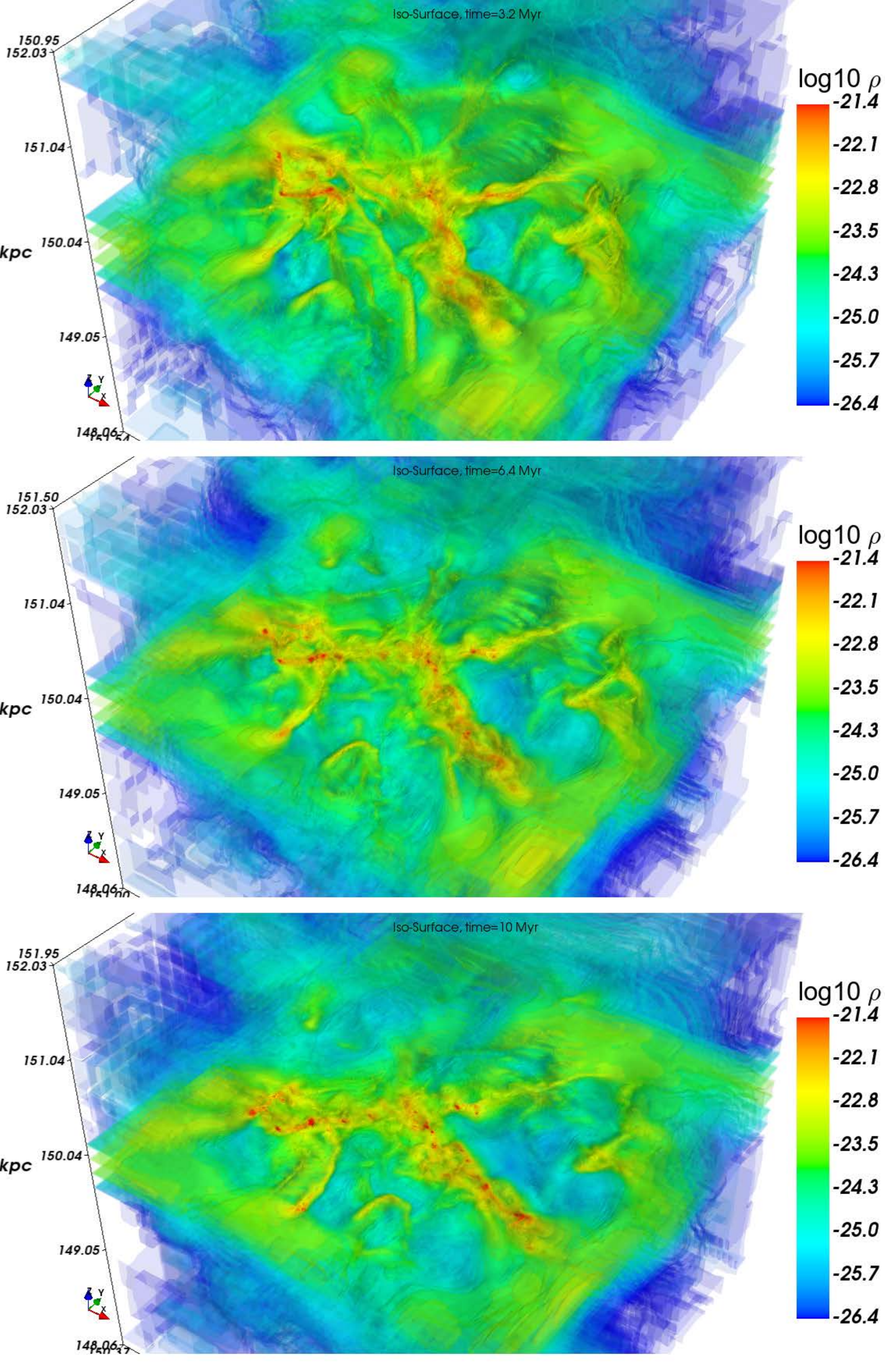}
\caption{Time evolution of the filaments in the active region. The domain is 
about 3~kpc in size.}
\label{Fig:evolActive}
\end{figure}

The time evolution of the active region 
is shown in Fig.~\ref{Fig:evolActive}. In comparing the first frame at $ t = 3.4$ Myr to the last at $ t = 10$ Myr, note that the low density regions between the filaments have expanded greatly, particularly the region to the right of the main filament. One also notes that dense gas, indicated in the red color $ \rm{log_{10}} \rho \ge  - 21.4 $, becomes progressively more filamentary structure in the last frame compared to the first.  Evidently compression and inescapable accretion onto the atomic filamentary structures in this active region occurs
as the feedback bubble expands.  We quantify these accretion rates in \S4.3 and 4.4.

At the last time point an increasing number of sites along the atomic filamentary structures 
form molecular condensations. 
Due to the limited image and regridding resolution, these condensations are 
shown as ``beads'' like structures dispersed along the filaments where the densities peak. This 
picture resembles the dense core formation along parsec scale 
molecular filaments, but here on much larger kpc scales in atomic, and not molecular gas.
However, as we enlarge the 100~pc region around the individual ``beads'', 
a variety of hierarchical molecular cloud complexes are revealed with 
detailed substructures.

\begin{figure*}
\includegraphics[width=\textwidth]{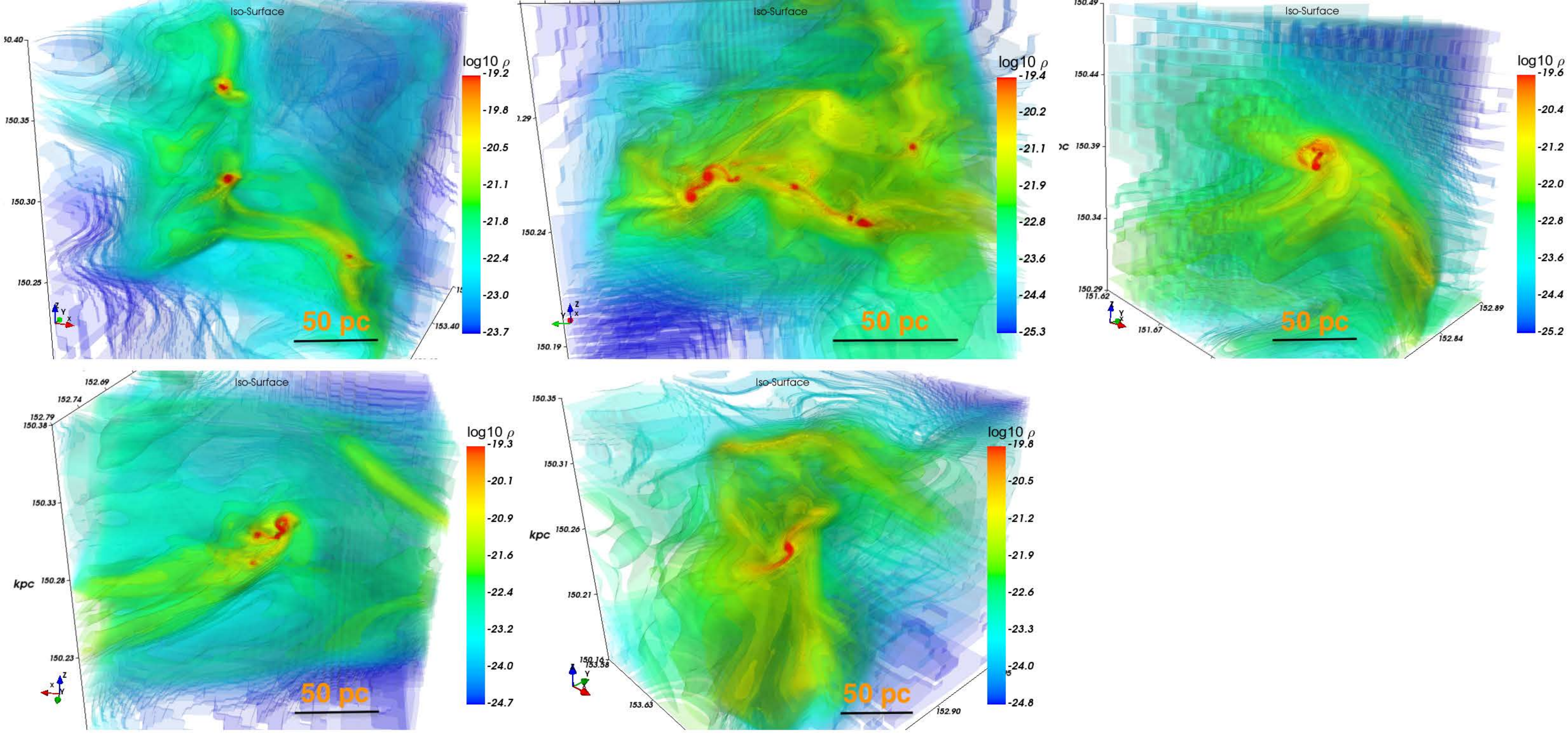}
\caption{The main molecular complexes formed in the active region at 6.4~Myr. 
All regions are plotted in a box of 200~pc size.}
\label{Fig:activeComplex}
\end{figure*}

We show the main molecular complexes of the active region at a 
time frame of 6.4~Myr in Fig.~\ref{Fig:activeComplex}. It is obvious 
that the ``beads'' along the kpc filaments are not simple spherical 
clumps, but rather contain multiple sub-clumps that group themselves 
in hierarchical structures. It is expected that such regions naturally
form star clusters. For the regions in the third and fourth panel, 
shear motions from the larger scale tend to yield a net angular momentum 
along certain direction, resulting in swirling structures tens of 
pc in size (we will discuss a more well-developed disk-like structure 
in the quiescent region in the next section). 
Even for the isolated structure in the last panel, the flows 
from larger scales are channelling towards the center from multiple 
directions. These flows would be incorrectly absent by design in isolated setups of 
molecular cloud formation from an $\sim$100~pc box.

The smallest, highest density regions in each of the complexes shown here
are of the order of a few pc and have number densities and $n \simeq 10^4 - 10^5$ cm$^{-3}$.  
These are the scales on which individual, massive star clusters form.   We interpret
the structures in these panels and their substructures as the formation molecular clouds and their internal
star clusters.  

\subsection{Accretion onto filaments, and flows within them}

The ultimate source of mass for a growing filament is accretion onto it from the surrounding medium.   This can be material pushed into it from the converging superbubble walls in the active region, or from the passage of the spiral wave through the background disk that is more typical of our quiescent region as an example.  Flows within the filament and parallel to its axis can arise from the component of the accretion flow that is parallel to the filament axis as well by gravitationally mediated flows that arise from the flow of filament material onto
growing mass fragments within them. As it is important to evaluate these flows quantitatively, we make a brief digression to develop the needed formulae for them. The general reader may skip ahead to section 4.4.

\begin{figure*}
    \centering
    \includegraphics[width=0.48\linewidth]{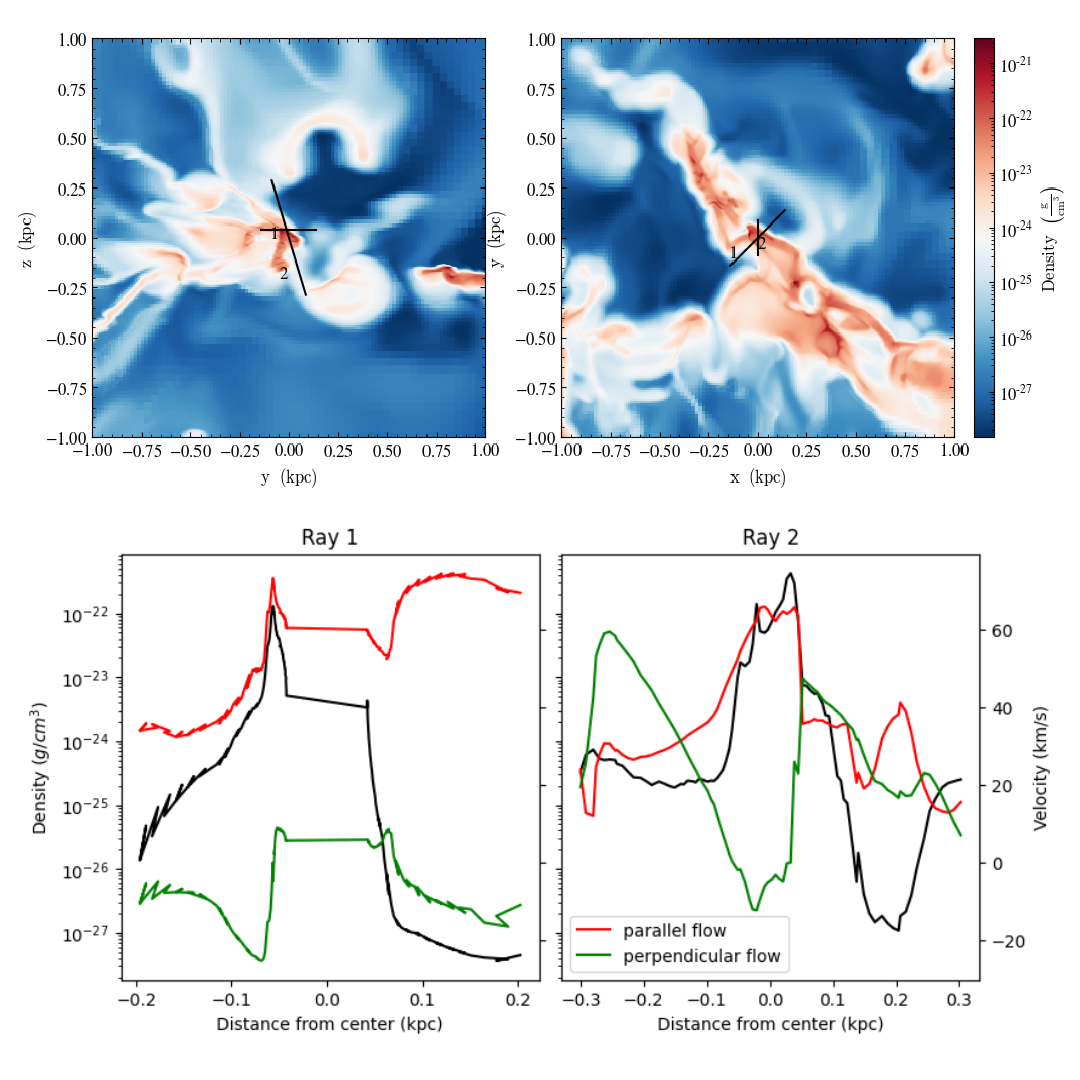}
    \includegraphics[width=0.48\linewidth]{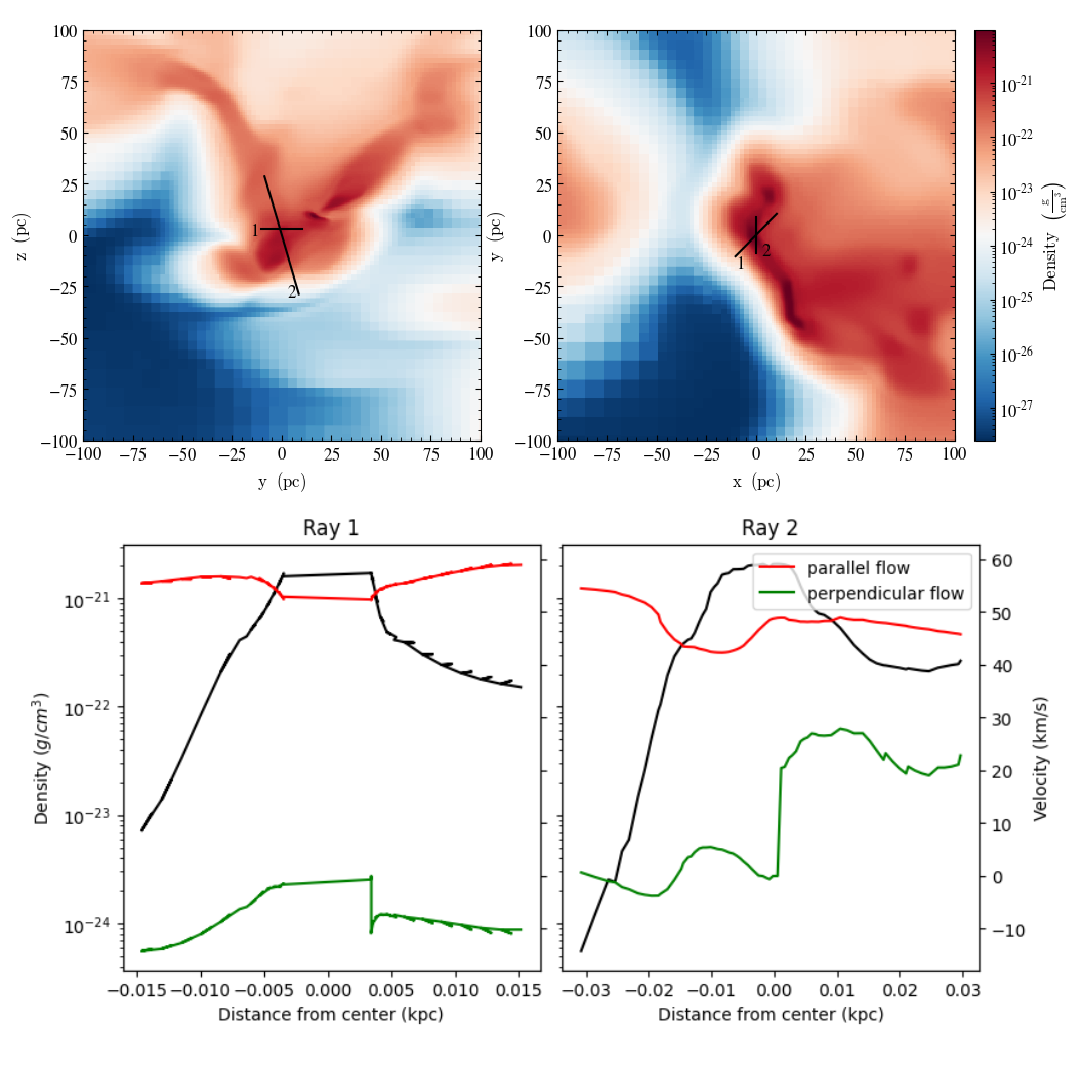}
    \caption{\textit{Top:} Density slices depicting filamentary areas of interest in the 3 kpc region (left) and our zoomed in 200 pc (right) region. The rays traced for flows shown via black lines, labelled 1 and 2 for first and second ray, respectively. \textit{Bottom:} Density and velocity profiles for each ray in both the kpc and zoomed pc scale. Black line gives the density profile of each ray, red lines show velocity along the filament, and green lines show the velocity onto the filament, respective to the direction of the ray.  }
    \label{fig:active_velprof}
\end{figure*}

Most of the analysis consists in evaluating mass conservation in flows onto and along cylinders. So, we begin with the continuity equation that relates the time rate of change of density in the filament to the net flux of material through it; 

\begin{equation}
\label{eqn:continuity}
    \frac{\partial\rho}{\partial t} +  \bm{\nabla} \cdot (\rho \underline{v}) = 0
\end{equation}
For a cylinder with axial symmetry, this equation states that the rate of change of gas density in the filament is the difference between the mass flux normal to its cylindrical surface, and that which is parallel to its axis and within it.  

The rate of change of a the cylinder's mass $\dot {M}$ is the difference between the accretion rate onto it $\dot {M}_{f}$ through the filament radius $r_f$, and the flow rate $\dot {M}_{f, \parallel}$ through it. This result follows from Equation \ref{eqn:continuity} by an application of Gauss's theorem \footnote{By integrating the continuity equation over the volume of the filament, Gauss's theorem allows us to equate the volume integral of the divergence of the mass flux with an integral of the normal component of that flux across the surface area of the cylinder.}    

\subsubsection{Accretion flows onto filaments}

Since the inflow onto a cylinder can vary along its length, we first consider a filament of length $dL$ and cross sectional area $\pi r_f^2$.  From an observational perspective, it is also sometimes more useful to measure the local accretion rate per unit length at point to point along the filament.  The results of the previous subsection then give; 

\begin{equation}
  \frac{d \dot M_{f}}{dL} = - 2 \pi r_{f} \rho_{e} v_{r} 
\end{equation}
where $\rho_e$ is the density of gas at the filament outer edge that is being accreted on the cylinder, and $v_{r}$ is the true radial velocity component of the external inflow perpendicular to the filament axis, at the cylinder's surface. Note, in cylindrical co-ordinates, a radial inflow into the cylinder has $v_r \le 0 $, which makes the growth rate positive.  Here we also assume that the filament accretes from a 3D volume (full $2 \pi$ radians). There is a geometric correction if the filament is accreting out of a 2D sheet.

For the case that the variations along $L$ are small, and the inflow velocity remains fairly constant,
the mass accretion rate $\dot {M}_{f}$ onto a uniform filament of length $L$ arising from perpendicular inflow through its surface is then
\begin{equation}
  \dot {M}_{f}= - 2 \pi r_{f} \rho_{e} v_{r} L
 \end{equation}
with the convention that radial inflow velocity is negative.

This accretion rate onto the filament can also be expressed in several different ways;
\begin{equation}
\dot {M}_f = \frac{ M_{acc}}{t_{cr}} = - 2 v_r (M_{acc}/r_f) = - \pi \Sigma_e v_r L
\end{equation} 
where we define 
$M_{acc}  =  \pi r_{f}^2 \rho_{e} L $  
as the external mass that has accreted onto the cylinder over a time;
$ t_{cr} =  { r_f / ( 2 v_{r} ) } $.  The second form on the right hand side has been used by \citet{Kirk+2013}. In the third expression,  $\Sigma_e = 2 \rho_e r_f $ is the equivalent column density of gas accreted onto to the filament - the source of the accreting material. 

The time $t_{cr}$ is the characteristic crossing time for the external radial converging accretion flow 
to cross the width of the filament.  It measures the accretion time scale onto the filament due to the external 
flow field.  It follows that for as long as the accretion flow is steady, the amount of mass that is added to the filament by flow through its radial surface grows linearly with time as,
\begin{equation}
M_f(t) \propto t/t_{cr}
\end{equation}

\subsubsection{Flows along filaments}

First we consider a small segment of height $dL$ of uniform cylinder of radius $r_f$.  There the net flux of mass along the cylinder depends upon the difference in speeds of the material through the two "caps" of the cylinder.  Hence, the mass flow rate per unit length depends upon the gradient of the aligned velocity $dv_{\parallel} / dL$;
\begin{equation}
   \frac{ d \dot {M}_{f, \parallel}}{dL}  = \pi \: r_f^2  \rho_f \frac{dv_{\parallel}}{dL}
   \end{equation}
Note that for the flow of a uniform "river" along into and out of a cylinder, there is no velocity gradient so that there is no change in the mass of the filament due to this aligned flow. Integrating along the length of the filament, the net flow along and through it depends on the velocity difference $\Delta v_{\parallel} $ long the filament length:
\begin{equation}
 \label{eqn:flowrate_parallel}
 \dot {M}_{f, \parallel}  = \pi \: r_f^2  \rho_f \
\Delta v_{\parallel} =\frac{\pi}{2} \Sigma_f \Delta v_{\parallel}r_f
 \end{equation}
 where $\Sigma_f = 2 \rho_f r_f $ is the column density of gas in the filament. It is also the peak column density one would observe for a radially varying density distribution in the filament( eg. \citep{Beuther+2020b}).

The filaments in our simulation are not uniform - they are well fit by Plummer profiles  consisting of a flat portion, and a power law envelope with power laws $ p = 1.5 - 2$.  Since the mass in the filament scales as $ M(r) \propto \rho r^2$, the cylinder mass increases with radius significantly as one goes from $R_{flat}$ to the filament boundary $r_f$ - which as we have seen is a factor of 10 larger for the filaments analyzed.  Integrating the filament mass contributions from both the flat and power law envelope, we find that the filament aligned mass flow rate through some cross section of a filament;
 \begin{equation}
 \label{eqn:flowrate_parallel}
 \dot {M}_{f, \parallel}  = \pi f_{env} \: r_f^2  \rho_f \Delta v_{\parallel} 
 \end{equation}
where   $\rho_f$ is the filament density in the flat region,  $ v_{\parallel} $ is the velocity within the filament
and parallel to its axis, and the correction factor for the contribution to the mass flow in the envelope is 
\begin{equation}
    f_{env} = \frac{1}{(1 - (p/2)) x_f^2} (x_f^{2-p} - (p/2))
\end{equation} 
for a power law index $p \neq 2$ and where $x_f = r_f / R_{flat}  $.  Note that for a uniform filament out to radius $r_f$; $p=0$ and $f_{env} = 1$.  From the plots of both these and smaller scale filaments in our simulations, an average value is $<p> = 1.75$.  We also find typical values of $x_f \simeq 10 $ for filaments on both kpc and 100pc scales; all of which gives $f_{env} = .076$.

The ratio of the aligned flow rate through some section of a filament, to the filamentary accretion rate is from Equations 8 and 11:

\begin{equation}
        \label{eqn: accretion_ratio}   
        \frac{\dot{M}_{f, \parallel}}{\dot{M}_f} 
    = \frac{f_{env}}{2} \frac{ \rho_f . \Delta v_{\parallel} . r_f} {\rho_{e} . v_{r} . L}
    \end{equation}
In steady state ($\partial \rho / \partial t = 0 $), if a filament were merely a "river" - a conduit of everything that was pushed into it by accretion - then this ratio would be unity; a simple consequence of mass conservation.  

We evaluate this ratio from our data.  Due to the density profile of the envelope, we have $f_{env} \simeq 0.08 $.  We also found that the ratio of the width to the length of the kpc filaments  is at least 10 $r_f / L \simeq 100 pc / 1000 pc = 0.1 $. From the density profiles we compare the value of $\rho_f$ in the flat region of the filament to that at the edge $\rho_e$, where it merges with the surroundings, and fin $\rho_f / \rho_e \simeq 30 $.  Finally, the velocity flows had $\Delta v_{\parallel} / v_r \le 1$.  From Equation \ref{eqn: accretion_ratio} we then deduce that  $\dot{M}_{f, \parallel} / {\dot{M}_f} \le 0.12$.  In other words, the accretion rates onto the filament is roughly 10 times greater than the mass flow rate along it. This is in agreement with the observations of relative accretion vs longitudinal flow rates for larger filaments \citep{Hacar+2022}.  It is interesting that this is also observed for flows associated with filaments in which a low mass star cluster is forming, as discussed in 
\S 5 \citep{Kirk+2013}.

The density and the column density profiles of filaments that become self gravitating build with time, as we clearly see in Fig.~\ref{Fig:evolActive}.
When line mass of a filament exceeds the critical value, it becomes prone to gravitational fragmentation.  At that point, a periodic contribution to the parallel velocity field appears that is a consequence of gravitational fragmentation. An example of this is the velocity field arising 
from the linear instability theory for equilibrium filaments with radial density profiles in the models of  \citet{Fiege_Pudritz2000b}. If mass differences between fragments grow, as expected when filaments undergo non-uniform accretion, then the most massive fragment will accrete over larger regions. The gravitational speeds that govern $\Delta v_{\parallel} $ can be much smaller than velocities $v_r$ characterizing the inflow rates.

\subsection{Numerical results: multi-scale filament flows in the active region}

\subsubsection{Large scale kpc atomic filaments: } 
We evaluate the accretion rate per unit length for the kpc atomic gas filaments using our numerical data.   For the velocity, we first use numerical values of the velocity field nearest the large scale kpc atomic filaments shown in the large scale view; Fig.~\ref{Fig:active_0083}.   The reader can also see this by using the colour designations for velocity amplitudes for the velocity vectors nearest the filament.  These vary along the length of the filament  but take a characteristic speed of $v_r \simeq 50 $ km s$^{-1}$ (mid aqua colour vectors) near to the highlighted zoom in region at the apex of three filaments.  The velocity vectors along the filament edge also vary in their directions but we will take this as a typical value.   The other two physical quantities also come from our numerical data - and can be read off the plots of the radial density profiles shown in the right panels of Fig.~\ref{Fig:lineMass_large}.  In all three panels, we see the profiles meet the background density at a scale of  $ r_f \simeq 100$ pc where $\rho_e \simeq 10^{-23} $ g cm$^{-3}$ under such a steady filamentary accretion flow.

Using these data, the accretion rate per pc onto the atomic filaments at the kpc scale is;
\begin{equation}
\label{eqn:accretion_rate}
\begin{split}
 \frac{d \dot M_{a,f}}{dL} = 4.47 \times 10^3  \left( \frac{r_f}{100 \rm{pc}}  \right)  \left( \frac{ \rho_{e}}{10^{-23} \rm{g.cm}^{-3}} \right) \\
 \left(\frac{v_{r}}{50 \rm{km.s}^{-1}} \right) \frac{\rm M_{\odot}}{\rm{Myr}.\rm{pc} }
\end{split}
\end{equation}
 
 The growth time for large scale atomic filament in our simulation is,
\begin{equation}
 t_{cr} \simeq  1.0  (r_f / 100pc) .(v_{r} / 50 \rm{km s}^{-1} )^{-1}  \rm{Myr}. 
 \end{equation} 
  The filament will gain enough mass to surpass its critical line mass in several million years.  The accretion flow cuts off as the region becomes quiescent.  Bubble lifetimes, as noted in the Introduction, are 7-42 Myr in NGC628.  The filament forming in our active region is a consequence of the superbubble collision there,  and the time scale for this collision to be completed is shown in the evolution sequence in Fig.~\ref{Fig:evolActive}, or about 7 Myr. That timescale is still long enough however to create massive atomic filaments and drive them over their stability thresholds.

This numerical result shows that at this rate, and over approximately 3 Myr (the snapshot featured in Fig.~\ref{Fig:active_0083}), the kpc filament reaches a line mass of $ m =  (d \dot {M}_{a,f} / dL) \:   t = 1.34 \times 10^5 \rm{M}_{\odot} \rm{pc}^{-1}$.   From the left most panel in Fig.~\ref{Fig:lineMass_large} we can see that this exceeds the average critical line mass for filaments on this kpc scale (note that the fluctuations reflect variations of the accretion velocity along the filament).   The entire filament of kpc in length has a total accretion rate  of 
\begin{equation}
\dot {M}_{a,f} = 4.47 \times 10^6  \rm{M_{\odot} Myr}^{-1}
\end{equation}
and will have accreted a total mass of $3.13 \times 10^7 \rm{M}_{\odot} $ over  the 7 Myr shown.   This is clearly sufficient to produce a number of GMC mass fragments that we are seeing in the simulation (see \S 5.2). 

We also evaluate the filament aligned mass flow rate  $\dot {M}_{a,f\parallel}$ in the kpc atomic filament. 

\begin{equation}
\label{eqn:flow_rate}
\begin{split}
\dot {M}_{a,f\parallel} = 2.68 \times 10^4 f_{env} \left(\frac{R_{flat,a}}{10 \rm{pc}}  \right)^2 \left( \frac{ \rho_{f}}{10^{-22} \rm{g.cm}^{-3} } \right) \\ \left(\frac{v_{\parallel}}{60 \rm{km.s}^{-1}} \right)
\frac{\rm{M}_{\odot}}{\rm{Myr} }.
\end{split}
\end{equation}
 This is the accretion rate onto the GMC mass fragment within the filament.

To summarize, our numerical data on the 3kpc scale shows that superbubble driven compression drives significant mass flows onto the forming kpc atomic filaments ultimately driving them over the critical line mass on that scale for the filaments to fragment into GMCs.  The time scale for this to occur is from 3 Myr and upward.

\subsubsection{Cluster forming clumps: 100 pc GMC scales}
  Following the same procedure as in Equations \ref{eqn:accretion_rate} and \ref{eqn:flow_rate}, we take our numerical data from 100 pc scale GMC filaments in Fig.~\ref{Fig:lineMass_small} and the filament velocity profile shown in Fig.~\ref{fig:active_velprof},which yields $v_{r} = 10 $ km s$^{-1}$.  We find that the filament radius and external density are $r_f \simeq 10 $ pc and $\rho_e \simeq 3 \times 10^{-22} $ g cm$^{-3}$, respectively.  These give an accretion rate per pc of the order  
\begin{equation}
d \dot {M}_{GMC,f} / dL = 2.68 \times 10^3  \rm{M_{\odot} / \rm{Myr} . \rm{pc} }
\end{equation} 
This implies that the total accretion rate onto 100 pc filaments that characterize filamentary molecular clouds in this active region is
\begin{equation}
\dot {M}_{GMC,f} = 2.68 \times 10^5  \rm{M_{\odot}} / \rm{Myr}
\end{equation}
namely several pc, clump masses that are the nurseries of star clusters will achieve $10^4 - 10^5 \rm{M}_{\odot}$ within the 5 Myr. 

For the GMC filament aligned flow, we use the filament density in the flat regime in the right hand panels of Fig.~\ref{Fig:lineMass_small}, where $R_{flat, GMC} \simeq 1 \rm{pc}$ with a density $\rho_{f} \simeq 10^{-21}\rm{g.cm}^{-3}$.  The filament profile for $v_{\parallel}$  shown in Fig.~\ref{fig:active_velprof} is a bit more complicated.  On these scales, we see that the velocity profile at larger radii than the density peak is rather flat and has a value consistent with that in the broader atomic gas.  We do see that there is a dip of the order $v_{\parallel} \simeq 5 $ km s$^{-1} $ in the densest region.  If we take the magnitude of the filament aligned flow in the GMC filament center as $v_{\parallel} \simeq 50 $ km s$^{-1} $, then the filament aligned flow rate in the fragmenting filament of the order, 
\begin{equation}
    \dot {M}_{GMC,f\parallel} = 2.23 \times 10^3 f_{env} 
\rm{M}_{\odot} / \rm{Myr}. 
\end{equation}
This suggests that these massive GMCs will assemble $ 2 \times 10^4 \rm{M_\odot} $ clumps in 10 Myr, just from the aligned flow rate.    On this scale too,  the accretion rate onto the GMC filament exceeds the flow rate along it. We note also that these being some of the most massive GMCs in our simulation, this process will also assemble the most massive clump masses (see also \citep{Howard+2018}.

\subsubsection{More detail: probing filament flows}

We take a closer look at the velocity field associated with the filaments.  In Figure\ref{fig:active_velprof} we show the radial profiles of the velocity components, $v_r$, $v_{\parallel}$, and density profiles of our selected filaments and of their surrounding medium along two rays that are both perpendicular to the filament axis (and that are are orthogonal to one another).  More detail about the method can be found in Appendix \ref{App.B}.  These are probes of the profiles at a particular point along the filament in both the 3kpc (left panels) and 200pc zooms (right panels) of the active region of our galactic disk. The top row of figures in each panel show the detailed maps while the bottom row shows the profiles. The differences in density and velocity profiles between rays 1 and 2 indicate the departure from cylindrical symmetry of the filament and its associated flows.   The radial density profiles of each are black lines in bottom row figures.  The co-ordinates shown on the figure are with respect to the main density peak (which is $0.0$ on the graphs). The red and green lines in the profiles are the radial profiles of the flow along and onto the filament, respectively named parallel and perpendicular flow.  We leave to a subsequent paper the more complete description of the velocity fields and accretion flows associated with filaments.

For both the 3kpc and 200pc cuts, we see clear evidence of the filaments and can determine their radii as $\sim 0.1$ kpc and $\sim 5 $ pc, respectively. Furthermore, one can see evidence of the denser molecular filament within the bounds of the atomic filament of our 3kpc region. This filament is not centered within the atomic filament, indicating an asymmetry of the filament's radial profile.  

The parallel velocities peaks with the filament density for both rays in the 3kpc scale, with and average value of $v_{\parallel} \simeq 60$ km s$^{-1}$.  The parallel velocity drops as the ray reaches into areas cleared by superbubble expansion, where gas density drops to densities on the order of $10^{-27}~g~cm^{-3}$.  We also note that in the zoomed in 200 pc scale, this component is $40-50 $ km s${-1}$ and is not much different than the lower density gas around it.  This is the flow component then that moves gas along the filament into the growing GMC.    

The sign of the perpendicular velocities indicate whether the filament is accreting gas or dissipating. Accretion is present when the signs between distance and velocity match, such that negative perpendicular velocity on the left of center will indicate accretion on the side, and vice verse for the right side. These again show evidence of asymmetry in the filaments in our 3kpc region, as Ray 1 shows the gas left of center flowing onto the filament, while right of center it is flowing away. This also correlates with the superbubble region, indicating that the feedback is actively destroying the filament on one side. On the other hand, Ray 2 shows both sides accreting gas onto the filament, such that the asymmetry of the filamentary flows is here entirely caused by the presence of the expanding bubble. 

The 200pc zoom-in of this region displays similar characteristics. Perpendicular velocity flows show the filament region is being dissipated on one side. In Ray 1, we see the flows right of center moving gas away from the filament, whereas Ray 2 depicts this happening on the left of center. However, a key difference is the density profiles, which do not show the rays extending into superbubble areas where these perpendicular flows move away from the filament. Instead, we see filament densities on the order of $10^{-22}~g~cm^{-3}$. We also note the presence of a dense clump north of center in our density slice plots. From this, we conclude the rays chosen here are close enough to a dense clump that we begin to see evidence of the clump's accretion, pulling gas out of the filament.

\subsection{Quiescent Region: Galactic Shear and Disk-Like GMCs}
\label{S.Quiet}

\begin{figure}
\includegraphics[width=\columnwidth]{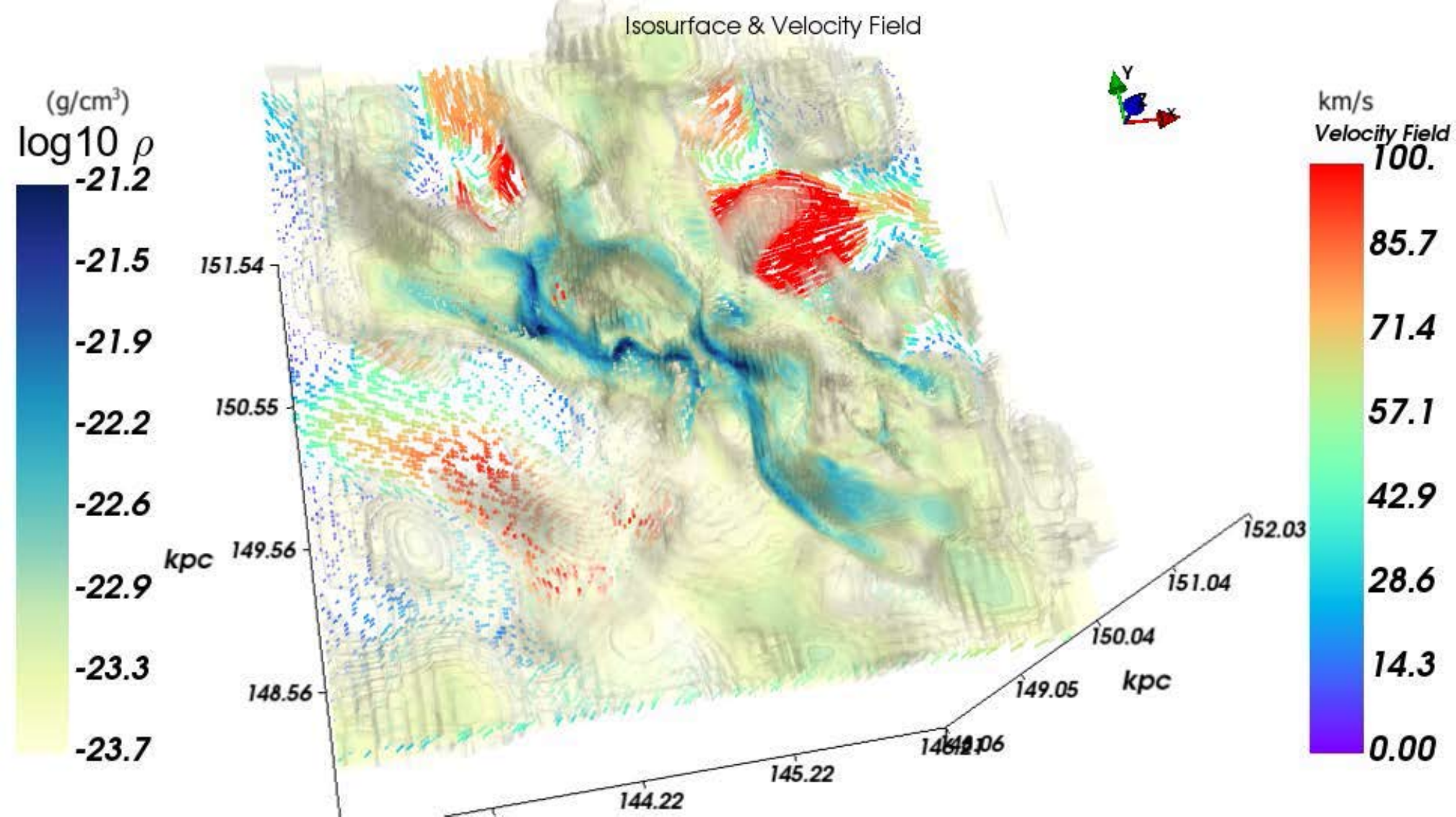}
\caption{Density isosurface and velocity field across the atomic filamentary 
structures in the quiescent region. The velocity vectors are drawn along x-y, 
parallel to the galactic plane.}
\label{Fig:Inactive_0089}
\end{figure}

We now investigate the quiescent 3 kpc region where the main structures 
are currently free from influences of previous feedback 
events. In Fig.~\ref{Fig:Inactive_0089}, we show the density and 
velocity distributions at 3.2~Myr after the restart of the 3~kpc 
zoom-in region. The  ``loop'' like structure comprises 
two filaments each of several kpc lengths, shearing toward each other along a plane perpendicular 
to the galactic plane (along x-y). Despite a small expanding region 
near the top right quadrant, the evolution of the central ``ring''
structure is more or less unaffected by that expanding motion. 

A close examination of the velocity field adjacent to the kpc filaments in Fig.~\ref{Fig:Inactive_0089} shows that the 
velocities are significantly smaller than in the active region, with $v_{r} \simeq 20 $ km s$^{-1}$.  This by itself will
reduce the accretion rate onto filaments by factors of 2-3 compared to the more superbubble driven flow in the active 
region.

In analogy with the results for the active region, we show velocity and density profiles on kpc and 100 pc scales for the quiescent region in Appendix B, Fig.~\ref{Fig:quiet_velprof} to which we refer the reader for all of the details.    We see that there are not large changes.  The flow field is much less complicated however.  

\begin{figure}
\includegraphics[width=\columnwidth]{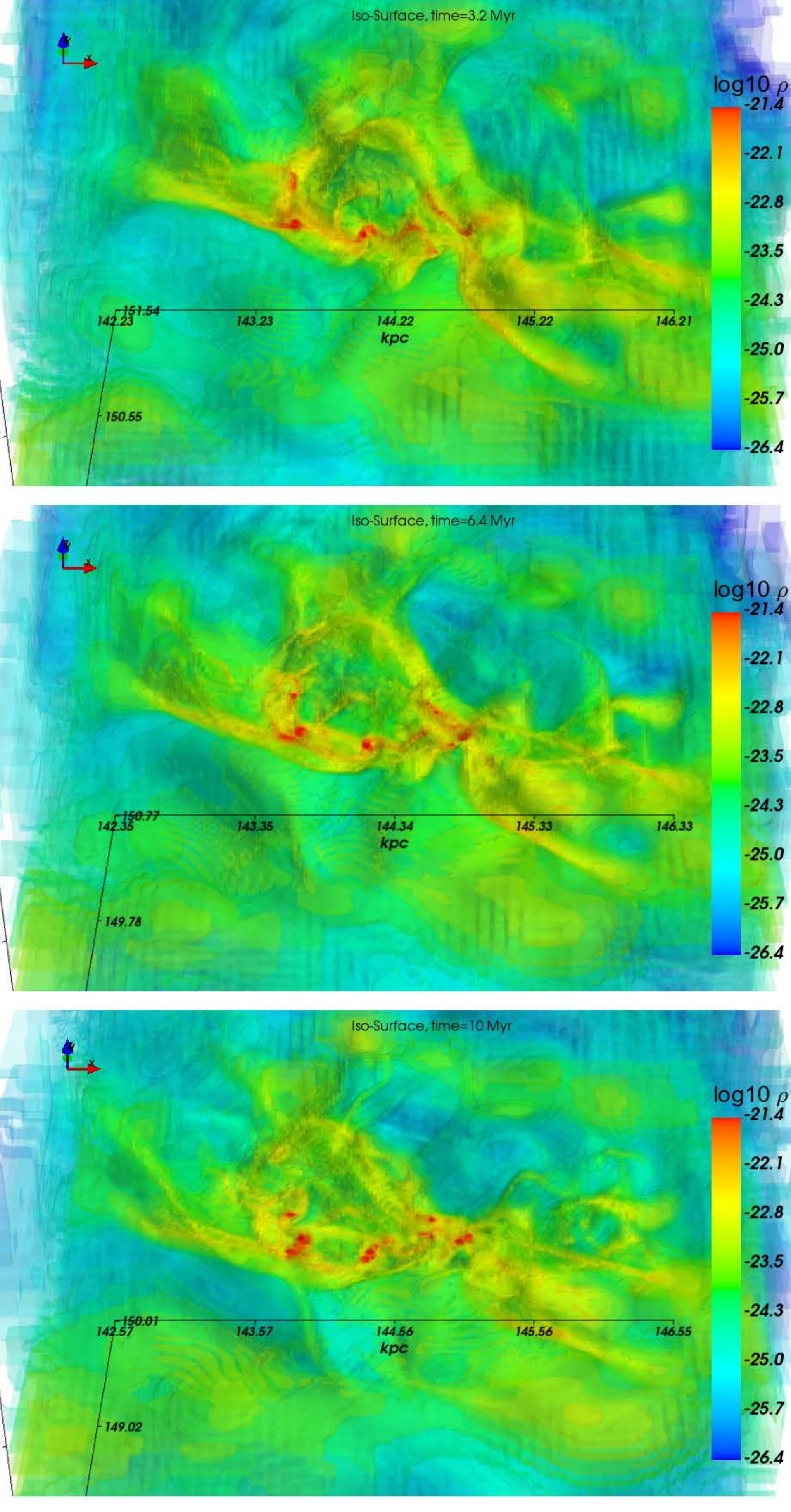}
\caption{Time evolution of the filaments in the quiescent region. 
The domain is about 3~kpc in size.}
\label{Fig:evolQuiet}
\end{figure}

Similar to the active region above, multiple condensations form along 
the ``loop'' like filamentary structures at later times. 
The evolution of this structure is shown in Fig.~\ref{Fig:evolQuiet}. As compared to the active region, 
the matter is more concentrated in the central ``loop'' structure 
in the quiescent region. The condensations already show substructures 
with the image resolution at the 3~kpc scale. 

\begin{figure*}
\includegraphics[width=\textwidth]{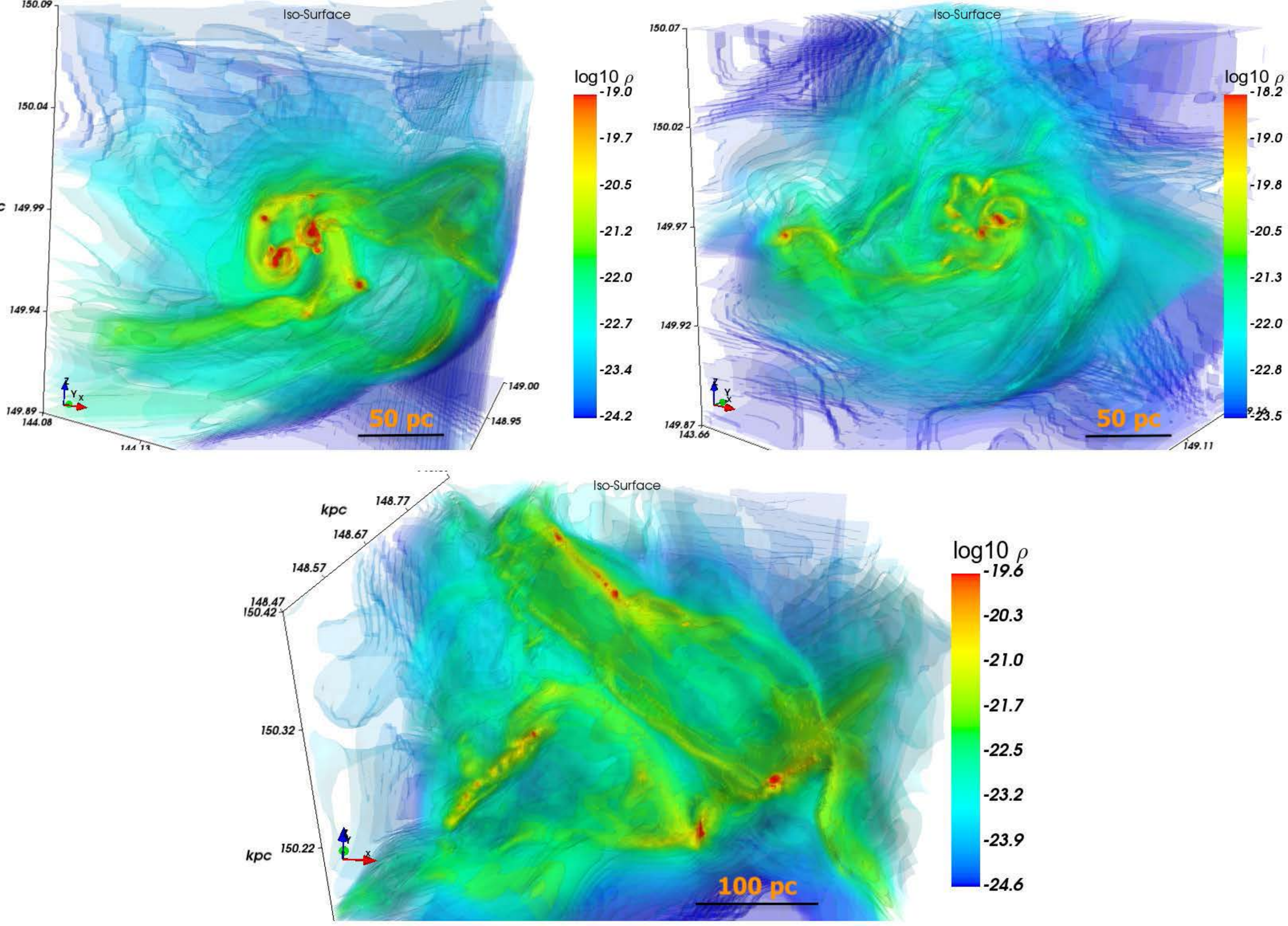}
\caption{The main molecular complexes formed in the quiescent region 
at 6.4~Myr. The first two panels have a box size of 200~pc, 
while the last box size is 400~pc. The galactic plane is along the 
x-y plane.}
\label{Fig:inactiveComplex}
\end{figure*}

In analogy with the zoom-ins of the individual clouds within the large scale filaments in 
the active region,  we show the main molecular complexes in the quiescent region at $t=6.4 \simeq \rm{Myr} $ 
after restart in Fig.~\ref{Fig:inactiveComplex}. Two of the 
molecular complexes show prominent disk and spiral structures 
perpendicular to the galactic disk plane (along x-y). Their angular 
momentum comes from the shear of the vertical ``loop'' structure 
(Fig.~\ref{Fig:inactiveComplex}) which is also oriented perpendicular 
to the galactic plane. The rotating disk resembles that around 
individual stars, except that they are much larger structures at 
50--100~pc scale and are the initial forms of GMCs. The third 
region in Fig.~\ref{Fig:inactiveComplex} encloses 3 sub complexes 
within a larger box of 400~pc, where the structures are mostly linear 
in space. Therefore, similar to the active region, a diversity 
in the morphology of dense GMC structures remains in the quiescent region 
as well.

We note that disk like features have been seen in other insufficiently resolved simulations.  Most modern codes however, resolve dense regions to below the local Jeans length, and this is certainly the case with our RAMSES simulations.  Technically, we resolve the Jeans length by a factor of $N=4$ cells (see Methods).  Thus our forming clouds are well resolved (down to 4.8 pc) on the initial galactic scale even before we dive down into the higher resolution regions. Our highest resolution regions - at a fraction of a parsec (down to 20 levels of refinement) - are reasonably immune against any carry over from possible disk-like artifacts.

\subsubsection{ Disk formation and stability}  

The extended spiral structures and accretion 
flows are not necessarily co-planar with the disk plane; they join the 
disk from different angles and bring in different angular momenta, 
which can perturb the disk structure and promote fragmentation. 
To examine the angular momentum assembly around the disk GMC, 
we transform the velocity components to the disk frame, and identify 
a disk normal direction using the cut plane mode in \textsc{mayavi} 
\citep{RamaVaro2011}. The disk normal direction is more or less 
perpendicular to that of the galactic plane. 

\begin{figure*}
\includegraphics[width=0.95\textwidth]{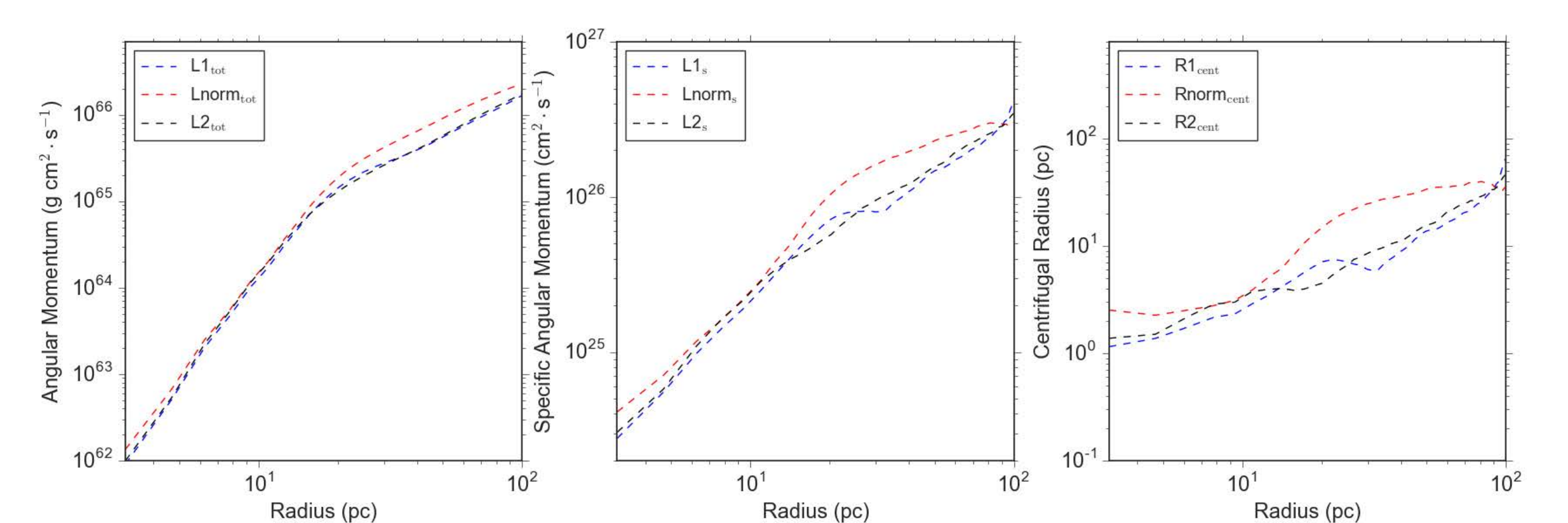}
\includegraphics[width=0.95\textwidth]{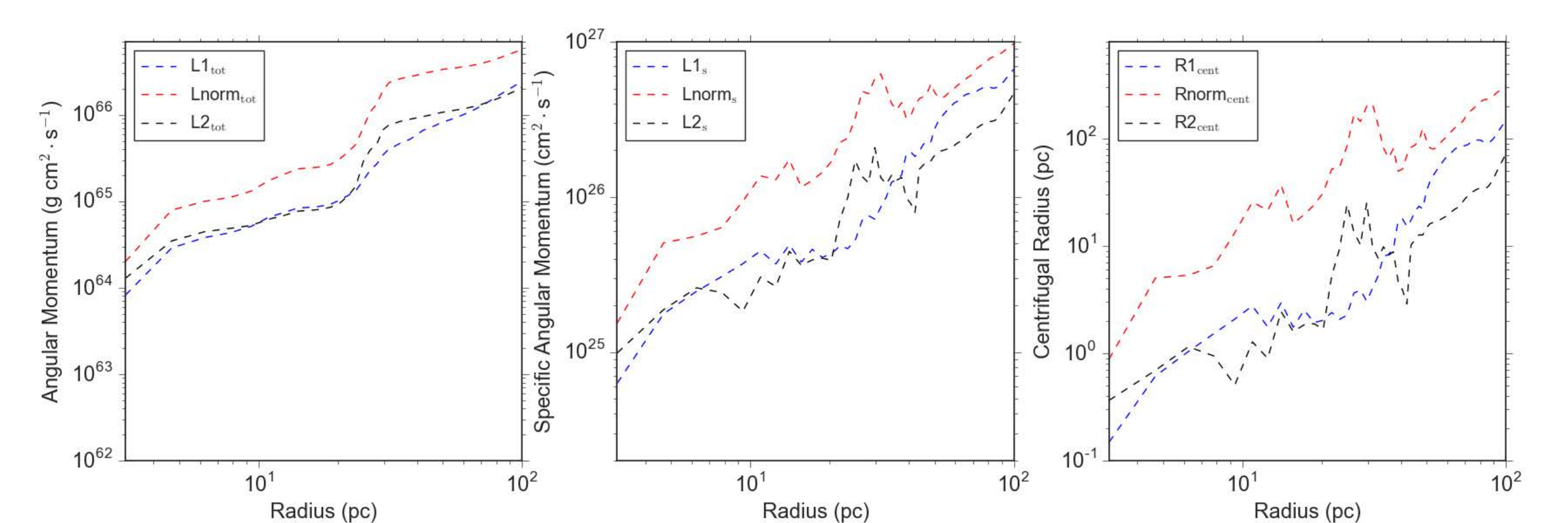}
\caption{Distribution of total (cumulative) angular momentum (left panels),
specific angular momentum (middle panels), and the expected centrifugal 
radius (right panels) of concentric spherical shells centered around the 
disk structure in the quiescent region, for the two frame at $t$=0~Myr (top panels) and $t$=6.4~Myr 
(bottom panels), respectively.} 
\label{Fig:AMcent}
\end{figure*}

In Fig.~\ref{Fig:AMcent}, we see that 
both the total and specific angular momenta along the normal direction 
dominate over that of the other orthogonal directions along the disk plane. 
As a result, the expected centrifugal radius $R_{\rm c}$, computed from 
${\bm{j}^2 / GM}$ is also the largest along the disk normal direction. 
As this disk GMC assembles mass from the surrounding flows, the 
difference in angular momenta between the normal direction and other 
orthogonal directions increases over time, indicating that the net 
effect of accretion flows with somewhat different angular momenta 
averages to a direction aligned with the normal direction of the disk GMC. 
Furthermore, the expected centrifugal radius $R_{\rm c}$ at $t$=0~Myr also 
shows a plateau of 20--30~pc for the infall matter at 20--100~pc scale.
However, at $t$=6.4~Myr, the expected $R_{\rm c}$ is a monotonic curve 
without a well-defined plateau indicating accretion flows at different 
distances will land at different centrifugal radii.  More precisely, the disk
is being built up in time by filamentary inflow of increasingly larger relative
angular momentum.  This late build up of disk structures has been also noted in the context of protostellar disk formation \citep{Kuffmeier+2023}.

\begin{figure}
\includegraphics[width=\columnwidth]{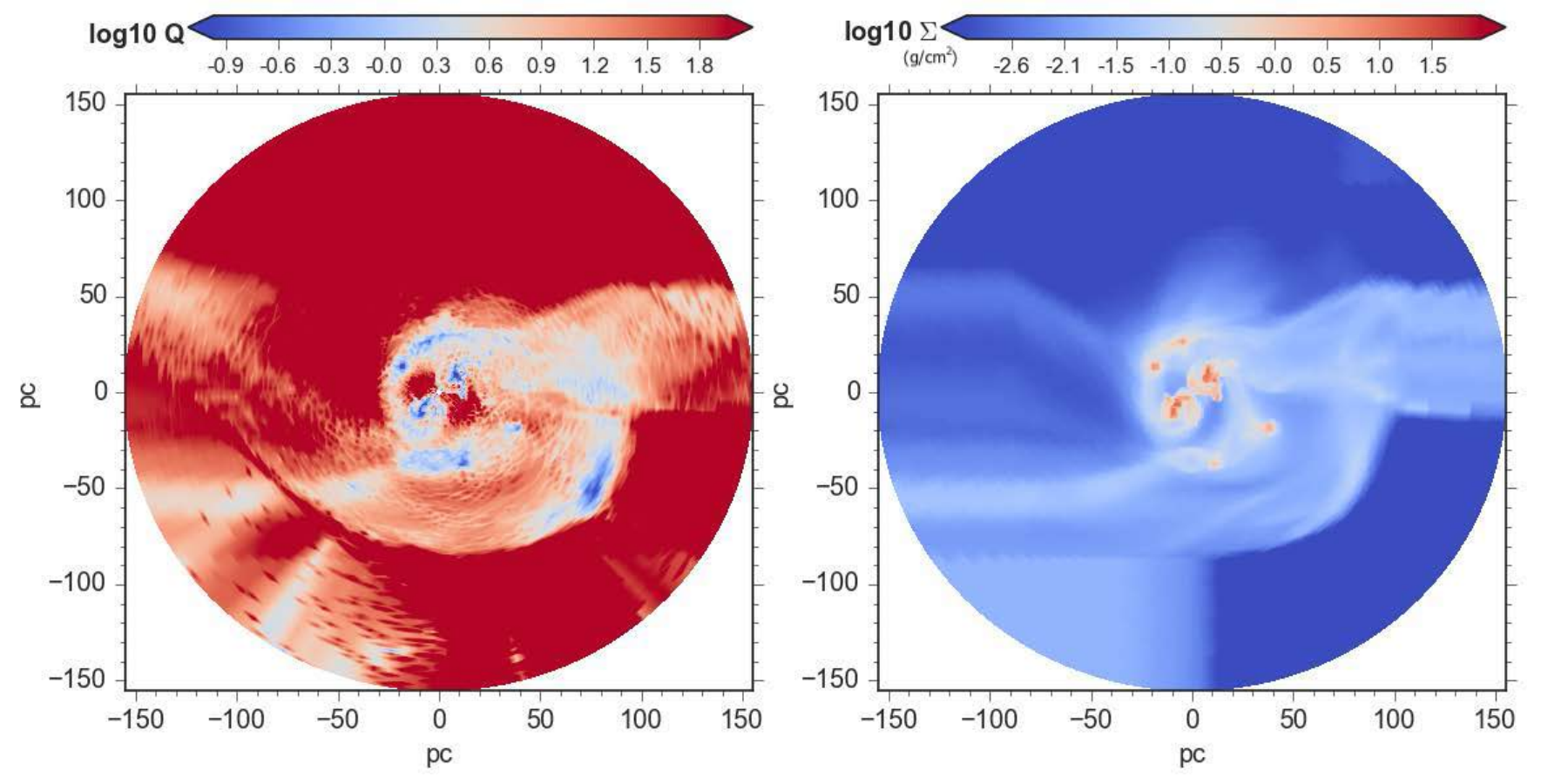}
\caption{Logarithmic distribution of the Toomre Q parameter along the 
plane of the disk GMC (left panels) and the column density 
projected along the disk normal direction (right panels), at the 
time frame $t$=6.4~Myr.}
\label{Fig:ToomreQ}
\end{figure}

In Fig.~\ref{Fig:ToomreQ} we investigate the stability 
of the disk-like GMC against gravitational fragmentation using the 
Toomre Q parameter \citep{Toomre1964}.  This is similar to analysis done on a much smaller scale, on the stability of protoplanetary disks leading
to massive star formation \citep{Klassen+2016, Ahmadi+2018, Ahmadi+2023}.  We take advantage of the disk normal 
direction identified above and interpolate the kinematic quantities into polar 
coordinates along the disk plane for computing the Toomre Q. 
The Toomre unstable locations are tightly correlated with the 
structures with high column density.  This suggests that the filaments that form in this
flattened, rotating disk like structure have arisen by gravitational instability in the disk at 50--100~pc scales.

\begin{figure*}
\includegraphics[width=\columnwidth]{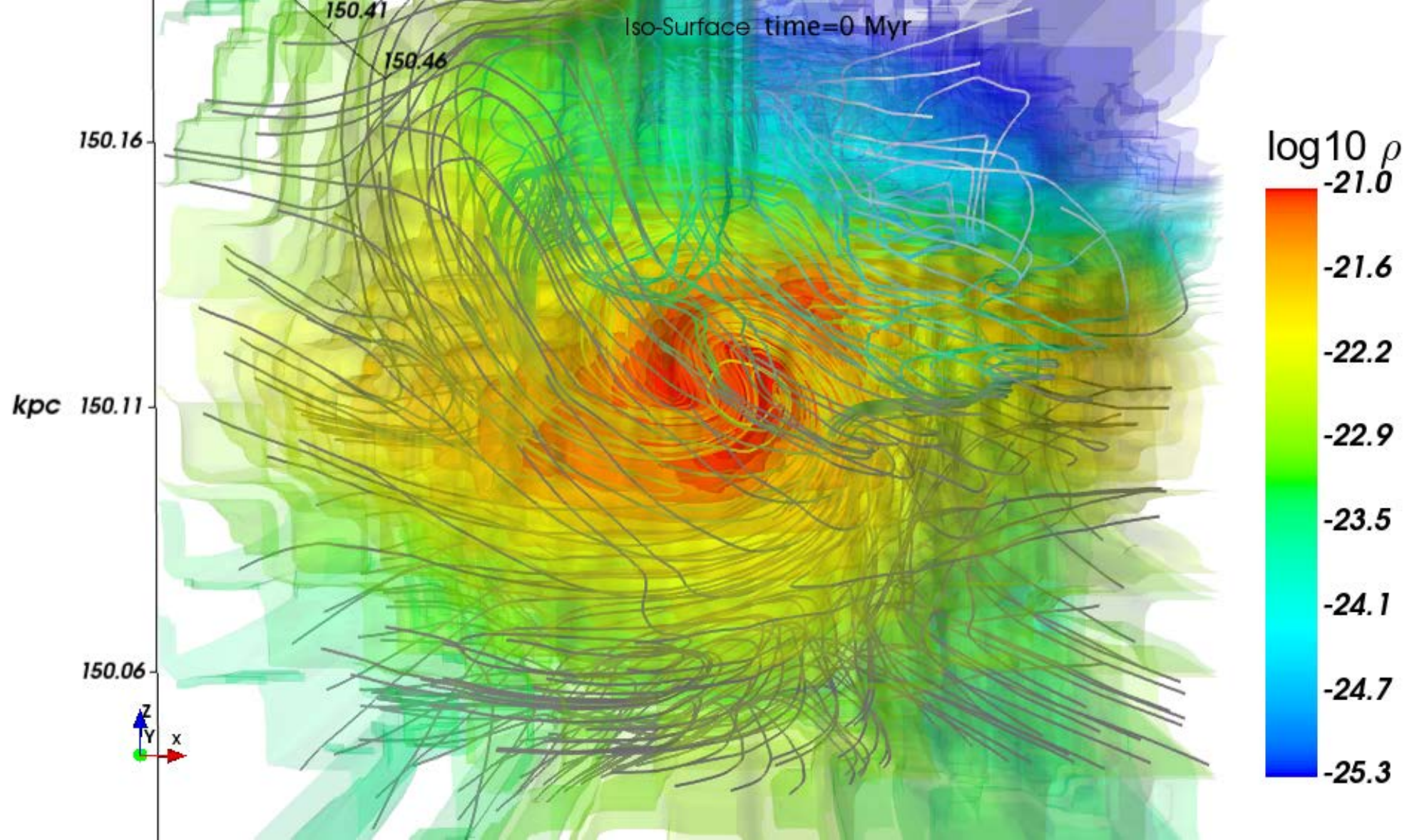}
\includegraphics[width=\columnwidth]{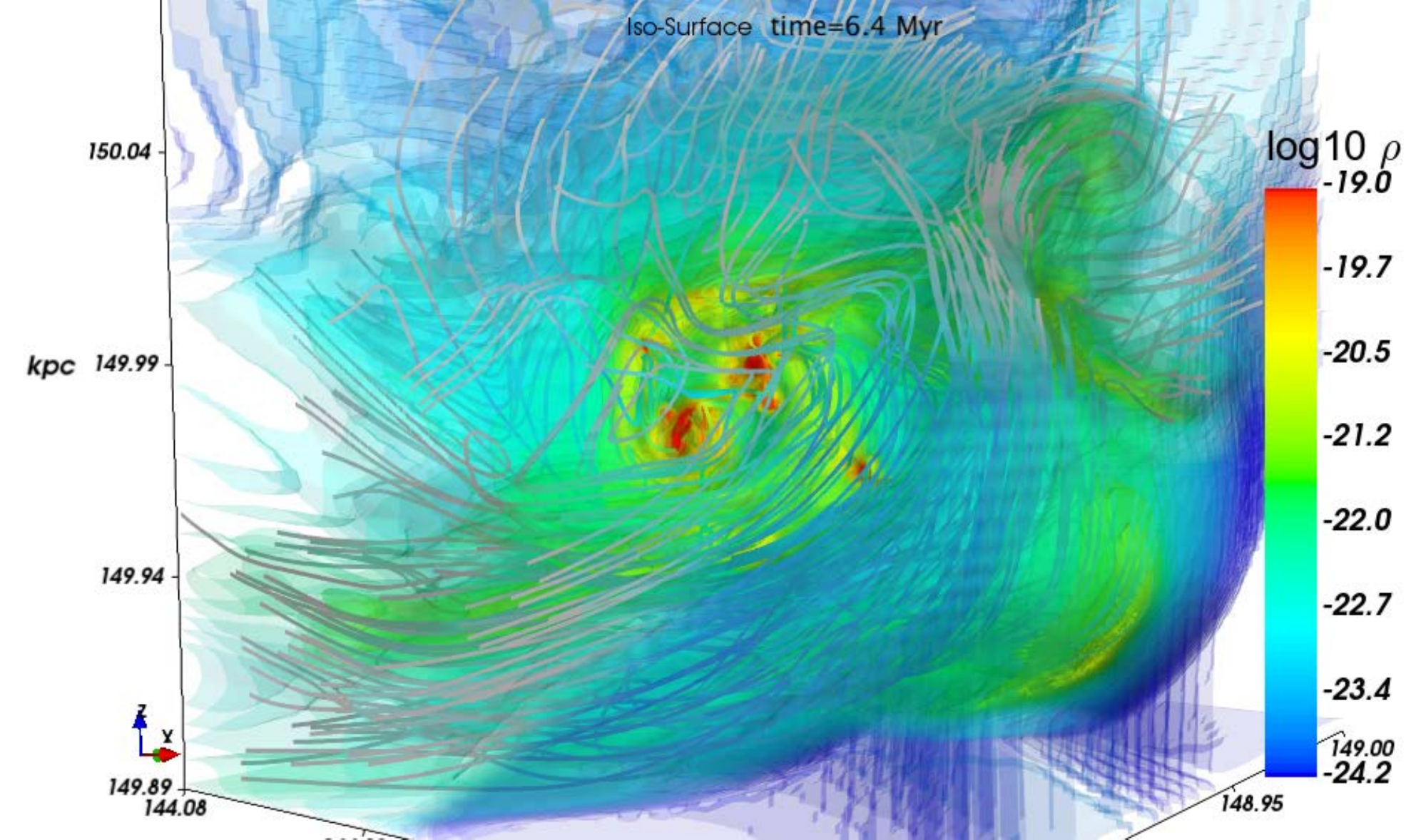}
\caption{Logarithmic density distribution of the central 200~pc molecular 
cloud complex in the quiescent region, with magnetic streamlines. Left: $t$=0~Myr. Right: $t$=6.4~Myr.}
\label{Fig:helicalB}
\end{figure*}

\begin{figure}
\includegraphics[width=\columnwidth]{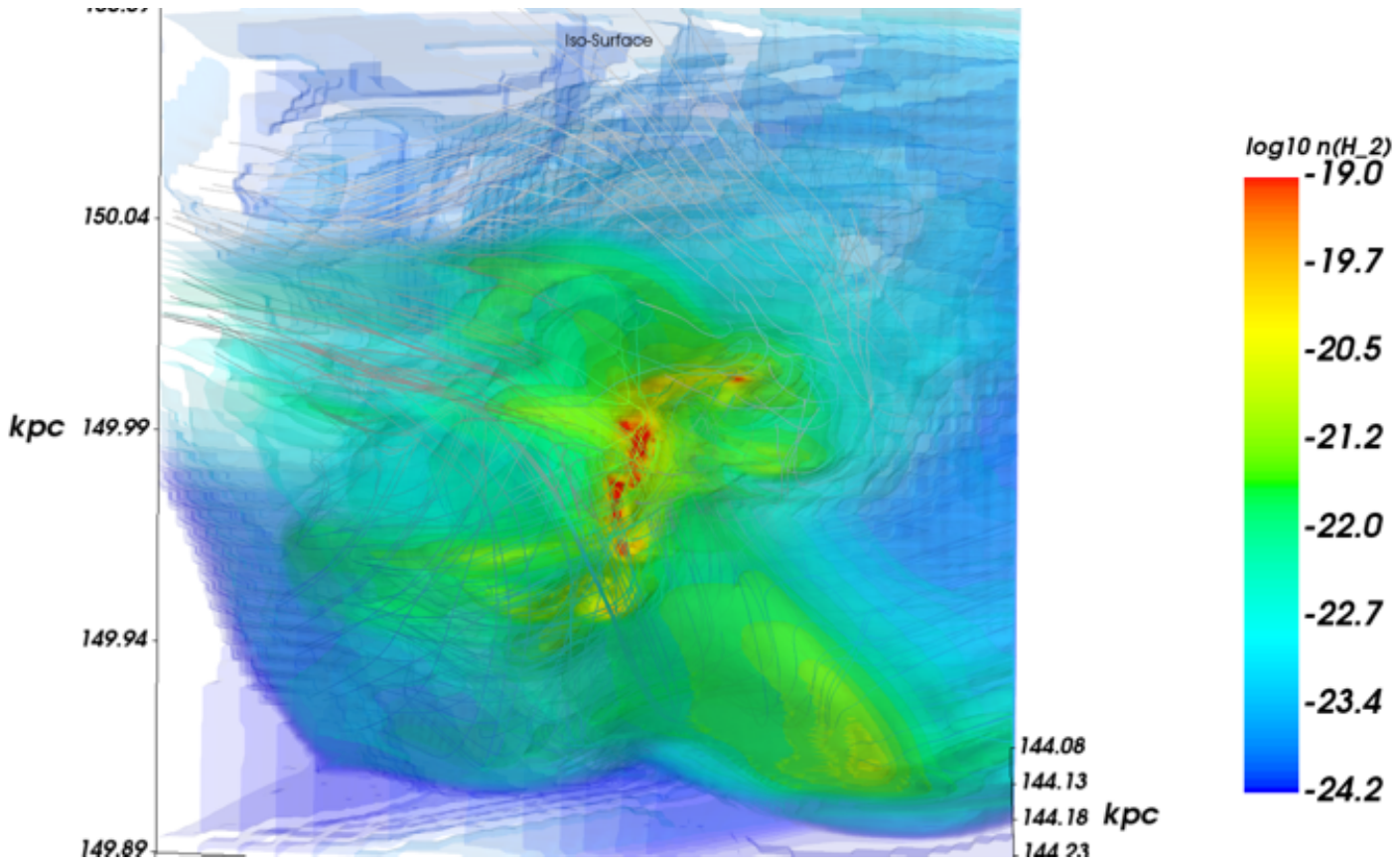}
\caption{Edge on view of disk- like logarithmic density distribution in right panel of Fig. 20.
Helical 
field lines appear to wind around the accreting filament in the lower right hand side of the figure.}
\label{Fig:helical_edgeon}
\end{figure}

\subsubsection{ Disk and helical magnetic field}

Fig.~\ref{Fig:helicalB} shows the magnetic field lines that are associated with the forming filament/disk system.  The left panel shows
the configuration of field and disk in
the central 200~pc region at the initial $t$=0~Myr of the zoom-in restart, while the right we see the region evolved over 6.4 Myr. 
As shown in the left panel of Fig.~\ref{Fig:helicalB}, a spiral arm like structure is clearly seen.  It is forming within a flattened 
disk-like region that is better seen in the more evolved state shown in the right panel.  The region not quiescent and is being fed by several filaments. The bulk density just reaches a few 100~cm$^{-3}$.  In the panel on the right, a disk like structure has become much more apparent after 6.4 Myr of evolution. 
As pointed out in Fig.~\ref{Fig:AMcent}, the expected centrifugal radius $R_{\rm c}$ at $t$=0~Myr 
shows a plateau of 20--30~pc for the infall matter at 20--100~pc scale which
is consistent with the disk radius shown in Fig.~\ref{Fig:helicalB}. 

The converging filamentary flows that are creating the disk drag magnetic field lines with them.  
The resulting toroidal magnetic field structure is clearly seen as the swirl of field lines that appears to wrap around, and in the disk
at t=6.4~Myr (right panel).  At this time, the disk has become unstable  and breaks up into 
multiple high density clumps of 10$^{4}$~cm$^{-3}$.  The disk- like GMC 
therefore acquires a
 sheared out toroidal field as a consequence, that appears to trace along more
 or less parallel to the curving filaments in the disk.
 
In Fig.~\ref{Fig:helical_edgeon} we show an edge-on snap shot of the 
disk and filaments  in the previous figure. Here we see a very interesting feature;
the filaments accreting onto the disk appear to be wrapped with field in a helical magnetic field.
In particular, the filament accreting onto it on the bottom right region of the panel
appears to be wrapped with lines.

Helical fields have been discussed in the context of observations of Zeeman measurements of some filamentary clouds, notably Orion  \citep{Heiles1997, Tahani+2019, Tahani+2022}. These observations have been interpreted as arising from the wrapping of magnetic field lines in the form of an arc around a shock-produced filament.  We are performing synthetic observations of our structures in our simulation, in a separate piece of work using the POLARIS code.  This code can distinguish between these two types of geometry \citep{Reissl+2018}.In more recent work, however, \citet{Kong+2021} present observations of structure in the Orion A filament that could be produced by magnetic reconnection of clouds with anti-parallel magnetic fields. 

Detailed models of the structure of equilibrium magnetized filamentas that include helical fields were computed by \citet{Fiege_Pudritz2000a}.  As a sidebar, we note that dynamical simulations of colliding clouds with anti-parrallel fields show that the collision induced reconnection of fields produces a stable toroidal field that wraps the filament \citep{Kong+2022, Kong+2023}.  These structures require high spatial resolution to follow the small scale physics of magnetic reconnection.  

The important new ingredient in our simulation is the inclusion of galactic shear which will naturally 
twist magnetic field lines in flows. The significance of a helical field, if this is indeed what we are seeing, is that they reduce the critical line mass - pushing the filament towards instability. We discuss this further in \S 5.1.

\subsubsection{Accetion onto and flows along disk filaments}

\begin{figure}
\includegraphics[width=\columnwidth]{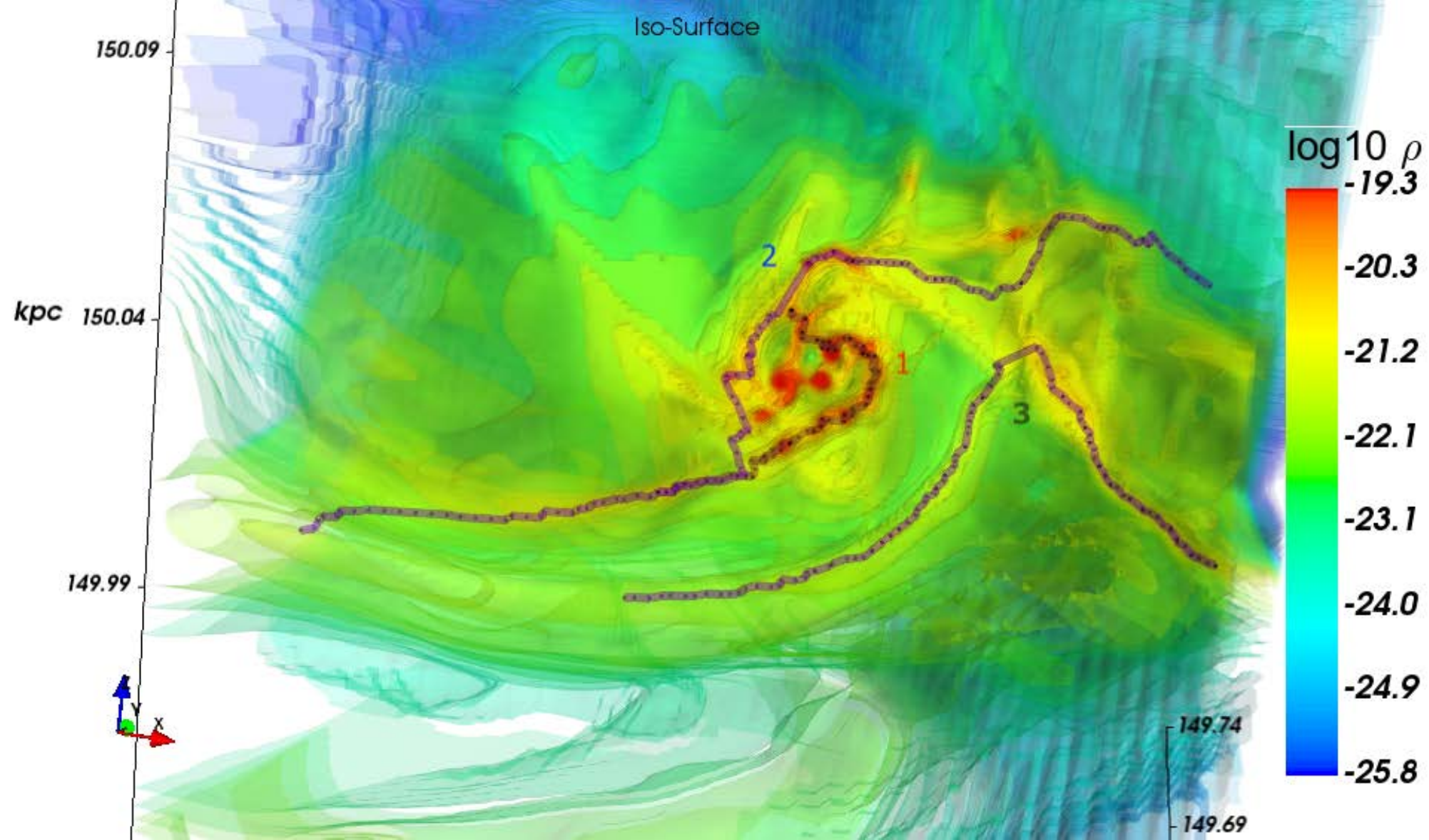}
\caption{Filament traced along the spiral structures of the 
disk GMC in the quiescent region, at $t=3.2$~Myr after the restart of the simulation.}
\label{Fig:traceSpiral_89}
\end{figure}

The spiral accretion flows wrapping around and feeding the disk are essentially 
filamentary structures, thus one can also apply the filament tracing 
method to examine the stability along such structures. As the 
clumps at $t$=6.4~Myr are already fully developed along the spirals, 
we apply the filament tracing to an early frame at $t$=3.2~Myr when 
the individual spiral arms are more spatially connected and less 
broken up by the clumps. 
Fig.~\ref{Fig:traceSpiral_89} demonstrates three filaments identified 
by our tracing method. The choice of joining different segments of filaments 
into one structure is less of a geometrical connectivity, but more for 
the convenience of our analysis. Filament 1 and 2 mostly trace the 
close-in spiral structures around the disk, whereas filament 3 traces 
the more extended accretion flows. 

\begin{figure*}
\includegraphics[width=\textwidth]{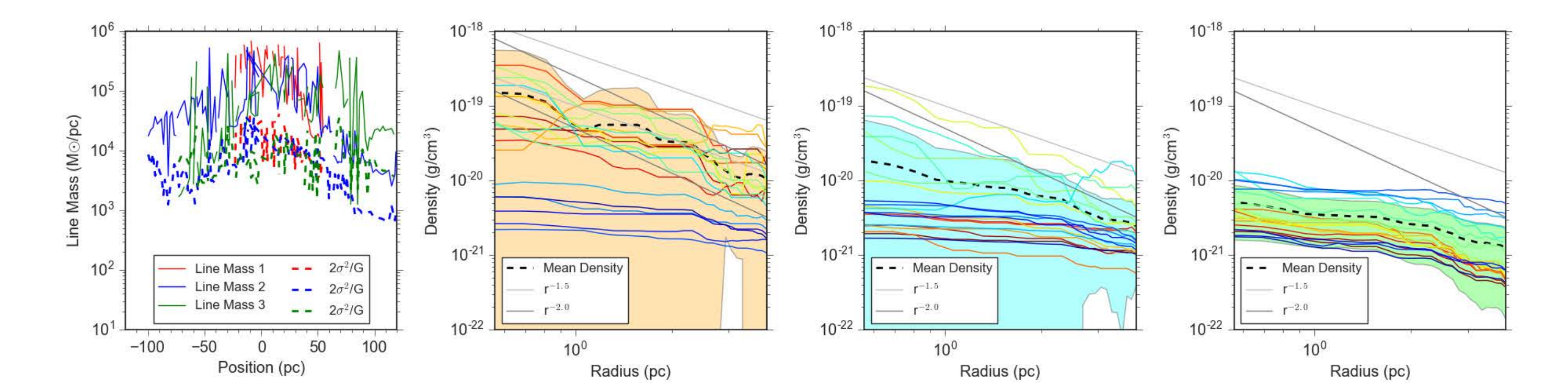}
\caption{Line mass along the spiral accreting flows (first panel) in
and the radial density profiles at the sampling locations 
(multi-colored lines) for flow 1 (second panel), flow 2 (third panel), 
and flow 3 (fourth panel) for the time frame at 3.2~Myr after the 
zoom-in restart. The shaded region in the density profile represents 
the standard deviation of the density spread.}
\label{Fig:lineMass_disk}
\end{figure*}

It is obvious in Fig.~\ref{Fig:lineMass_disk} that the three spiral 
structures are predominantly unstable against the local turbulent support, 
with typical line mass reaches between 10$^4$--10$^5$~M$_{\sun}$. 
However, the main source of support in the disk structure is the 
rotational support, hence fragmentation occurs while the accretion 
flows spiral around the disk. The density distributions in the 
cross-sections along the sampling points of the three filaments are 
shown in the second to fourth panels for filament 1, 2, and 3, respectively. 
Because filament 1 traces the densest spiral, the central density along 
the spiral tends to be as large as 10$^4$--10$^5$~cm$^{-3}$. The density 
spread is also the largest as the cross-sections along filament 1 
also cut through some low density gaps between the inner and outer 
spiral structures. On the other hand, the bulk densities in the 
cross-sections of filament 3 barely pass $\sim$10$^3$~cm$^{-3}$, 
with a low density spread as it mostly traces the extended region. 
The density properties of filament 2 lie in between filament 1 and 3. 
For all three spiral structures, a $r^{-1.5}$ or a flatter density 
profile fits better the density distributions in the cross-sections 
of the filamentary spiral structures.

On 100 pc scales, the accretion rate onto these disk filaments can be computed, as before, by using 
data from Fig.~\ref{Fig:lineMass_disk}.   From the right 3 panels of this figure we find that the 
filament radii are now of the order $ r_f \simeq 3 $ pc with an external density at that radius of 
$\rho_e \simeq 10^{-21} $ g cm$^{-3}$.  We show detailed velocity cut associated with these filaments in Appendix B, Fig.~\ref{Fig:quiet_velprof}, where we find the accretion velocity onto the filaments to be $ v_r \simeq  20 $  km s$^{-1}$ giving an accretion rate onto a 100 pc filament of
\begin{equation}
    \dot {M}_{GMC,f} = 5.36 \times 10^5 \rm{M_{\odot}} / \rm{Myr}\end{equation}

The multi-scale analysis of this disk region poses an important question: at what scale does the transition from 
galactic shear and turbulence to local gravity and rotation take place? 
We can get some indication of this by examining the scale behaviour of the turbulence in this region.
The analysis carried out in Appendix C shows that there appears to be a transition scale at which local effects 
of the forming disk take over 
the turbulence cascade.   This is around 50--100~pc scale when the 
gas number density reaches few 100 to 1000~cm$^{-3}$.  We note that this is of the order of the scale height of the galactic disk, which may be playing a role. 

Finally, the existence of a flattened, disk like structure for a GMC in the quiescent region may be of interest in connection with the discovery 
of a sheet like geometry for some molecular clouds, mentioned in the Introduction \cite{RezaeiKh+2020, RezaeiKh_Kainulainen2022, Zucker+2021}.  
We suggest that the flattened geometry of the filaments in these observations may have some connection to disk-like rotating structure containing filamentary GMCs, discovered here.  

\section{Discussion \& Conclusions}
\label{Chap.Discuss}

We have developed a simulation approach  using the RAMSES code for tracking the formation of structure from 
galactic to sub-parsec cluster scales in a magnetized spiral galaxy undergoing supernova feedback from large OB associations
(star particles).   On galactic scales our simulations reach an initial resolution of 4.85 pc  and the galaxy is fully dynamic:  all
turbulence or spiral wave dynamics are naturally occurring and self-consistent.   We utilize a 
zoom-in approach in which regions of interest in the galactic disk are further refined  within 
a 3 kpc corotating box to a resolution reaching 20 levels of refinement, or 0.286 pc, while de-refining the rest of the galaxy.   As our simulated box still keeps the background galactic disk, the 
large scale flows remain connected to smaller structures as we evolve the zoomed-in region.   Supernova driven bubbles
on kpc scales dominate the feedback processes, whose walls compress lower density gas into kpc scale atomic filament
structures.   Among the wide variety of structures available, we chose to zoom into two, qualitatively different, 3 kpc subregions:
one is active in having at least 3 superbubbles from previous events compressing lower density gas, and a quieter region in which 
feedback is less prevalent and the action of a spiral wave segment is important.  

\subsection{Critical Line Masses: Filament Scaling Relations and Structure Formation}
\label{S.scaling_relations}

We have found in our simulations that supersonic turbulence on scales exceeding 30 pc ( Fig.~\ref{Fig:power_spec}  )  follows a Burger's relation, $ \sigma_{nt}^2 \propto L $,  which agrees with the re-analysis of the size-line width relation by  \citet{Hacar+2022}.   

The critical line mass  in a medium with $\sigma \propto L^{1/2}$ becomes a function of the filament length scale $L$ \citep{Hacar+2022}: 
\begin{equation}
\begin{split}
m_{crit} =  \frac{ 2 \sigma^2}{G}
 = \frac{ 2 c_s^2}{G} \left ( 1 + \frac{ L}{L_o} \right ) \\
 = 16.6   \left ( 1 + \frac{ L}{L_o} \right ).
\left ( \frac{T}{10\, \rm{K}} \right ) \rm{M_{\odot} pc}^{-1} 
 \end{split}
\end{equation}
where $L_o$ is that scale at which the turbulent speeds begin to exceed the sound speed; observations give $L_o \simeq 0.5 $ pc.    On scales much exceeding this thermal scale $L_o$,  the critical lines mass becomes $m_{crit} \propto L$.   For filaments near to their critical line masses, this gives the relation quoted in the Introduction $L \propto M^{0.5}$  \citep{Hacar+2022}.   On kpc scales then, the critical line mass becomes of the order $m_{crit, kpc} \simeq 3.32 \times 10^4 \rm{M_{\odot} pc}^{-1} $.  This corresponds well with the critical line masses for the massive atomic filaments in our simulation (see Fig.~\ref{Fig:lineMass_large}).   

We have not yet compiled the statistics of filaments in our galactic simulations in order to determine the distribution of their  lengths and line masses - and leave this for future works.  Such a procedure has been done for the observed Orion B molecular cloud  \citep{Orkisz+2019} who showed that the probability distribution function for line masses is lognormal.

\subsubsection{Magnetized filaments}

As noted in the introduction, the expression for the critical line mass has a correction factor that depends upon the magnetic field strength and geometry.  For a uniform filament.  From equation 28 in \citet{Fiege_Pudritz2000a}, it is easily shown that the critical line mass is modified by a magnetic correction factor that accounts for both a poloidal field (along the axis) and as well as  possibly wrapped helical field components;

\begin{equation}
    m_{crit, B} = f_B . m_{crit} 
\end{equation}
where $f_B$ is a magnetic field correction factor for the filament, given by 

\begin{equation}
f_B = \frac{1 + (v_A/\sigma)^2}{1 + (v_{A,\phi} / \sigma_c)^2}.
\end{equation}
Here, $v_A = B_{\parallel} / \sqrt{4 \pi \rho}$ is the Alfv\'en speed along the field in the filament, $v_{A,\phi} = B_{\phi} / \sqrt{4 \pi \rho}$ is the Alfv\'en speed in the wrapping helical field, and $\sigma_c^2 = 4 \pi G \rho r_f^2$ is the square of the velocity dispersion for the self-gravitating filament.  

In the absence of a helical filament field ( $v_{A,\phi}= 0$ ), the relation simply says that the critical mass of the filament is increased by the magnetic pressure of the filament field.  For turbulence that is trans-Alfv\'enic $(v_A/\sigma \simeq 1)$, then $f_B \simeq 2$.  Magnetic effects therefore could be contributing to the scatter observed in the \citet{Hacar+2022} plots.  We also note that in a self-gravitating filament in which flows along the filament axis occur, the field lines that are perpendicular to the filament axis on the filament surface will, upon penetrating through to the core of the filament, be dragged along with the gravitationally driven flow onto the fragment(s) - thus creating the $B_{\parallel}$ field directed along the filament axis (for simulations that show this, see \citet{Klassen+2017, Beuther+2020a}).

A helical field will reduce the critical line mass due to its compressing effect on the filament.  It acts in concert with gravity in driving instability by reducing the critical line mass.  As an example, for the case of trans Alv\'enic turbulence in which the filament also has a significant toroidal field $v_{A,\phi} \simeq \sigma_c$, then the magnetic effects cancel leaving $f_B \simeq 1$ 

\subsection{Filament Fragmentation: GMC and Cluster Masses}

Our simulations show that the filaments are undergoing fragmentation processes into substructures on varying scales (GMCs and cluster scale clumps). While this is a dynamic and non-linear process, it  is still useful to compare these fragmentation scales with predicted by linear instability theory for filaments in equilibrium.  \citet{Fiege_Pudritz2000b} presented numerical 
calculations of the instability of magnetized equilibrium filaments whose radial density profiles are similar to those found in our dynamical simulations. 

Two general regimes of instability are identified for magnetized filaments with a general mix of poloidal and toroidal field:  gravity driven fragmentation which occurs on longer scales for filaments whose poloidal field dominates the toroidal field, and MHD driven modes on small scales if the toroidal field is stronger than the poloidal filament field.  We adopt the former in order to compare fragmentation with the fragmentation we observe in our simulation on multiple scales.
   
The wavenumber of the fastest growing modes parallel to the filament (z) axis for filaments $ k_{z,max}$ were computed for a wide range of filament magnetizations \citep{Fiege_Pudritz2000b}.   For filaments that have line masses from $0.2 - 1.0 $ of their critical line mass, Figs 2 and 6,  Table 2 in that paper, show that for filament mass to flux   ( $ \equiv \Gamma_z^{-1} $ in that paper) greater than unity ( $\Gamma_z^{-1} \ge 1 $ ) then;

\begin{equation}
r_o k_{z, max} =  0.462
\end{equation}
where $r_o$ is the core radius of the filament, and $\lambda_{max} = 2 \pi / k_{z, max} $ is the associated spacing between density peaks of this most unstable mode.  In the limit of very  with mass to flux ratios much less than unity ( $\Gamma_z^{-1} \le 1 $);
\begin{equation}
r_o k_{z, max} =  0.214
\end{equation} 
for which the spacing between fragment peaks becomes a factor of at least 2 greater because of magnetic support against fragmentation.  

The point to these linear instability results is that they can be scaled to filaments of any width.   
For the high mass to flux filaments then, the linear instability results suggest that the ratio of fragment spacing to width of the filament is $ \lambda_{max} = 2 \pi r_o / (0.462 ) = 13.6 r_o$.  We note that this result is similar to that found for other fragmentation schemes.  As an example, \citet{Hacar+2022} review a number of earlier models, for which $ \lambda_{max} \simeq 2 \times \lambda_{crit} = 7.86 R_{flat} = 22.2 r_o $

We compare the predictions of \citep{Fiege_Pudritz2000b} with our simulations.  We noted that the typical spacing between density peaks for the kpc scale filaments in our  simulations was 200 pc.  We also noted that the typical core-radius $r_o \simeq 15 $ pc for these, which then would predict a spacing of $\lambda_{max} \simeq 13.6 \times 15 \rm{pc} = 204 \rm{pc} $, which is in good agreement with our numerical data.
The variation of the spacing of fragments along the filament can arise from a number of effects, one of them being magnetic pressure variations. We will follow this up in a future study. 

For filaments near to their critical line mass,  the fragment mass that  one expects is the critical line mass times the fragment length scale.  For magnetized filaments, this gives;

\begin{equation}
M_{frag} = m_{crit,B} \: . \: \lambda_{max}  
\end{equation}
 
Assuming that gravity and not magnetic fields are responsible for the fragmentation, then we may substitute  $ \lambda_{max} = 13.6 r_o$ to derive a general scaling for filament fragments, where the values of $m_{crit}$ and $r_o$ are used for the appropriate scale. Thus from linear instability theory:

\begin{equation}
    M_{frag} \simeq 13.6 \: m_{crit} \: r_o
\end{equation}

We apply this to our kpc and molecular scale filaments.  Introducing the critical line mass and core radius for the kpc atomic filament in the active region:  $m_{crit} \simeq 4 \times 10^4 \rm{M}_{\odot};  r_o \simeq 10 pc$, we find that 
\begin{equation}
M_{frag,a}= 5.4 \times 10^6 \rm{M}_{\odot} 
\end{equation}
This is on the more massive side of the GMC mass pdf.  This suggests that the filaments being produced via superbubble convergence produce some of the most massive GMCs in our simulation.  

For the GMC, 200pc scale, we have from our data $m_{crit} \simeq  10^3 \rm{M}_{\odot};  r_o \simeq 1.0 pc$, so that the fragmentation of GMC filaments, 
\begin{equation}
M_{frag, GMC}= 1.4 \times 10^4 \rm{M}_{\odot}. 
\end{equation}
This is typical of the lower end of the spectrum of massive star clusters, which are capable of producing small OB associations. 

Summarizing, gravitational fragmentation of filaments near their critical line mass produces GMC and cluster masses in atromic and GMC filaments.  Fragment masses depend on the critical line mass and core radius at that scale.  While the critical line mass scales with L, the scaling of the core radius $r_o$ is yet to be determined.  We simply use values taken from our simulations.  

\subsection{Scaling Relations for Filament Accretion and Aligned Flows}

Our numerical results show that accretion rates onto filaments exceed filament aligned mass flow rates by an order of magnitude or more over the time scales that we are examining.  This shows that filamentary hierarchies are highly efficient in gathering mass and moving it from scale to scale. There is more than enough mass flux available to make GMCs on Myr time scales, addressesing the general question of how the large amounts of gas needed to make molecular clouds is collected \citep{McKee_Ostriker2007}. Supershell driven mass flow results in high rates of accretion flow onto filament, and to a lesser extent, spiral wave driven accretion seems to perform the same for the outer filaments we have examined.  

More importantly, we have found that the filament aligned flow rates on larger scales exceed those on the 100pc scales by an order of magnitude. Recall that in \S 4.4a and \S 4.4b: $ \dot {M}_{a,f\parallel} = 2.68 \times 10^4 f_{env} 
\rm{M}_{\odot} / \rm{Myr} $ and $\dot {M}_{GMC,f\parallel} = 2.23 \times 10^3 f_{env} 
\rm{M}_{\odot} / \rm{Myr} $, respectively.
These results also suggest that there is a scaling relation between filament aligned flow rates and filament scale.  

To investigate this, we note that the equation for the field aligned accretion rate takes a simple form when we express it in terms of a line mass.  Specifically,  Equation~\ref{eqn:flowrate_parallel} can be rewritten as 
\begin{equation}
    \dot M_{f,\parallel} = M_{f, \parallel} (\Delta v_{\parallel} / L) = m \:. \: \Delta v_{\parallel}
\end{equation}
where $m$ is the line mass $M_f / L$.  For a filament  near to its critical line mass, then
\begin{equation}
    \dot M_{f,\parallel} = m_{crit,B} \:. \: \Delta v_{\parallel} \propto f_B\: . \: \Delta v_{\parallel} \: . \: L
\:    \end{equation}
where for the last inequality we consider the regime where we are considerably larger than the thermal scale $L \ge L_o$.   For small scale thermal filaments $L \leq L_o$ characterizing the formation of individual stars, this relation predicts that the flow rate is relatively independent of $L$.  

This result then provides a theoretical argument that accretion rates in larger filaments scale with filament scale (or equivalently, line mass).  This is an important result that arises from the scaling relation in Burgers turbulence. A caveat here is whether or not there is some scaling of $v_{\parallel}$.  Our numerical results seem to indicate that if there is, it is not strong on the scales we are simulating.  

\subsection{Correspondence with Local GMCs in the Milky Way}

Our simulations (see Fig.~\ref{Fig:active_0083} )  produce cloud structures that are impacted by superbubble interaction. The structures in the active 3kpc region strongly resemble the observations of the 3D structure own local
group of molecular clouds which are strongly associated with bubbles \cite{RezaeiKh+2020,RezaeiKh_Kainulainen2022, Zucker+2021, Zucker+2022}. This correspondence may even extend to the traveling, kpc scale Radcliffe wave disturbance in the gas density and molecular cloud distribution identified in Milky Way structure studies \citep{Konietzka+2024}. A detailed study of the associated velocity field of our filament will appear in our subsequent paper. 

We have discovered that some GMCs may also appear as flattened, rotating disk like structures in which filaments form
by Toomre instabilities.   These rotating 100 pc structures appear in the quiescent region in our simulations where shear 
effects may not be overcome by forcing by expanding bubble walls.   These results for the quiescent region are compatible with recent observations of 3D GMC structures which are filament like or more
sheet-like structures  \citep{RezaeiKh+2020}.  We suggest that it would be interesting to search for larger scale rotation
of the observed flattened structures.

\subsection{Scaling to Observed Low Mass Cluster Formation?}
 
For both the atomic and GMC filaments in our simulations, we have seen that the accretion rates onto the filaments exceed the flow rates through the filaments by at least an order of magnitude. 

It is interesting to compare our results  with observations of filamentary flows for the formation of the small scale and low mass Serpens South cluster studied by \citet{Kirk+2013}.  This is a forming young, low mass cluster with a dense grouping of about 600 stars, embedded in a filament.  The inferred accretion rate onto the filament is $\dot M_f \simeq 130 - 280 \: \rm{M_{\odot} Myr^{-1}}$, and a much lower filament aligned flow of $\dot M_{f, \parallel} \simeq 28 \: \rm{M_{\odot} Myr^{-1}}$.  The ratio of these two rates is about a factor of 10 as we see in our simulations, summarized in \S 5.3.  The authors also report that the critical line mass in Serpens south filament, which is 0.33 pc long, is $ m_ \simeq 60.3 \rm{M_{\odot}} \rm{pc}^{-1}$.  This value is three times greater than the critical (thermal) line mass in the filament, or about $20 \rm{M_{\odot}} \rm{pc}^{-1}$. 

Are our simulation results pertinent to systems like these? We note that multiplying the Serpens filament length scale by a factor of 300 to scale to 100 pc filaments, then our scaling relation would predict a critical line mass of the order $ 6 \times 10^3 \rm{M_{\odot}} \rm{pc}^{-1}$ and filament aligned flow rates at the much higher values of 300 times greater, or nearly $ \dot M_{f, \parallel} \simeq 8.4 \times 10^3 \: \rm{M_{\odot} Myr^{-1}}$.  This is larger but of the order of the flow rate in our simulation results. The difference here is likely to reflect differences in the value of $v_{\parallel} $ in the filament, which is dependent on the surrounding environment.

\subsection{Structure of the Magnetic Field}
\label{S.BField}

Although we initialized a relatively weak toroidal magnetic field at the 
galactic scale, we find that turbulent dynamo processes produce galactic fields at strengths of a few $ \mu \rm{G} $ in the diffuse atomic gas. Compression likely produces the $\simeq 10 \mu \rm{G} $ field we see in the denser molecular gas, following the roughly $B \propto \rho^{2/3}$ relation.  Superbubbles disrupt magnetically channeled 
flows arising from the Parker instability \ct{Kortgen+2018, Kortgen+2019}) and drive outflow of fields and hot gas into 
the galactic halo.  In the galactic plane, magnetic
 fields are swept up into superbubble walls. Here, one sees regions where the field is parallel to the filament as well those where it is perpendicular.  Fields will be stretched, strengthened, and made more orderly as a consequence of being stretched by the expanding shells.  This may be why molecular filaments are often observed to have field lines perpendicular to the filament axis.  
 
We leave to another paper the task of computing the critical line masses for magnetized filaments - to see if filament fields parallel to the filament axis significantly affect the critical line mass.

\subsection{Conclusions}
We summarize our key findings below.  Our most general findings are:

\begin{itemize}

\item Our simulations show that a spiral galaxy with active steller feedback creates a hierarchy of filamentary and superbubble structures. The gravitational fragmentation of filaments that exceed their critical line mass produce a hierarchy of structure: GMCs within kpc atomic filaments,  and cluster masses within filamentary GMCs.  
\item Fragmentation sets in after accretion onto subcritical filaments (whether by expanding superbubbles or spiral waves) drives them over the critical line mass at the relevant physical scale. 
\item Our results bear a number of close similarities with observed multiscale structures,  including: the association of nearby molecular clouds in the Milky Way with  large scale bubble features;  the properties of kpc length filaments observed in both atomic and molecular gas; the nearly periodic fragmentation of some filaments into molecular clouds and on smaller scales, star clusters;  the 3D structure of GMCs which may be both filamentary or disk like in the organization; and possibly the origin of helical magnetic fields which have been suggested but not seen in other simulations.

\end{itemize}

Our more detailed results are:

\begin{itemize}

\item Large scale filaments found in the galactic halo out to a galactic radius of 10 kpc are oriented perpendicular to the disk. This is a consequence of superbubble eruptions out of the disk plane. 
\item The mass spectrum of GMCs follows a Schechter function in which the cutoff of the most massive GMCs arises from supernova feedback.
\item  Supernova driven superbubbles and spiral waves create supersonic turbulence that results in a hierarchy of filamentary structure; from galactic scales
to the resolution limits of our simulations (0.285 pc)
\item The associated turbulence is characteristic of a shocked medium - Burger's turbulence where  $ \sigma_{nt} \propto L^{1/2} $.  
\item The radial structure of filaments on all scales is similar, and well fit with a Plummer profile with power laws $p = 1.5 - 2.0$ in accord with observations. 
\item The core radii of filaments measured by $R_{flat}$ are not fixed, but appear to scale with L.  Values for  $R_{flat}$ are limited by the resolution of our simulations.
\item The Burger's relation found in our simulations predicts the value of critical line mass with scale $L$.   Our simulations show that fragmentation occurs at these predicted values. 
\item Filament accretion rates exceed filament aligned mass flow rates by about a factor of 10 for both kpc filaments, as well as on the 100 pc scale GMC filaments.
\item The filament aligned mass flow rate scales with the critical line mass, and for atomic and large scale GMC filaments, are proportional to the scale $L$, as predicted by Burgers scalings. 
\item  In active regions, large scale kpc filamentary structures are produced in the walls of converging superbubbles.  Such filaments
become gravitationally unstable when filamentary accretion driven by the expanding shells drives their line mass over the critical value, with filament accretion rates of the order  $4.5 \times 10^6  \rm{M_{\odot} yr}^{-1} $.

\item Flattened rotating clouds of 100 pc scales are formed in quieter regions in the outer regions of our simulation, due to the convergence of filamentary flows.  Here, galactic shear appears to play a more important role.   Filaments within 
such a forming disk like cloud arise from Toomre instabilities.  Fragmentation
of such filaments into clusters by gravitational instability.
\item We have evidence for helical fields that arise in association with with filaments that are flowing into the disk like structure.  The disk itself has a toroidal magnetic field structure.  
\end{itemize}
 
There are many aspects of our simulations that will be explored in future work.  As an example, we are using the Polaris code \citep{Reissl+2016} to perform synthetic observations of dust polarization features as signatures of MHD structures in our simulations, in part to test for helical magnetic fields.
Also, a statistical analysis of a complete sample of filaments on galactic and zoomed in scales, as well as the characterization of their structure is nearing completion and will be presented in a subsequent paper (Pillsworth et al, 2024, in preparation).  This work will also include detailed analyses of filamentary flows onto filaments and their fragments (GMCs, cluster clumps).

\section*{Acknowledgements}
We are happy to acknowledge interesting discussions with Henrik Beuther, Ralf Klessen, Alyssa Goodman, Thomas Henning, Johnny Henshaw,  No\'e Brucy,  David Whitworth, Stefan Reissl,  Shanghuo Li, Sara Rezaei Kh., Molly Wells, Catherine Zucker, Eric Koch, Marta Reino-Campos, Oleg Gnedin, and Jerry Sellwood.  We thank Romain Teyssier for help in tuning the RAMSES code used for the simulations.  We also thank an anonymous referee whose report helped improve the paper.  REP is grateful for the support and hospitality extended to him during his sabbatical leave (2022-2023) at ITA and MPIA in Heidelberg, where the manuscript was completed.  REP and JW are supported by Discovery Grants from NSERC of Canada. RP is partially supported by an OGS graduate scholarship, and HR is supported by an NSERC postgraduate scholarship.  Computational resources for this project were enabled by a grant to REP from Compute Canada/Digital Alliance Canada and carried out on Cedar and Graham computing clusters.

\bibliographystyle{mnras}
\bibliography{references}

\begin{thebibliography}{}
\makeatletter
\relax
\def\mn@urlcharsother{\let\do\@makeother \do\$\do\&\do\#\do\^\do\_\do\%\do\~}
\def\mn@doi{\begingroup\mn@urlcharsother \@ifnextchar [ {\mn@doi@}
  {\mn@doi@[]}}
\def\mn@doi@[#1]#2{\def\@tempa{#1}\ifx\@tempa\@empty \href
  {http://dx.doi.org/#2} {doi:#2}\else \href {http://dx.doi.org/#2} {#1}\fi
  \endgroup}
\def\mn@eprint#1#2{\mn@eprint@#1:#2::\@nil}
\def\mn@eprint@arXiv#1{\href {http://arxiv.org/abs/#1} {{\tt arXiv:#1}}}
\def\mn@eprint@dblp#1{\href {http://dblp.uni-trier.de/rec/bibtex/#1.xml}
  {dblp:#1}}
\def\mn@eprint@#1:#2:#3:#4\@nil{\def\@tempa {#1}\def\@tempb {#2}\def\@tempc
  {#3}\ifx \@tempc \@empty \let \@tempc \@tempb \let \@tempb \@tempa \fi \ifx
  \@tempb \@empty \def\@tempb {arXiv}\fi \@ifundefined
  {mn@eprint@\@tempb}{\@tempb:\@tempc}{\expandafter \expandafter \csname
  mn@eprint@\@tempb\endcsname \expandafter{\@tempc}}}

\bibitem[\protect\citeauthoryear{{Abreu-Vicente}, {Ragan}, {Kainulainen},
  {Henning}, {Beuther}  \& {Johnston}}{{Abreu-Vicente}
  et~al.}{2016}]{Abreu-Vicente+2016}
{Abreu-Vicente} J.,  {Ragan} S.,  {Kainulainen} J.,  {Henning} T.,  {Beuther}
  H.,   {Johnston} K.,  2016, \mn@doi [\aap] {10.1051/0004-6361/201527674},
  \href {https://ui.adsabs.harvard.edu/abs/2016A&A...590A.131A} {590, A131}

\bibitem[\protect\citeauthoryear{{Agertz} \& {Kravtsov}}{{Agertz} \&
  {Kravtsov}}{2015}]{Agertz+2015}
{Agertz} O.,  {Kravtsov} A.~V.,  2015, \mn@doi [\apj]
  {10.1088/0004-637X/804/1/18}, \href
  {https://ui.adsabs.harvard.edu/abs/2015ApJ...804...18A} {804, 18}

\bibitem[\protect\citeauthoryear{{Agertz}, {Teyssier}  \& {Moore}}{{Agertz}
  et~al.}{2011}]{Agertz+2011}
{Agertz} O.,  {Teyssier} R.,   {Moore} B.,  2011, \mn@doi [\mnras]
  {10.1111/j.1365-2966.2010.17530.x}, \href
  {https://ui.adsabs.harvard.edu/abs/2011MNRAS.410.1391A} {410, 1391}

\bibitem[\protect\citeauthoryear{{Ahmadi} et~al.,}{{Ahmadi}
  et~al.}{2018}]{Ahmadi+2018}
{Ahmadi} A.,  et~al., 2018, \mn@doi [\aap] {10.1051/0004-6361/201732548}, \href
  {https://ui.adsabs.harvard.edu/abs/2018A&A...618A..46A} {618, A46}

\bibitem[\protect\citeauthoryear{{Ahmadi} et~al.,}{{Ahmadi}
  et~al.}{2023}]{Ahmadi+2023}
{Ahmadi} A.,  et~al., 2023, \mn@doi [arXiv e-prints]
  {10.48550/arXiv.2305.00020}, \href
  {https://ui.adsabs.harvard.edu/abs/2023arXiv230500020A} {p. arXiv:2305.00020}

\bibitem[\protect\citeauthoryear{{Andr{\'e}} et~al.,}{{Andr{\'e}}
  et~al.}{2010}]{Andre+2010}
{Andr{\'e}} P.,  et~al., 2010, \mn@doi [\aap] {10.1051/0004-6361/201014666},
  \href {https://ui.adsabs.harvard.edu/abs/2010A&A...518L.102A} {518, L102}

\bibitem[\protect\citeauthoryear{{Andr{\'e}}, {Di Francesco}, {Ward-Thompson},
  {Inutsuka}, {Pudritz}  \& {Pineda}}{{Andr{\'e}} et~al.}{2014}]{Andre+2014}
{Andr{\'e}} P.,  {Di Francesco} J.,  {Ward-Thompson} D.,  {Inutsuka} S.~I.,
  {Pudritz} R.~E.,   {Pineda} J.~E.,  2014, in {Beuther} H.,  {Klessen} R.~S.,
  {Dullemond} C.~P.,   {Henning} T.,  eds, Protostars and Planets VI. p.~27
  (\mn@eprint {arXiv} {1312.6232}),
  \mn@doi{10.2458/azu\_uapress\_9780816531240-ch002}

\bibitem[\protect\citeauthoryear{{Andr{\'e}}, {Arzoumanian}, {K{\"o}nyves},
  {Shimajiri}  \& {Palmeirim}}{{Andr{\'e}} et~al.}{2019}]{Andre+2019}
{Andr{\'e}} P.,  {Arzoumanian} D.,  {K{\"o}nyves} V.,  {Shimajiri} Y.,
  {Palmeirim} P.,  2019, \mn@doi [\aap] {10.1051/0004-6361/201935915}, \href
  {https://ui.adsabs.harvard.edu/abs/2019A&A...629L...4A} {629, L4}

\bibitem[\protect\citeauthoryear{{Arzoumanian} et~al.,}{{Arzoumanian}
  et~al.}{2011}]{Arzoumanian+2011}
{Arzoumanian} D.,  et~al., 2011, \mn@doi [\aap] {10.1051/0004-6361/201116596},
  \href {https://ui.adsabs.harvard.edu/abs/2011A&A...529L...6A} {529, L6}

\bibitem[\protect\citeauthoryear{{Barnes} et~al.,}{{Barnes}
  et~al.}{2023}]{Barnes+2023}
{Barnes} A.~T.,  et~al., 2023, \mn@doi [\apjl] {10.3847/2041-8213/aca7b9},
  \href {https://ui.adsabs.harvard.edu/abs/2023ApJ...944L..22B} {944, L22}

\bibitem[\protect\citeauthoryear{{Beuther} et~al.,}{{Beuther}
  et~al.}{2018}]{Beuther+2018}
{Beuther} H.,  et~al., 2018, \mn@doi [\aap] {10.1051/0004-6361/201833021},
  \href {https://ui.adsabs.harvard.edu/abs/2018A&A...617A.100B} {617, A100}

\bibitem[\protect\citeauthoryear{{Beuther} et~al.,}{{Beuther}
  et~al.}{2020a}]{Beuther+2020b}
{Beuther} H.,  et~al., 2020a, \mn@doi [\aap] {10.1051/0004-6361/202037950},
  \href {https://ui.adsabs.harvard.edu/abs/2020A&A...638A..44B} {638, A44}

\bibitem[\protect\citeauthoryear{{Beuther} et~al.,}{{Beuther}
  et~al.}{2020b}]{Beuther+2020a}
{Beuther} H.,  et~al., 2020b, \mn@doi [\apj] {10.3847/1538-4357/abc019}, \href
  {https://ui.adsabs.harvard.edu/abs/2020ApJ...904..168B} {904, 168}

\bibitem[\protect\citeauthoryear{{Beuther} et~al.,}{{Beuther}
  et~al.}{2022}]{Beuther+2022}
{Beuther} H.,  et~al., 2022, \mn@doi [\aap] {10.1051/0004-6361/202244040},
  \href {https://ui.adsabs.harvard.edu/abs/2022A&A...665A..63B} {665, A63}

\bibitem[\protect\citeauthoryear{{Bleuler}, {Teyssier}, {Carassou}  \&
  {Martizzi}}{{Bleuler} et~al.}{2015}]{Bleuler+2015}
{Bleuler} A.,  {Teyssier} R.,  {Carassou} S.,   {Martizzi} D.,  2015, \mn@doi
  [Computational Astrophysics and Cosmology] {10.1186/s40668-015-0009-7}, \href
  {https://ui.adsabs.harvard.edu/abs/2015ComAC...2....5B} {2, 5}

\bibitem[\protect\citeauthoryear{{Bournaud}, {Elmegreen}, {Teyssier}, {Block}
  \& {Puerari}}{{Bournaud} et~al.}{2010}]{Bournaud+2010}
{Bournaud} F.,  {Elmegreen} B.~G.,  {Teyssier} R.,  {Block} D.~L.,   {Puerari}
  I.,  2010, \mn@doi [\mnras] {10.1111/j.1365-2966.2010.17370.x}, \href
  {https://ui.adsabs.harvard.edu/abs/2010MNRAS.409.1088B} {409, 1088}

\bibitem[\protect\citeauthoryear{{Chira}, {Kainulainen},
  {Ib{\'a}{\~n}ez-Mej{\'\i}a}, {Henning}  \& {Mac Low}}{{Chira}
  et~al.}{2018}]{Chira+2018}
{Chira} R.~A.,  {Kainulainen} J.,  {Ib{\'a}{\~n}ez-Mej{\'\i}a} J.~C.,
  {Henning} T.,   {Mac Low} M.~M.,  2018, \mn@doi [\aap]
  {10.1051/0004-6361/201731836}, \href
  {https://ui.adsabs.harvard.edu/abs/2018A&A...610A..62C} {610, A62}

\bibitem[\protect\citeauthoryear{{Clarke} \& {Whitworth}}{{Clarke} \&
  {Whitworth}}{2015}]{Clarke_Whitworth2015}
{Clarke} S.~D.,  {Whitworth} A.~P.,  2015, \mn@doi [\mnras]
  {10.1093/mnras/stv393}, \href
  {https://ui.adsabs.harvard.edu/abs/2015MNRAS.449.1819C} {449, 1819}

\bibitem[\protect\citeauthoryear{{Crutcher} \& {Kemball}}{{Crutcher} \&
  {Kemball}}{2019}]{Crutcher2019}
{Crutcher} R.~M.,  {Kemball} A.~J.,  2019, \mn@doi [Frontiers in Astronomy and
  Space Sciences] {10.3389/fspas.2019.00066}, \href
  {https://ui.adsabs.harvard.edu/abs/2019FrASS...6...66C} {6, 66}

\bibitem[\protect\citeauthoryear{{Dale}, {Bonnell}, {Clarke}  \& {Bate}}{{Dale}
  et~al.}{2005}]{Dale+2005}
{Dale} J.~E.,  {Bonnell} I.~A.,  {Clarke} C.~J.,   {Bate} M.~R.,  2005, \mn@doi
  [\mnras] {10.1111/j.1365-2966.2005.08806.x}, \href
  {https://ui.adsabs.harvard.edu/abs/2005MNRAS.358..291D} {358, 291}

\bibitem[\protect\citeauthoryear{{Dale}, {Ercolano}  \& {Bonnell}}{{Dale}
  et~al.}{2012}]{Dale+2012}
{Dale} J.~E.,  {Ercolano} B.,   {Bonnell} I.~A.,  2012, \mn@doi [\mnras]
  {10.1111/j.1365-2966.2012.21205.x}, \href
  {https://ui.adsabs.harvard.edu/abs/2012MNRAS.424..377D} {424, 377}

\bibitem[\protect\citeauthoryear{{Dobbs} \& {Wurster}}{{Dobbs} \&
  {Wurster}}{2021}]{Dobbs+2021}
{Dobbs} C.~L.,  {Wurster} J.,  2021, \mn@doi [\mnras] {10.1093/mnras/stab150},
  \href {https://ui.adsabs.harvard.edu/abs/2021MNRAS.502.2285D} {502, 2285}

\bibitem[\protect\citeauthoryear{{Federrath} \& {Klessen}}{{Federrath} \&
  {Klessen}}{2012}]{Federrath_Klessen2012}
{Federrath} C.,  {Klessen} R.~S.,  2012, \mn@doi [\apj]
  {10.1088/0004-637X/761/2/156}, \href
  {https://ui.adsabs.harvard.edu/abs/2012ApJ...761..156F} {761, 156}

\bibitem[\protect\citeauthoryear{{Fiege} \& {Pudritz}}{{Fiege} \&
  {Pudritz}}{2000a}]{Fiege_Pudritz2000a}
{Fiege} J.~D.,  {Pudritz} R.~E.,  2000a, \mn@doi [\mnras]
  {10.1046/j.1365-8711.2000.03066.x}, \href
  {https://ui.adsabs.harvard.edu/abs/2000MNRAS.311...85F} {311, 85}

\bibitem[\protect\citeauthoryear{{Fiege} \& {Pudritz}}{{Fiege} \&
  {Pudritz}}{2000b}]{Fiege_Pudritz2000b}
{Fiege} J.~D.,  {Pudritz} R.~E.,  2000b, \mn@doi [\mnras]
  {10.1046/j.1365-8711.2000.03067.x}, \href
  {https://ui.adsabs.harvard.edu/abs/2000MNRAS.311..105F} {311, 105}

\bibitem[\protect\citeauthoryear{{Fischera} \& {Martin}}{{Fischera} \&
  {Martin}}{2012}]{Fischera_Martin2012}
{Fischera} J.,  {Martin} P.~G.,  2012, \mn@doi [\aap]
  {10.1051/0004-6361/201218961}, \href
  {https://ui.adsabs.harvard.edu/abs/2012A&A...542A..77F} {542, A77}

\bibitem[\protect\citeauthoryear{{Fujii} et~al.,}{{Fujii}
  et~al.}{2021}]{Fujii+2021}
{Fujii} K.,  et~al., 2021, \mn@doi [\mnras] {10.1093/mnras/stab1202}, \href
  {https://ui.adsabs.harvard.edu/abs/2021MNRAS.505..459F} {505, 459}

\bibitem[\protect\citeauthoryear{{Gehman}, {Adams}  \& {Watkins}}{{Gehman}
  et~al.}{1996}]{Gehman+1996}
{Gehman} C.~S.,  {Adams} F.~C.,   {Watkins} R.,  1996, \mn@doi [\apj]
  {10.1086/178098}, \href
  {https://ui.adsabs.harvard.edu/abs/1996ApJ...472..673G} {472, 673}

\bibitem[\protect\citeauthoryear{{Goodman} et~al.,}{{Goodman}
  et~al.}{2014}]{Goodman+2014}
{Goodman} A.~A.,  et~al., 2014, \mn@doi [\apj] {10.1088/0004-637X/797/1/53},
  \href {https://ui.adsabs.harvard.edu/abs/2014ApJ...797...53G} {797, 53}

\bibitem[\protect\citeauthoryear{{Grisdale}}{{Grisdale}}{2021}]{Grisdale2021}
{Grisdale} K.,  2021, \mn@doi [\mnras] {10.1093/mnras/staa3524}, \href
  {https://ui.adsabs.harvard.edu/abs/2021MNRAS.500.3552G} {500, 3552}

\bibitem[\protect\citeauthoryear{{Hacar}, {Clark}, {Heitsch}, {Kainulainen},
  {Panopoulou}, {Seifried}  \& {Smith}}{{Hacar} et~al.}{2022}]{Hacar+2022}
{Hacar} A.,  {Clark} S.,  {Heitsch} F.,  {Kainulainen} J.,  {Panopoulou} G.,
  {Seifried} D.,   {Smith} R.,  2022, \mn@doi [arXiv e-prints]
  {10.48550/arXiv.2203.09562}, \href
  {https://ui.adsabs.harvard.edu/abs/2022arXiv220309562H} {p. arXiv:2203.09562}

\bibitem[\protect\citeauthoryear{{Harris} \& {Pudritz}}{{Harris} \&
  {Pudritz}}{1994}]{Harris_Pudritz1994}
{Harris} W.~E.,  {Pudritz} R.~E.,  1994, \mn@doi [\apj] {10.1086/174310}, \href
  {https://ui.adsabs.harvard.edu/abs/1994ApJ...429..177H} {429, 177}

\bibitem[\protect\citeauthoryear{{Heiles}}{{Heiles}}{1997}]{Heiles1997}
{Heiles} C.,  1997, \mn@doi [\apjs] {10.1086/313010}, \href
  {https://ui.adsabs.harvard.edu/abs/1997ApJS..111..245H} {111, 245}

\bibitem[\protect\citeauthoryear{{Hennebelle} \& {Chabrier}}{{Hennebelle} \&
  {Chabrier}}{2011}]{Hennebelle_Chabrier2011}
{Hennebelle} P.,  {Chabrier} G.,  2011, \mn@doi [\apjl]
  {10.1088/2041-8205/743/2/L29}, \href
  {https://ui.adsabs.harvard.edu/abs/2011ApJ...743L..29H} {743, L29}

\bibitem[\protect\citeauthoryear{{Henning}, {Linz}, {Krause}, {Ragan},
  {Beuther}, {Launhardt}, {Nielbock}  \& {Vasyunina}}{{Henning}
  et~al.}{2010}]{Henning+2010}
{Henning} T.,  {Linz} H.,  {Krause} O.,  {Ragan} S.,  {Beuther} H.,
  {Launhardt} R.,  {Nielbock} M.,   {Vasyunina} T.,  2010, \mn@doi [\aap]
  {10.1051/0004-6361/201014635}, \href
  {https://ui.adsabs.harvard.edu/abs/2010A&A...518L..95H} {518, L95}

\bibitem[\protect\citeauthoryear{{Henshaw} et~al.,}{{Henshaw}
  et~al.}{2020}]{Henshaw+2020}
{Henshaw} J.~D.,  et~al., 2020, \mn@doi [Nature Astronomy]
  {10.1038/s41550-020-1126-z}, \href
  {https://ui.adsabs.harvard.edu/abs/2020NatAs...4.1064H} {4, 1064}

\bibitem[\protect\citeauthoryear{{Howard}, {Pudritz}  \& {Harris}}{{Howard}
  et~al.}{2017}]{Howard+2017}
{Howard} C.~S.,  {Pudritz} R.~E.,   {Harris} W.~E.,  2017, \mn@doi [\mnras]
  {10.1093/mnras/stx1363}, \href
  {https://ui.adsabs.harvard.edu/abs/2017MNRAS.470.3346H} {470, 3346}

\bibitem[\protect\citeauthoryear{{Howard}, {Pudritz}  \& {Harris}}{{Howard}
  et~al.}{2018}]{Howard+2018}
{Howard} C.~S.,  {Pudritz} R.~E.,   {Harris} W.~E.,  2018, \mn@doi [Nature
  Astronomy] {10.1038/s41550-018-0506-0}, \href
  {https://ui.adsabs.harvard.edu/abs/2018NatAs...2..725H} {2, 725}

\bibitem[\protect\citeauthoryear{{Inutsuka} \& {Miyama}}{{Inutsuka} \&
  {Miyama}}{1997}]{Inutsuka+1997}
{Inutsuka} S.-i.,  {Miyama} S.~M.,  1997, \mn@doi [\apj] {10.1086/303982},
  \href {https://ui.adsabs.harvard.edu/abs/1997ApJ...480..681I} {480, 681}

\bibitem[\protect\citeauthoryear{{Jeffreson}, {Kruijssen}, {Keller}, {Chevance}
   \& {Glover}}{{Jeffreson} et~al.}{2020}]{Jeffreson+2020}
{Jeffreson} S. M.~R.,  {Kruijssen} J.~M.~D.,  {Keller} B.~W.,  {Chevance} M.,
  {Glover} S. C.~O.,  2020, \mn@doi [\mnras] {10.1093/mnras/staa2127}, \href
  {https://ui.adsabs.harvard.edu/abs/2020MNRAS.498..385J} {498, 385}

\bibitem[\protect\citeauthoryear{{Kashiwagi} \& {Tomisaka}}{{Kashiwagi} \&
  {Tomisaka}}{2021}]{Kashiwagi_Tomisaka2021}
{Kashiwagi} R.,  {Tomisaka} K.,  2021, \mn@doi [\apj]
  {10.3847/1538-4357/abea7a}, \href
  {https://ui.adsabs.harvard.edu/abs/2021ApJ...911..106K} {911, 106}

\bibitem[\protect\citeauthoryear{{Kim}, {Ostriker}  \& {Stone}}{{Kim}
  et~al.}{2002}]{Kim+2002}
{Kim} W.-T.,  {Ostriker} E.~C.,   {Stone} J.~M.,  2002, \mn@doi [\apj]
  {10.1086/344367}, \href
  {https://ui.adsabs.harvard.edu/abs/2002ApJ...581.1080K} {581, 1080}

\bibitem[\protect\citeauthoryear{{Kim} et~al.,}{{Kim} et~al.}{2016}]{Kim+2016}
{Kim} J.-h.,  et~al., 2016, \mn@doi [\apj] {10.3847/1538-4357/833/2/202}, \href
  {https://ui.adsabs.harvard.edu/abs/2016ApJ...833..202K} {833, 202}

\bibitem[\protect\citeauthoryear{{Kim} et~al.,}{{Kim}
  et~al.}{2022}]{KimJWST+2022}
{Kim} J.,  et~al., 2022, \mn@doi [arXiv e-prints] {10.48550/arXiv.2211.15698},
  \href {https://ui.adsabs.harvard.edu/abs/2022arXiv221115698K} {p.
  arXiv:2211.15698}

\bibitem[\protect\citeauthoryear{{Kirk}, {Myers}, {Bourke}, {Gutermuth},
  {Hedden}  \& {Wilson}}{{Kirk} et~al.}{2013}]{Kirk+2013}
{Kirk} H.,  {Myers} P.~C.,  {Bourke} T.~L.,  {Gutermuth} R.~A.,  {Hedden} A.,
  {Wilson} G.~W.,  2013, \mn@doi [\apj] {10.1088/0004-637X/766/2/115}, \href
  {https://ui.adsabs.harvard.edu/abs/2013ApJ...766..115K} {766, 115}

\bibitem[\protect\citeauthoryear{{Kirk}, {Klassen}, {Pudritz}  \&
  {Pillsworth}}{{Kirk} et~al.}{2015}]{Kirk+2015}
{Kirk} H.,  {Klassen} M.,  {Pudritz} R.,   {Pillsworth} S.,  2015, \mn@doi
  [\apj] {10.1088/0004-637X/802/2/75}, \href
  {https://ui.adsabs.harvard.edu/abs/2015ApJ...802...75K} {802, 75}

\bibitem[\protect\citeauthoryear{{Klassen}, {Pudritz}, {Kuiper}, {Peters}  \&
  {Banerjee}}{{Klassen} et~al.}{2016}]{Klassen+2016}
{Klassen} M.,  {Pudritz} R.~E.,  {Kuiper} R.,  {Peters} T.,   {Banerjee} R.,
  2016, \mn@doi [\apj] {10.3847/0004-637X/823/1/28}, \href
  {https://ui.adsabs.harvard.edu/abs/2016ApJ...823...28K} {823, 28}

\bibitem[\protect\citeauthoryear{{Klassen}, {Pudritz}  \& {Kirk}}{{Klassen}
  et~al.}{2017}]{Klassen+2017}
{Klassen} M.,  {Pudritz} R.~E.,   {Kirk} H.,  2017, \mn@doi [\mnras]
  {10.1093/mnras/stw2889}, \href
  {https://ui.adsabs.harvard.edu/abs/2017MNRAS.465.2254K} {465, 2254}

\bibitem[\protect\citeauthoryear{{Klessen} \& {Glover}}{{Klessen} \&
  {Glover}}{2016}]{Klessen_Glover2016}
{Klessen} R.~S.,  {Glover} S. C.~O.,  2016, in {Revaz} Y.,  {Jablonka} P.,
  {Teyssier} R.,   {Mayer} L.,  eds,  Saas-Fee Advanced Course Vol. 43,
  Saas-Fee Advanced Course. p.~85 (\mn@eprint {arXiv} {1412.5182}),
  \mn@doi{10.1007/978-3-662-47890-5_2}

\bibitem[\protect\citeauthoryear{{Kong} et~al.,}{{Kong}
  et~al.}{2021}]{Kong+2021}
{Kong} S.,  et~al., 2021, \mn@doi [\apj] {10.3847/1538-4357/abc687}, \href
  {https://ui.adsabs.harvard.edu/abs/2021ApJ...906...80K} {906, 80}

\bibitem[\protect\citeauthoryear{{Kong}, {Whitworth}, {Smith}  \&
  {Hamden}}{{Kong} et~al.}{2022}]{Kong+2022}
{Kong} S.,  {Whitworth} D.~J.,  {Smith} R.~J.,   {Hamden} E.~T.,  2022, \mn@doi
  [\mnras] {10.1093/mnras/stac2932}, \href
  {https://ui.adsabs.harvard.edu/abs/2022MNRAS.517.4679K} {517, 4679}

\bibitem[\protect\citeauthoryear{{Kong}, {Ossenkopf-Okada}, {Arce}, {Klessen}
  \& {Xu}}{{Kong} et~al.}{2023}]{Kong+2023}
{Kong} S.,  {Ossenkopf-Okada} V.,  {Arce} H.~G.,  {Klessen} R.~S.,   {Xu} D.,
  2023, \mn@doi [\apjs] {10.3847/1538-4365/acbfb0}, \href
  {https://ui.adsabs.harvard.edu/abs/2023ApJS..265...58K} {265, 58}

\bibitem[\protect\citeauthoryear{{Konietzka} et~al.,}{{Konietzka}
  et~al.}{2024}]{Konietzka+2024}
{Konietzka} R.,  et~al., 2024, \mn@doi [\nat] {10.1038/s41586-024-07127-3},
  \href {https://ui.adsabs.harvard.edu/abs/2024Natur.628...62K} {628, 62}

\bibitem[\protect\citeauthoryear{{K{\"o}rtgen}, {Banerjee}, {Pudritz}  \&
  {Schmidt}}{{K{\"o}rtgen} et~al.}{2018}]{Kortgen+2018}
{K{\"o}rtgen} B.,  {Banerjee} R.,  {Pudritz} R.~E.,   {Schmidt} W.,  2018,
  \mn@doi [\mnras] {10.1093/mnrasl/sly094}, \href
  {https://ui.adsabs.harvard.edu/abs/2018MNRAS.479L..40K} {479, L40}

\bibitem[\protect\citeauthoryear{{K{\"o}rtgen}, {Banerjee}, {Pudritz}  \&
  {Schmidt}}{{K{\"o}rtgen} et~al.}{2019}]{Kortgen+2019}
{K{\"o}rtgen} B.,  {Banerjee} R.,  {Pudritz} R.~E.,   {Schmidt} W.,  2019,
  \mn@doi [\mnras] {10.1093/mnras/stz2491}, \href
  {https://ui.adsabs.harvard.edu/abs/2019MNRAS.489.5004K} {489, 5004}

\bibitem[\protect\citeauthoryear{{Kraljic}, {Renaud}, {Bournaud}, {Combes},
  {Elmegreen}, {Emsellem}  \& {Teyssier}}{{Kraljic}
  et~al.}{2014}]{Kraljic+2014}
{Kraljic} K.,  {Renaud} F.,  {Bournaud} F.,  {Combes} F.,  {Elmegreen} B.,
  {Emsellem} E.,   {Teyssier} R.,  2014, \mn@doi [\apj]
  {10.1088/0004-637X/784/2/112}, \href
  {https://ui.adsabs.harvard.edu/abs/2014ApJ...784..112K} {784, 112}

\bibitem[\protect\citeauthoryear{{Kritsuk}, {Norman}, {Padoan}  \&
  {Wagner}}{{Kritsuk} et~al.}{2007}]{Kritsuk+2007}
{Kritsuk} A.~G.,  {Norman} M.~L.,  {Padoan} P.,   {Wagner} R.,  2007, \mn@doi
  [\apj] {10.1086/519443}, \href
  {https://ui.adsabs.harvard.edu/abs/2007ApJ...665..416K} {665, 416}

\bibitem[\protect\citeauthoryear{{Krumholz}}{{Krumholz}}{2014}]{Krumholz+2014}
{Krumholz} M.~R.,  2014, \mn@doi [\physrep] {10.1016/j.physrep.2014.02.001},
  \href {https://ui.adsabs.harvard.edu/abs/2014PhR...539...49K} {539, 49}

\bibitem[\protect\citeauthoryear{{Krumholz} \& {McKee}}{{Krumholz} \&
  {McKee}}{2005}]{Krumholz_McKee2005}
{Krumholz} M.~R.,  {McKee} C.~F.,  2005, \mn@doi [\apj] {10.1086/431734}, \href
  {https://ui.adsabs.harvard.edu/abs/2005ApJ...630..250K} {630, 250}

\bibitem[\protect\citeauthoryear{{Kuffmeier}, {Jensen}  \&
  {Haugb{\o}lle}}{{Kuffmeier} et~al.}{2023}]{Kuffmeier+2023}
{Kuffmeier} M.,  {Jensen} S.~S.,   {Haugb{\o}lle} T.,  2023, \mn@doi [European
  Physical Journal Plus] {10.1140/epjp/s13360-023-03880-y}, \href
  {https://ui.adsabs.harvard.edu/abs/2023EPJP..138..272K} {138, 272}

\bibitem[\protect\citeauthoryear{{Kumar}, {Palmeirim}, {Arzoumanian}  \&
  {Inutsuka}}{{Kumar} et~al.}{2020}]{Kumar+2020}
{Kumar} M.~S.~N.,  {Palmeirim} P.,  {Arzoumanian} D.,   {Inutsuka} S.~I.,
  2020, \mn@doi [\aap] {10.1051/0004-6361/202038232}, \href
  {https://ui.adsabs.harvard.edu/abs/2020A&A...642A..87K} {642, A87}

\bibitem[\protect\citeauthoryear{{Larson}}{{Larson}}{1981}]{Larson1981}
{Larson} R.~B.,  1981, \mn@doi [\mnras] {10.1093/mnras/194.4.809}, \href
  {https://ui.adsabs.harvard.edu/abs/1981MNRAS.194..809L} {194, 809}

\bibitem[\protect\citeauthoryear{{Lee}, {Hennebelle}  \& {Chabrier}}{{Lee}
  et~al.}{2017}]{Lee_Hennebelle+2017}
{Lee} Y.-N.,  {Hennebelle} P.,   {Chabrier} G.,  2017, \mn@doi [\apj]
  {10.3847/1538-4357/aa898f}, \href
  {https://ui.adsabs.harvard.edu/abs/2017ApJ...847..114L} {847, 114}

\bibitem[\protect\citeauthoryear{{Leike}, {Glatzle}  \& {En{\ss}lin}}{{Leike}
  et~al.}{2020}]{Leike+2020}
{Leike} R.~H.,  {Glatzle} M.,   {En{\ss}lin} T.~A.,  2020, \mn@doi [\aap]
  {10.1051/0004-6361/202038169}, \href
  {https://ui.adsabs.harvard.edu/abs/2020A&A...639A.138L} {639, A138}

\bibitem[\protect\citeauthoryear{{Mac Low} \& {Klessen}}{{Mac Low} \&
  {Klessen}}{2004}]{Maclow_Klessen2004}
{Mac Low} M.-M.,  {Klessen} R.~S.,  2004, \mn@doi [Reviews of Modern Physics]
  {10.1103/RevModPhys.76.125}, \href
  {https://ui.adsabs.harvard.edu/abs/2004RvMP...76..125M} {76, 125}

\bibitem[\protect\citeauthoryear{{McKee} \& {Ostriker}}{{McKee} \&
  {Ostriker}}{2007}]{McKee_Ostriker2007}
{McKee} C.~F.,  {Ostriker} E.~C.,  2007, \mn@doi [\araa]
  {10.1146/annurev.astro.45.051806.110602}, \href
  {https://ui.adsabs.harvard.edu/abs/2007ARA&A..45..565M} {45, 565}

\bibitem[\protect\citeauthoryear{{Men'shchikov} et~al.,}{{Men'shchikov}
  et~al.}{2010}]{Men'shchikov+2010}
{Men'shchikov} A.,  et~al., 2010, \mn@doi [\aap] {10.1051/0004-6361/201014668},
  \href {https://ui.adsabs.harvard.edu/abs/2010A&A...518L.103M} {518, L103}

\bibitem[\protect\citeauthoryear{{Myers}}{{Myers}}{2009}]{Myers2009}
{Myers} P.~C.,  2009, \mn@doi [\apj] {10.1088/0004-637X/700/2/1609}, \href
  {https://ui.adsabs.harvard.edu/abs/2009ApJ...700.1609M} {700, 1609}

\bibitem[\protect\citeauthoryear{{Navarro}, {Frenk}  \& {White}}{{Navarro}
  et~al.}{1997}]{Navarro+1997}
{Navarro} J.~F.,  {Frenk} C.~S.,   {White} S. D.~M.,  1997, \mn@doi [\apj]
  {10.1086/304888}, \href
  {https://ui.adsabs.harvard.edu/abs/1997ApJ...490..493N} {490, 493}

\bibitem[\protect\citeauthoryear{{Nordlund} \& {Padoan}}{{Nordlund} \&
  {Padoan}}{1999}]{Nordlund_Padoan1999}
{Nordlund} {\r{A}}.~K.,  {Padoan} P.,  1999, in {Franco} J.,  {Carraminana} A.,
   eds, Interstellar Turbulence. p.~218 (\mn@eprint {arXiv}
  {astro-ph/9810074}), \mn@doi{10.48550/arXiv.astro-ph/9810074}

\bibitem[\protect\citeauthoryear{{Offner}, {Clark}, {Hennebelle}, {Bastian},
  {Bate}, {Hopkins}, {Moraux}  \& {Whitworth}}{{Offner}
  et~al.}{2014}]{Offner+2014}
{Offner} S.~S.~R.,  {Clark} P.~C.,  {Hennebelle} P.,  {Bastian} N.,  {Bate}
  M.~R.,  {Hopkins} P.~F.,  {Moraux} E.,   {Whitworth} A.~P.,  2014, in
  {Beuther} H.,  {Klessen} R.~S.,  {Dullemond} C.~P.,   {Henning} T.,  eds,
  Protostars and Planets VI. pp 53--75 (\mn@eprint {arXiv} {1312.5326}),
  \mn@doi{10.2458/azu_uapress_9780816531240-ch003}

\bibitem[\protect\citeauthoryear{{Orkisz} et~al.,}{{Orkisz}
  et~al.}{2019}]{Orkisz+2019}
{Orkisz} J.~H.,  et~al., 2019, \mn@doi [\aap] {10.1051/0004-6361/201833410},
  \href {https://ui.adsabs.harvard.edu/abs/2019A&A...624A.113O} {624, A113}

\bibitem[\protect\citeauthoryear{{Ostriker}}{{Ostriker}}{1964}]{Ostriker1964}
{Ostriker} J.,  1964, \mn@doi [\apj] {10.1086/148005}, \href
  {https://ui.adsabs.harvard.edu/abs/1964ApJ...140.1056O} {140, 1056}

\bibitem[\protect\citeauthoryear{{Padoan} \& {Nordlund}}{{Padoan} \&
  {Nordlund}}{2011}]{Padoan_Nordlund2011}
{Padoan} P.,  {Nordlund} {\r{A}}.,  2011, \mn@doi [\apj]
  {10.1088/0004-637X/730/1/40}, \href
  {https://ui.adsabs.harvard.edu/abs/2011ApJ...730...40P} {730, 40}

\bibitem[\protect\citeauthoryear{{Padoan}, {Federrath}, {Chabrier}, {Evans},
  {Johnstone}, {J{\o}rgensen}, {McKee}  \& {Nordlund}}{{Padoan}
  et~al.}{2014}]{Padoan+2014}
{Padoan} P.,  {Federrath} C.,  {Chabrier} G.,  {Evans} N.~J. I.,  {Johnstone}
  D.,  {J{\o}rgensen} J.~K.,  {McKee} C.~F.,   {Nordlund} {\r{A}}.,  2014, in
  {Beuther} H.,  {Klessen} R.~S.,  {Dullemond} C.~P.,   {Henning} T.,  eds,
  Protostars and Planets VI. pp 77--100 (\mn@eprint {arXiv} {1312.5365}),
  \mn@doi{10.2458/azu_uapress_9780816531240-ch004}

\bibitem[\protect\citeauthoryear{{Palmeirim} et~al.,}{{Palmeirim}
  et~al.}{2013}]{Palmeirim+2013}
{Palmeirim} P.,  et~al., 2013, \mn@doi [\aap] {10.1051/0004-6361/201220500},
  \href {https://ui.adsabs.harvard.edu/abs/2013A&A...550A..38P} {550, A38}

\bibitem[\protect\citeauthoryear{{Pillai} et~al.,}{{Pillai}
  et~al.}{2020}]{Pillai+2020}
{Pillai} T. G.~S.,  et~al., 2020, \mn@doi [Nature Astronomy]
  {10.1038/s41550-020-1172-6}, \href
  {https://ui.adsabs.harvard.edu/abs/2020NatAs...4.1195P} {4, 1195}

\bibitem[\protect\citeauthoryear{{Polychroni} et~al.,}{{Polychroni}
  et~al.}{2013}]{Polychroni+2013}
{Polychroni} D.,  et~al., 2013, \mn@doi [\apjl] {10.1088/2041-8205/777/2/L33},
  \href {https://ui.adsabs.harvard.edu/abs/2013ApJ...777L..33P} {777, L33}

\bibitem[\protect\citeauthoryear{{Pudritz} \& {Kevlahan}}{{Pudritz} \&
  {Kevlahan}}{2013}]{Pudritz_Kevlahan2013}
{Pudritz} R.~E.,  {Kevlahan} N.~K.~R.,  2013, \mn@doi [Philosophical
  Transactions of the Royal Society of London Series A]
  {10.1098/rsta.2012.0248}, \href
  {https://ui.adsabs.harvard.edu/abs/2013RSPTA.37120248P} {371, 20120248}

\bibitem[\protect\citeauthoryear{{Ragan}, {Henning}, {Tackenberg}, {Beuther},
  {Johnston}, {Kainulainen}  \& {Linz}}{{Ragan} et~al.}{2014}]{Ragan+2014}
{Ragan} S.~E.,  {Henning} T.,  {Tackenberg} J.,  {Beuther} H.,  {Johnston}
  K.~G.,  {Kainulainen} J.,   {Linz} H.,  2014, \mn@doi [\aap]
  {10.1051/0004-6361/201423401}, \href
  {https://ui.adsabs.harvard.edu/abs/2014A&A...568A..73R} {568, A73}

\bibitem[\protect\citeauthoryear{{Ramachandran} \& {Varoquaux}}{{Ramachandran}
  \& {Varoquaux}}{2012}]{RamaVaro2011}
{Ramachandran} P.,  {Varoquaux} G.,  2012, {Mayavi2: 3D Scientific Data
  Visualization and Plottin'}, Astrophysics Source Code Library, record
  ascl:1205.008 (\mn@eprint {ascl} {1205.008})

\bibitem[\protect\citeauthoryear{{Recchi}, {Hacar}  \& {Palestini}}{{Recchi}
  et~al.}{2014}]{Recchi+2014}
{Recchi} S.,  {Hacar} A.,   {Palestini} A.,  2014, \mn@doi [\mnras]
  {10.1093/mnras/stu1566}, \href
  {https://ui.adsabs.harvard.edu/abs/2014MNRAS.444.1775R} {444, 1775}

\bibitem[\protect\citeauthoryear{{Reina-Campos}, {Keller}, {Kruijssen},
  {Gensior}, {Trujillo-Gomez}, {Jeffreson}, {Pfeffer}  \&
  {Sills}}{{Reina-Campos} et~al.}{2022}]{Reina-Campos+2022}
{Reina-Campos} M.,  {Keller} B.~W.,  {Kruijssen} J.~M.~D.,  {Gensior} J.,
  {Trujillo-Gomez} S.,  {Jeffreson} S. M.~R.,  {Pfeffer} J.~L.,   {Sills} A.,
  2022, \mn@doi [\mnras] {10.1093/mnras/stac1934}, \href
  {https://ui.adsabs.harvard.edu/abs/2022MNRAS.517.3144R} {517, 3144}

\bibitem[\protect\citeauthoryear{{Reissl}, {Wolf}  \& {Brauer}}{{Reissl}
  et~al.}{2016}]{Reissl+2016}
{Reissl} S.,  {Wolf} S.,   {Brauer} R.,  2016, \mn@doi [\aap]
  {10.1051/0004-6361/201424930}, \href
  {https://ui.adsabs.harvard.edu/abs/2016A&A...593A..87R} {593, A87}

\bibitem[\protect\citeauthoryear{{Reissl}, {Stutz}, {Brauer}, {Pellegrini},
  {Schleicher}  \& {Klessen}}{{Reissl} et~al.}{2018}]{Reissl+2018}
{Reissl} S.,  {Stutz} A.~M.,  {Brauer} R.,  {Pellegrini} E.~W.,  {Schleicher}
  D. R.~G.,   {Klessen} R.~S.,  2018, \mn@doi [\mnras] {10.1093/mnras/sty2415},
  \href {https://ui.adsabs.harvard.edu/abs/2018MNRAS.481.2507R} {481, 2507}

\bibitem[\protect\citeauthoryear{{Renaud} et~al.,}{{Renaud}
  et~al.}{2013}]{Renaud+2013}
{Renaud} F.,  et~al., 2013, \mn@doi [\mnras] {10.1093/mnras/stt1698}, \href
  {https://ui.adsabs.harvard.edu/abs/2013MNRAS.436.1836R} {436, 1836}

\bibitem[\protect\citeauthoryear{{Rezaei Kh.} \& {Kainulainen}}{{Rezaei Kh.} \&
  {Kainulainen}}{2022}]{RezaeiKh_Kainulainen2022}
{Rezaei Kh.} S.,  {Kainulainen} J.,  2022, \mn@doi [\apjl]
  {10.3847/2041-8213/ac67db}, \href
  {https://ui.adsabs.harvard.edu/abs/2022ApJ...930L..22R} {930, L22}

\bibitem[\protect\citeauthoryear{{Rezaei Kh.}, {Bailer-Jones}, {Soler}  \&
  {Zari}}{{Rezaei Kh.} et~al.}{2020}]{RezaeiKh+2020}
{Rezaei Kh.} S.,  {Bailer-Jones} C. A.~L.,  {Soler} J.~D.,   {Zari} E.,  2020,
  \mn@doi [\aap] {10.1051/0004-6361/202038708}, \href
  {https://ui.adsabs.harvard.edu/abs/2020A&A...643A.151R} {643, A151}

\bibitem[\protect\citeauthoryear{{Robinson} \& {Wadsley}}{{Robinson} \&
  {Wadsley}}{2023}]{Robinson_Wadsley2023}
{Robinson} H.,  {Wadsley} J.,  2023, \mn@doi [arXiv e-prints]
  {10.48550/arXiv.2310.15244}, \href
  {https://ui.adsabs.harvard.edu/abs/2023arXiv231015244R} {p. arXiv:2310.15244}

\bibitem[\protect\citeauthoryear{{Rosolowsky}, {Pineda}, {Kauffmann}  \&
  {Goodman}}{{Rosolowsky} et~al.}{2008}]{Rosolowsky+2008}
{Rosolowsky} E.~W.,  {Pineda} J.~E.,  {Kauffmann} J.,   {Goodman} A.~A.,  2008,
  \mn@doi [\apj] {10.1086/587685}, \href
  {https://ui.adsabs.harvard.edu/abs/2008ApJ...679.1338R} {679, 1338}

\bibitem[\protect\citeauthoryear{{Rosolowsky} et~al.,}{{Rosolowsky}
  et~al.}{2021}]{Rosolowsky+2021}
{Rosolowsky} E.,  et~al., 2021, \mn@doi [\mnras] {10.1093/mnras/stab085}, \href
  {https://ui.adsabs.harvard.edu/abs/2021MNRAS.502.1218R} {502, 1218}

\bibitem[\protect\citeauthoryear{{Schechter}}{{Schechter}}{1976}]{Schechter1976}
{Schechter} P.,  1976, \mn@doi [\apj] {10.1086/154079}, \href
  {https://ui.adsabs.harvard.edu/abs/1976ApJ...203..297S} {203, 297}

\bibitem[\protect\citeauthoryear{{Schleicher} \& {Stutz}}{{Schleicher} \&
  {Stutz}}{2018}]{Schleicher_Stutz2018}
{Schleicher} D. R.~G.,  {Stutz} A.,  2018, \mn@doi [\mnras]
  {10.1093/mnras/stx2975}, \href
  {https://ui.adsabs.harvard.edu/abs/2018MNRAS.475..121S} {475, 121}

\bibitem[\protect\citeauthoryear{{Semenov}, {Kravtsov}  \& {Gnedin}}{{Semenov}
  et~al.}{2018}]{Semenov+2018}
{Semenov} V.~A.,  {Kravtsov} A.~V.,   {Gnedin} N.~Y.,  2018, \mn@doi [\apj]
  {10.3847/1538-4357/aac6eb}, \href
  {https://ui.adsabs.harvard.edu/abs/2018ApJ...861....4S} {861, 4}

\bibitem[\protect\citeauthoryear{{Smith}, {Glover}, {Klessen}  \&
  {Fuller}}{{Smith} et~al.}{2016}]{Smith+2016}
{Smith} R.~J.,  {Glover} S. C.~O.,  {Klessen} R.~S.,   {Fuller} G.~A.,  2016,
  \mn@doi [\mnras] {10.1093/mnras/stv2559}, \href
  {https://ui.adsabs.harvard.edu/abs/2016MNRAS.455.3640S} {455, 3640}

\bibitem[\protect\citeauthoryear{{Smith} et~al.,}{{Smith}
  et~al.}{2020}]{Smith+2020}
{Smith} R.~J.,  et~al., 2020, \mn@doi [\mnras] {10.1093/mnras/stz3328}, \href
  {https://ui.adsabs.harvard.edu/abs/2020MNRAS.492.1594S} {492, 1594}

\bibitem[\protect\citeauthoryear{{Soler} et~al.,}{{Soler}
  et~al.}{2020}]{Soler+2020}
{Soler} J.~D.,  et~al., 2020, \mn@doi [\aap] {10.1051/0004-6361/202038882},
  \href {https://ui.adsabs.harvard.edu/abs/2020A&A...642A.163S} {642, A163}

\bibitem[\protect\citeauthoryear{{Soler} et~al.,}{{Soler}
  et~al.}{2022}]{Soler+2022}
{Soler} J.~D.,  et~al., 2022, \mn@doi [\aap] {10.1051/0004-6361/202243334},
  \href {https://ui.adsabs.harvard.edu/abs/2022A&A...662A..96S} {662, A96}

\bibitem[\protect\citeauthoryear{{Springel}}{{Springel}}{2010}]{Springel2010}
{Springel} V.,  2010, \mn@doi [\mnras] {10.1111/j.1365-2966.2009.15715.x},
  \href {https://ui.adsabs.harvard.edu/abs/2010MNRAS.401..791S} {401, 791}

\bibitem[\protect\citeauthoryear{{Stod{\'o}lkiewicz}}{{Stod{\'o}lkiewicz}}{1963}]{Stodolkiewicz1963}
{Stod{\'o}lkiewicz} J.~S.,  1963, \actaa, \href
  {https://ui.adsabs.harvard.edu/abs/1963AcA....13...30S} {13, 30}

\bibitem[\protect\citeauthoryear{{Syed} et~al.,}{{Syed}
  et~al.}{2022}]{Syed+2022}
{Syed} J.,  et~al., 2022, \mn@doi [\aap] {10.1051/0004-6361/202141265}, \href
  {https://ui.adsabs.harvard.edu/abs/2022A&A...657A...1S} {657, A1}

\bibitem[\protect\citeauthoryear{{Tahani}, {Plume}, {Brown}, {Soler}  \&
  {Kainulainen}}{{Tahani} et~al.}{2019}]{Tahani+2019}
{Tahani} M.,  {Plume} R.,  {Brown} J.~C.,  {Soler} J.~D.,   {Kainulainen} J.,
  2019, \mn@doi [\aap] {10.1051/0004-6361/201936280}, \href
  {https://ui.adsabs.harvard.edu/abs/2019A&A...632A..68T} {632, A68}

\bibitem[\protect\citeauthoryear{{Tahani} et~al.,}{{Tahani}
  et~al.}{2022}]{Tahani+2022}
{Tahani} M.,  et~al., 2022, \mn@doi [\aap] {10.1051/0004-6361/202141170}, \href
  {https://ui.adsabs.harvard.edu/abs/2022A&A...660A..97T} {660, A97}

\bibitem[\protect\citeauthoryear{{Teyssier}}{{Teyssier}}{2002}]{Teyssier2002}
{Teyssier} R.,  2002, \mn@doi [\aap] {10.1051/0004-6361:20011817}, \href
  {https://ui.adsabs.harvard.edu/abs/2002A&A...385..337T} {385, 337}

\bibitem[\protect\citeauthoryear{{Thilker} et~al.,}{{Thilker}
  et~al.}{2023}]{Thilker+2023}
{Thilker} D.~A.,  et~al., 2023, \mn@doi [arXiv e-prints]
  {10.48550/arXiv.2301.00881}, \href
  {https://ui.adsabs.harvard.edu/abs/2023arXiv230100881T} {p. arXiv:2301.00881}

\bibitem[\protect\citeauthoryear{{Tilley} \& {Pudritz}}{{Tilley} \&
  {Pudritz}}{2003}]{Tilley_Pudritz2003}
{Tilley} D.~A.,  {Pudritz} R.~E.,  2003, \mn@doi [\apj] {10.1086/376357}, \href
  {https://ui.adsabs.harvard.edu/abs/2003ApJ...593..426T} {593, 426}

\bibitem[\protect\citeauthoryear{{Tomisaka}}{{Tomisaka}}{2014}]{Tomisaka2014}
{Tomisaka} K.,  2014, \mn@doi [\apj] {10.1088/0004-637X/785/1/24}, \href
  {https://ui.adsabs.harvard.edu/abs/2014ApJ...785...24T} {785, 24}

\bibitem[\protect\citeauthoryear{{Toomre}}{{Toomre}}{1964}]{Toomre1964}
{Toomre} A.,  1964, \mn@doi [\apj] {10.1086/147861}, \href
  {http://adsabs.harvard.edu/abs/1964ApJ...139.1217T} {139, 1217}

\bibitem[\protect\citeauthoryear{{Tress}, {Smith}, {Sormani}, {Glover},
  {Klessen}, {Mac Low}  \& {Clark}}{{Tress} et~al.}{2020}]{Tress+2020}
{Tress} R.~G.,  {Smith} R.~J.,  {Sormani} M.~C.,  {Glover} S. C.~O.,  {Klessen}
  R.~S.,  {Mac Low} M.-M.,   {Clark} P.~C.,  2020, \mn@doi [\mnras]
  {10.1093/mnras/stz3600}, \href
  {https://ui.adsabs.harvard.edu/abs/2020MNRAS.492.2973T} {492, 2973}

\bibitem[\protect\citeauthoryear{{Tress}, {Sormani}, {Smith}, {Glover},
  {Klessen}, {Mac Low}, {Clark}  \& {Duarte-Cabral}}{{Tress}
  et~al.}{2021}]{Tress+2021}
{Tress} R.~G.,  {Sormani} M.~C.,  {Smith} R.~J.,  {Glover} S. C.~O.,  {Klessen}
  R.~S.,  {Mac Low} M.-M.,  {Clark} P.,   {Duarte-Cabral} A.,  2021, \mn@doi
  [\mnras] {10.1093/mnras/stab1683}, \href
  {https://ui.adsabs.harvard.edu/abs/2021MNRAS.505.5438T} {505, 5438}

\bibitem[\protect\citeauthoryear{{Truelove}, {Klein}, {McKee}, {Holliman},
  {Howell}  \& {Greenough}}{{Truelove} et~al.}{1997}]{Truelove+1997}
{Truelove} J.~K.,  {Klein} R.~I.,  {McKee} C.~F.,  {Holliman} John~H. I.,
  {Howell} L.~H.,   {Greenough} J.~A.,  1997, \mn@doi [\apjl] {10.1086/310975},
  \href {https://ui.adsabs.harvard.edu/abs/1997ApJ...489L.179T} {489, L179}

\bibitem[\protect\citeauthoryear{{Vazquez-Semadeni}}{{Vazquez-Semadeni}}{1994}]{Vazquez-Semadeni1994}
{Vazquez-Semadeni} E.,  1994, \mn@doi [\apj] {10.1086/173847}, \href
  {https://ui.adsabs.harvard.edu/abs/1994ApJ...423..681V} {423, 681}

\bibitem[\protect\citeauthoryear{{V{\'a}zquez-Semadeni}, {Ballesteros-Paredes}
  \& {Klessen}}{{V{\'a}zquez-Semadeni} et~al.}{2003}]{Vazquez-Semadeni+2003}
{V{\'a}zquez-Semadeni} E.,  {Ballesteros-Paredes} J.,   {Klessen} R.~S.,  2003,
  \mn@doi [\apjl] {10.1086/374325}, \href
  {https://ui.adsabs.harvard.edu/abs/2003ApJ...585L.131V} {585, L131}

\bibitem[\protect\citeauthoryear{{Veena} et~al.,}{{Veena}
  et~al.}{2021}]{Veena+2021}
{Veena} V.~S.,  et~al., 2021, \mn@doi [\apjl] {10.3847/2041-8213/ac341f}, \href
  {https://ui.adsabs.harvard.edu/abs/2021ApJ...921L..42V} {921, L42}

\bibitem[\protect\citeauthoryear{{Watkins} et~al.,}{{Watkins}
  et~al.}{2022}]{Watkins+2022}
{Watkins} E.~J.,  et~al., 2022, \mn@doi [arXiv e-prints]
  {10.48550/arXiv.2212.00811}, \href
  {https://ui.adsabs.harvard.edu/abs/2022arXiv221200811W} {p. arXiv:2212.00811}

\bibitem[\protect\citeauthoryear{{Zucker}, {Battersby}  \& {Goodman}}{{Zucker}
  et~al.}{2018}]{Zucker+2018}
{Zucker} C.,  {Battersby} C.,   {Goodman} A.,  2018, \mn@doi [\apj]
  {10.3847/1538-4357/aacc66}, \href
  {https://ui.adsabs.harvard.edu/abs/2018ApJ...864..153Z} {864, 153}

\bibitem[\protect\citeauthoryear{{Zucker} et~al.,}{{Zucker}
  et~al.}{2021}]{Zucker+2021}
{Zucker} C.,  et~al., 2021, \mn@doi [\apj] {10.3847/1538-4357/ac1f96}, \href
  {https://ui.adsabs.harvard.edu/abs/2021ApJ...919...35Z} {919, 35}

\bibitem[\protect\citeauthoryear{{Zucker}, {Alves}, {Goodman}, {Meingast}  \&
  {Galli}}{{Zucker} et~al.}{2022}]{Zucker+2022}
{Zucker} C.,  {Alves} J.,  {Goodman} A.,  {Meingast} S.,   {Galli} P.,  2022,
  \mn@doi [arXiv e-prints] {10.48550/arXiv.2212.00067}, \href
  {https://ui.adsabs.harvard.edu/abs/2022arXiv221200067Z} {p. arXiv:2212.00067}

\makeatother
\end{thebibliography}



\appendix

\section{ISM Phase Diagrams}
\label{App.A}

\begin{figure}
    \centering
    \includegraphics[width=0.49\linewidth]{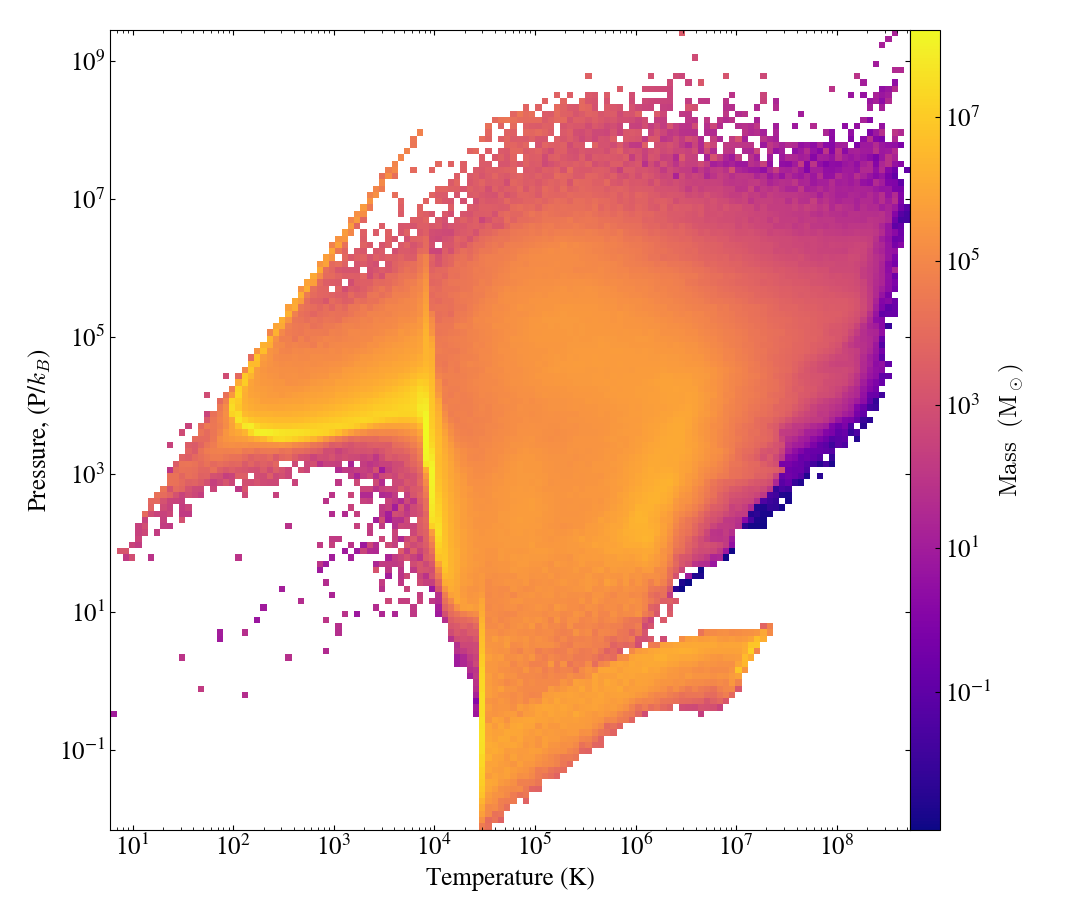}
    \includegraphics[width=0.49\linewidth]{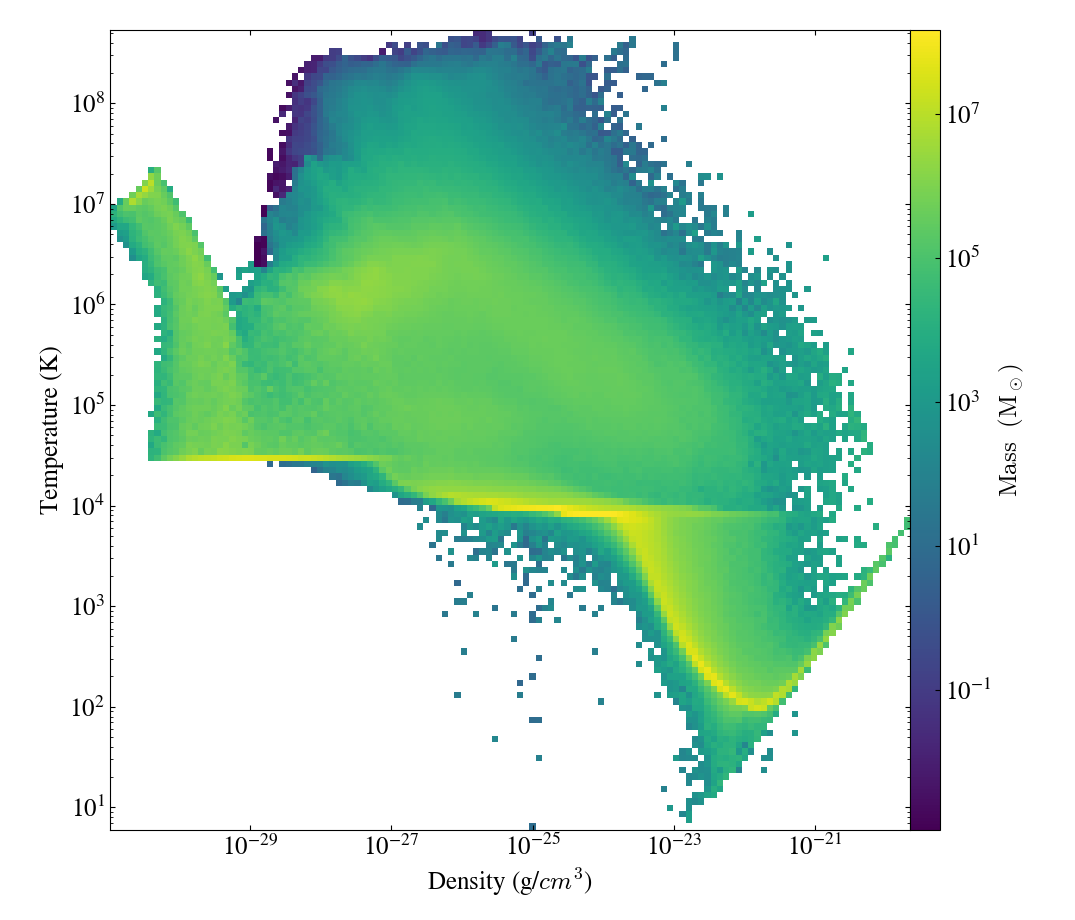}
    \caption{Pressure-temperature \textit{(left)} and Temperature-density (\textit{right}) phase plots. Colourbars represent mass per bin such that the yellow colours on both plots depict where the majority of the gas sits.}
    \label{Fig:gas_phases}
\end{figure}

\begin{figure}
    \centering
    \includegraphics[width=0.5\linewidth]{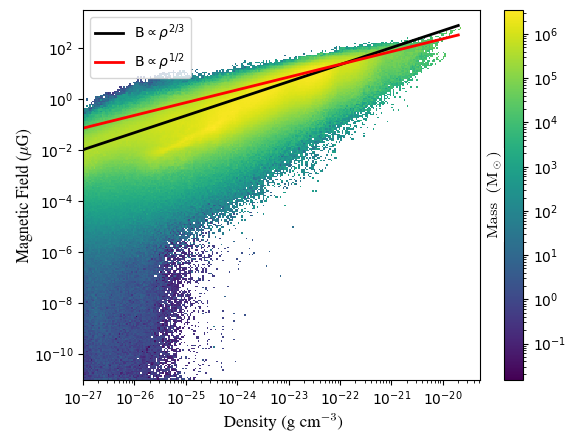}
    \caption{Magnetic field-density phase plot. Colourbars represent mass per bin, similar to figure \ref{Fig:gas_phases}. Scaling relations for $B \propto \rho^{1/2}$ and $B \propto \rho^{2/3}$ are given in red and black lines, respectively.}
    \label{Fig:magfield_density}
\end{figure}

We present an analysis of the physical state of the interstellar medium on the full global galactic scale, visualized in the phase plots presented in Figures \ref{Fig:gas_phases} and \ref{Fig:magfield_density}. Phase plots allow us to visualize where the bulk of the gas by mass resides within a certain parameter space in pressure (or density) and temperature. In the pressure-temperature space in the left panel of Figure \ref{Fig:gas_phases}, we plot the total mass in gas cells residing at a particular pressure and temperature. The mass is color coded. These plots are often used to describe gas phase information in simulations, especially large galactic-scale simulations \citep{Reina-Campos+2022}. As such, the areas in which the mass (represented by the colour bars) is highest show the typical state of the gas in our simulations.  As noted in the text, our phase diagrams resemble those in galactic simulations \citep{Kim+2016, Reina-Campos+2022}. 
 
Similarly, in Figure \ref{Fig:magfield_density} we plot the magnetic field strength versus the density for gas in our global ISM.  It shows the magnetic field strength associated with the majority of the gas (again, see color bar). In this plot we also superimpose two  popularly used power law scaling of the magnetic field with density, as discussed in \citet{Crutcher2019}.  Recall that our initial conditions used a very weak toroidal field with a 2/3 power law scaling.  Turbulent dynamo action would have erased initial memory of such a weak field very quickly, so the fact that our result finds a roughly 2/3 power law for the denser gas is significant.   

\section{Fourier Analysis and Turbulence Spectra}
\label{App.B}

The turbulence kinetic power spectrum $P_i(\bm{k})$ for the 
velocity components $\varv_x$, $\varv_y$, and $\varv_z$ are defined as,
\begin{equation}
P_i(\bm{k}) = \frac{1}{L_x L_y L_z} |\Bar{V}_i(\bm{k})|^2,~ i=x, y, z~,
\end{equation}
where $L_x$, $L_y$, and $L_z$ are the lengths of the box along the three 
directions, and $\Bar{V}_i(\bm{k})$ is the Fourier transform of the 
respective velocity component:
\begin{equation}
\Bar{V}_i(\bm{k})= \frac{1}{(2\pi)^{3/2}} \int \varv_i(\bm{x}) exp(-i\bm{k}\cdot\bm{x}) {\rm d}^3\bm{x},~ i=x, y, z~.
\end{equation}
The total turbulence power spectrum 
$P(k)=P_x(k) + P_y(k) + P_z(k)$ can be obtained by averaging the 
corresponding $P_i(\bm{k})$ at a constant magnitude $k=|\bm{k}|$. 
The energy spectrum is thus computed as $E(k)=4 \pi k^2 P(k)$. 

\begin{figure*}
\includegraphics[width=0.5\columnwidth]{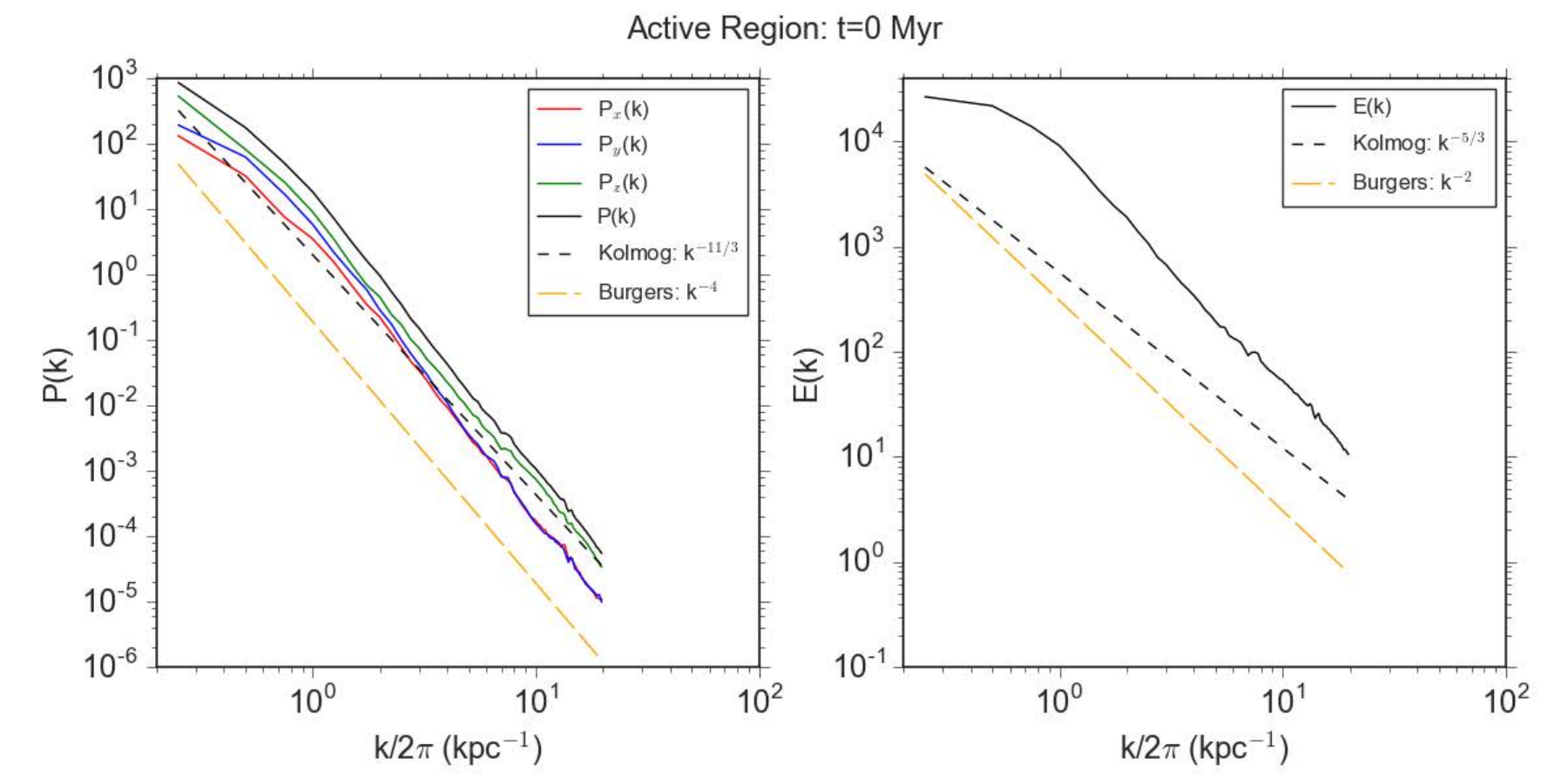}
\includegraphics[width=0.5\columnwidth]{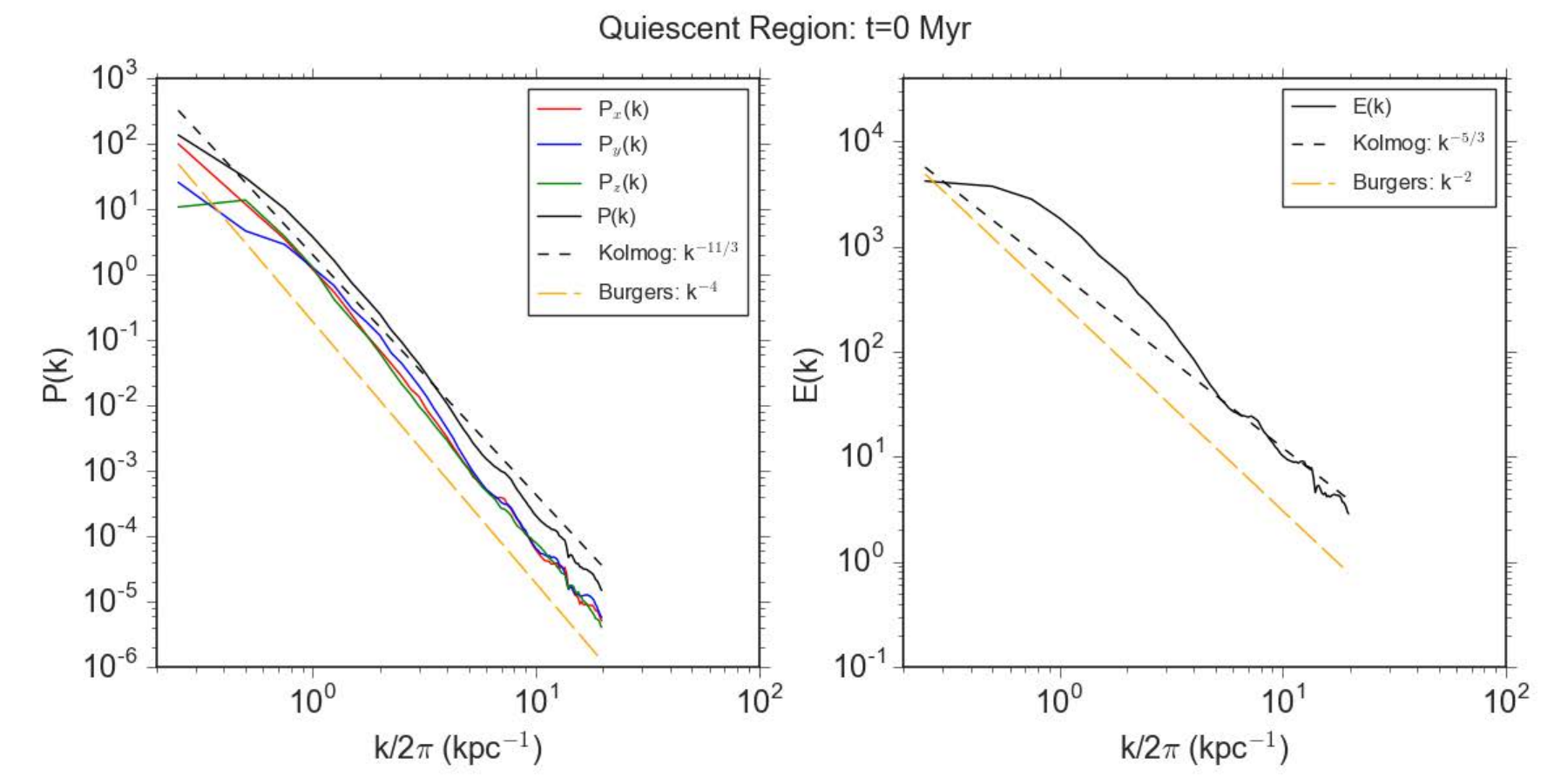}
\includegraphics[width=0.5\columnwidth]{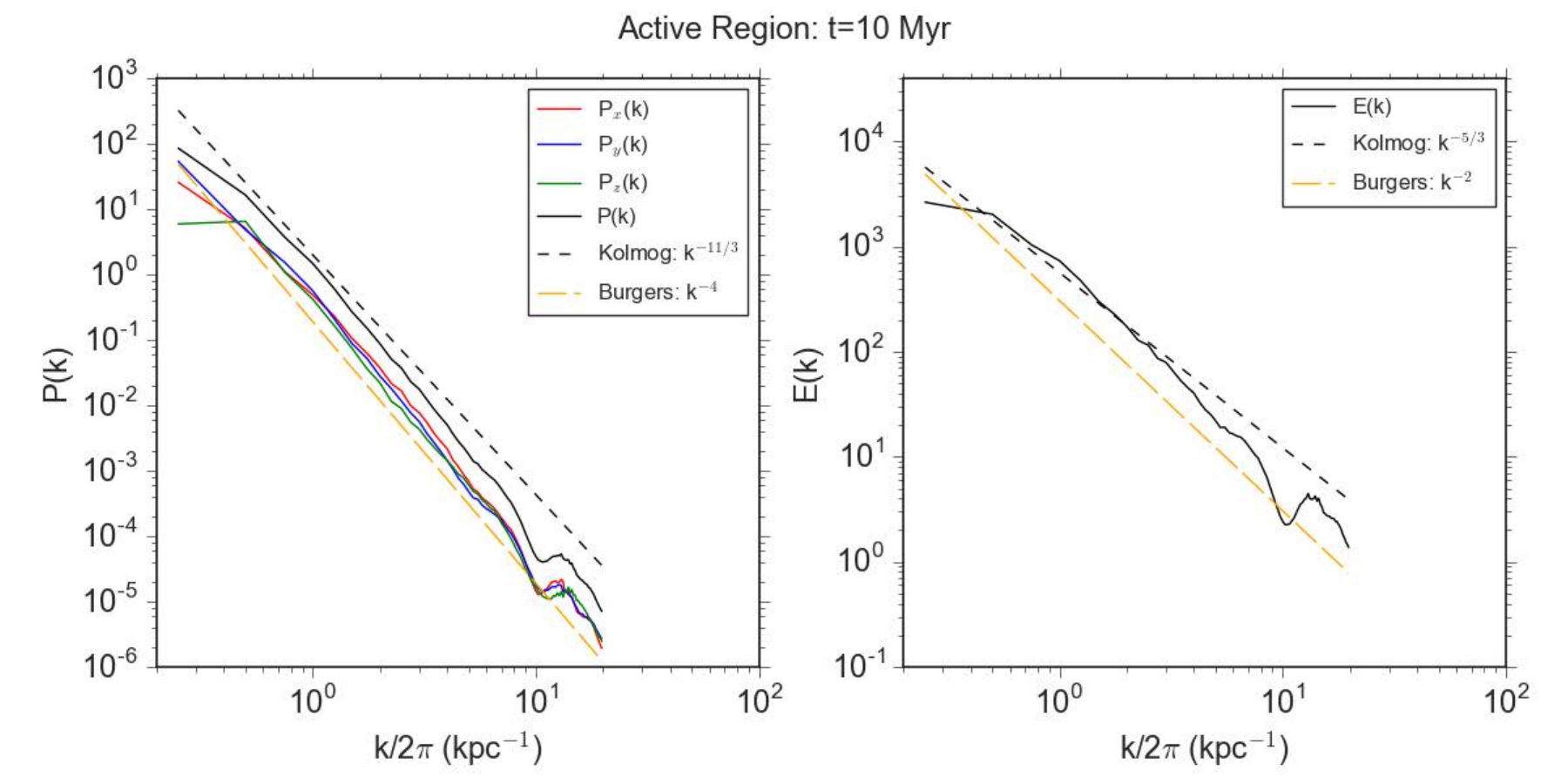}
\includegraphics[width=0.5\columnwidth]{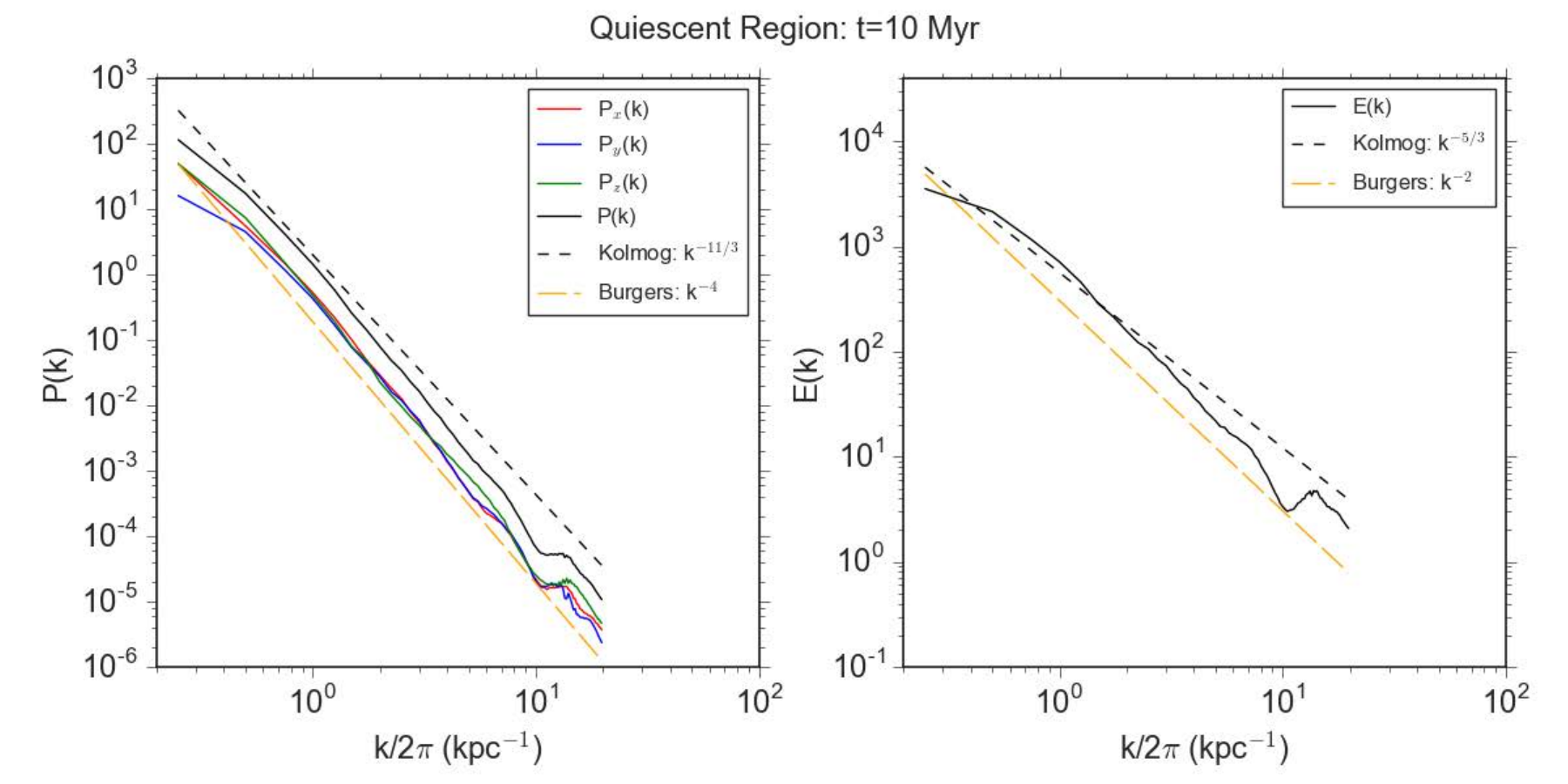}
\caption{Turbulence power $P(k)$ and energy spectrum $E(k)$ in the 3~kpc zoomed-in regions. Left panels show these spectra for the active region at two different times in the simulation; the initial time (after 200-300 Myr of galactic evolution), and at the end of 10 Myr evolution in the 3 kpc box.  The right hand panels show the same except for the quiescent region.}
\label{Fig:power_spec}
\end{figure*}

We note from the Figure that there is a marked change in 
the amplitudes of energy and power in the turbulent spectra between the active and
quiescent regions.  In particular,  
at $t=0$ the total amplitude of $P(k)$ and $E(k)$ in the active region exceed those
of their counterparts in the quiescent region by two and one order of magnitude, respectively.  The 
spectral shapes are unchanged, however.  Over the course of 10~Myr 
the turbulent power $P(k)$ decreases by $\sim$1 order of magnitude and 
the turbulent energy $E(k)$ decreases by nearly  $\sim$2 orders of magnitude to levels
similar to those in the quiescent region.  
In comparison, both $P(k)$ and $E(k)$ in the quiescent 
region decrease only by a factor of a few. 

One explanation for the effect is the difference in losses in the mass reservoir of the active region as
the bubbles from the previous generation of OB star formation expand.  Mass is pushed into the quiescent region so that turbulent decay is somewhat offset by the inflow. 
The existence of a large feedback bubble in the active region also provides a large initial power 
of the vertical turbulence motions ($V_z$) as compared to that in the quiescent 
region; however, the vertical turbulence is likely damped after 10~Myr.  It is also possible that the decay time scale for the turbulence in the active region is shorter than that in the quiescent one. 

\section{Filament Tracing Method}
\label{App.C}

Our method to trace the filamentary structures are based on the visualization 
of the 3D structures. Any prominent filaments can be represented by a 
starting position ($x_0$, $y_0$, $z_0$) and an initial normal direction 
$\bm{n}_0$. A 3D plane perpendicular to $\bm{n}_0$ can be constructed. 
We then scan with a given spatial increment $\delta n$ along the normal 
direction and move the plane forward in space. A maximum density point along 
the new plane can be located (should also in the vicinity of the density maximum 
point in the previous step). By differencing the position of the density maximum 
point on the new plane and that of the previous plane, one can obtain a 
new normal vector for the next scanning progression. In this fashion, a serious 
of sampling points can be located along the dense filament in space. 
The method is computational fast and accurate to the grid resolution of the 
visualization data. 

\subsection{Velocity Profiles}

To create the velocity profiles of our filaments, we use the normal directions from our filament tracer, and create two rays perpendicular to that vector, such that we cut through the filament in two perpendicular directions. This gives two rays, intersecting different sides of the filament, in order to depict the asymmetry of filaments.

\begin{figure*}
    \centering
    \includegraphics[width=0.48\linewidth]{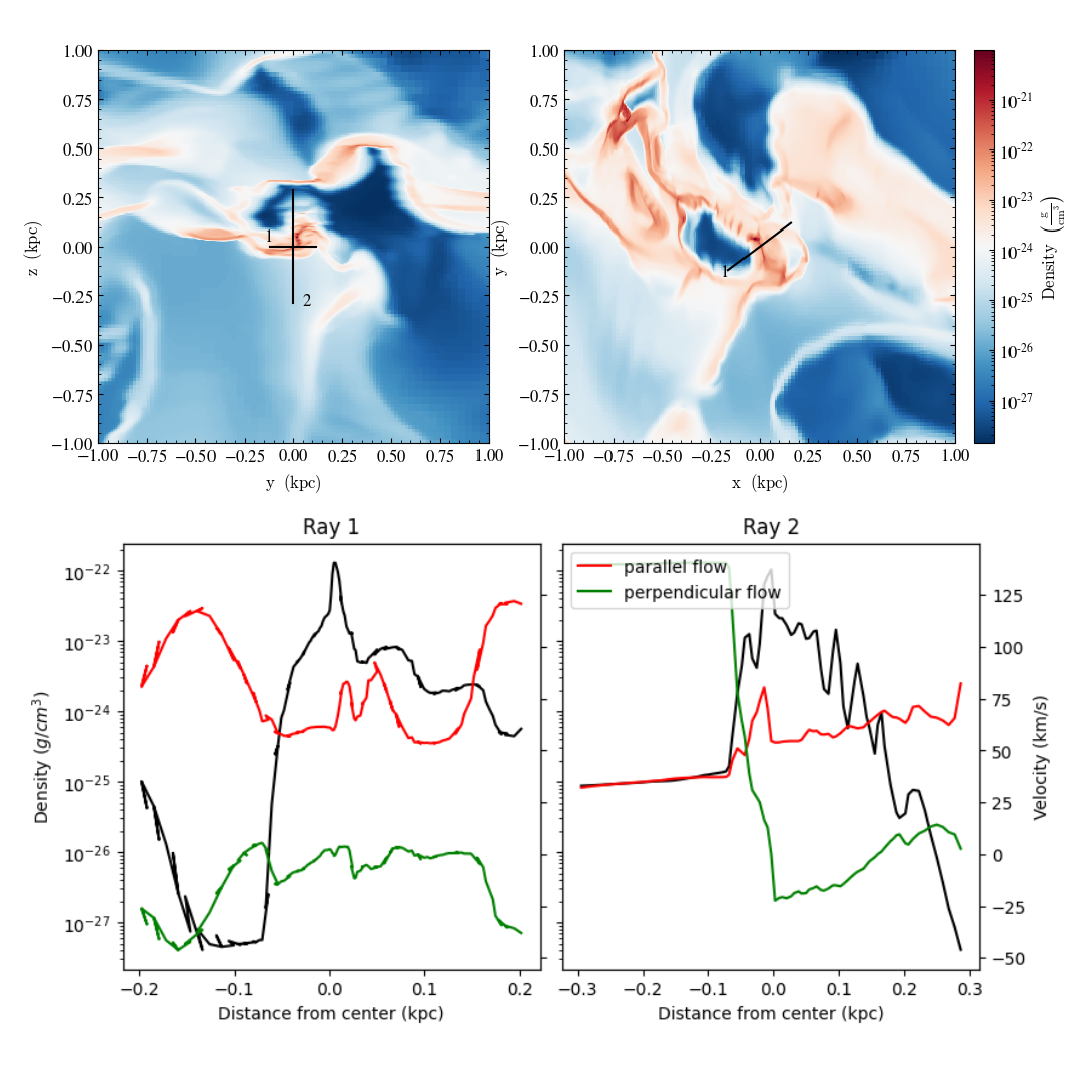}
    \includegraphics[width=0.48\linewidth]{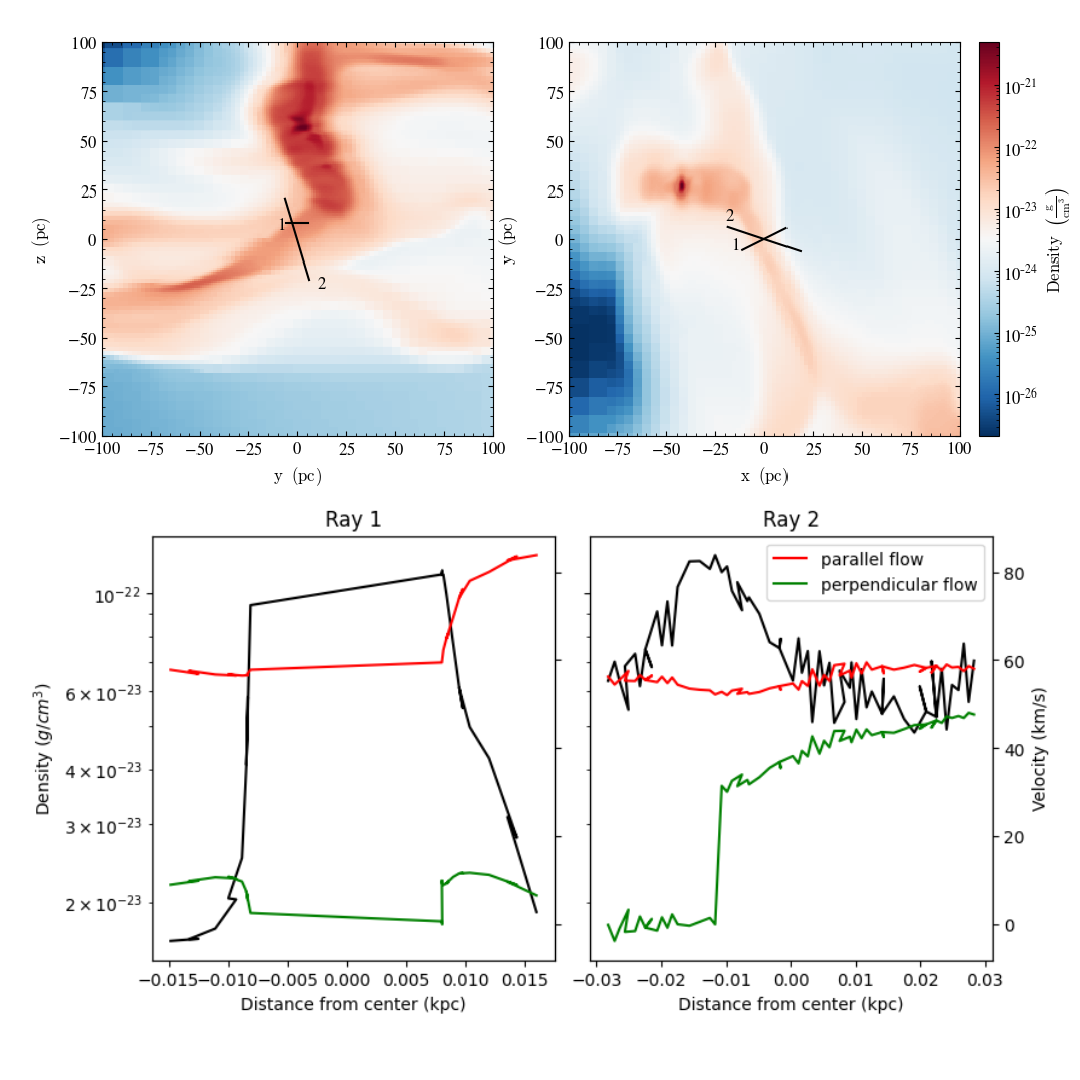}
    \caption{\textit{Top:} Density slices depicting filamentary areas of interest in the quiet 3 kpc region (left) and our zoomed in 200 pc (right) region. The rays traced for flows shown via black lines, labelled 1 and 2 for first and second ray, respectively. \textit{Bottom:} Density and velocity profiles for each ray in both the kpc and zoomed pc scale. Black line gives the density profile of each ray, red lines show velocity along the filament, and green lines show the velocity onto the filament, respective to the direction of the ray.}
    \label{Fig:quiet_velprof}
\end{figure*}

In Fig.~\ref{Fig:quiet_velprof}, we show velocity and density profiles of the featured filaments in the quiescent region. For each of these rays we profile the gas density and all three components of the gas velocity. We plot density as is, but separate our velocities into parallel and perpendicular components via cross and dot products. We note here that the use of parallel and perpendicular is in reference to the direction of the filament, not the direction of the ray, such that parallel flows are the velocity flow along the filament. To ascertain the flow along the filament, we take the cross product of the velocities with the direction of the ray and take its magnitude. Perpendicular flow onto the filament is ascertained via the dot product of the velocity and direction of the ray. An important aspect here is the significance of the sign of the perpendicular flow. Since the sign is with respect to the direction of the ray, then matching signs between distance and perpendicular velocity indicate flow onto the filament. If the signs are opposite from each other, the gas is flowing out of the filament, indicating the filament is being torn apart.

\section{Transition Scale from Galactic Shear to Local Gravity and Rotation}
\label{App.D}

\begin{figure}
\includegraphics[width=\columnwidth]{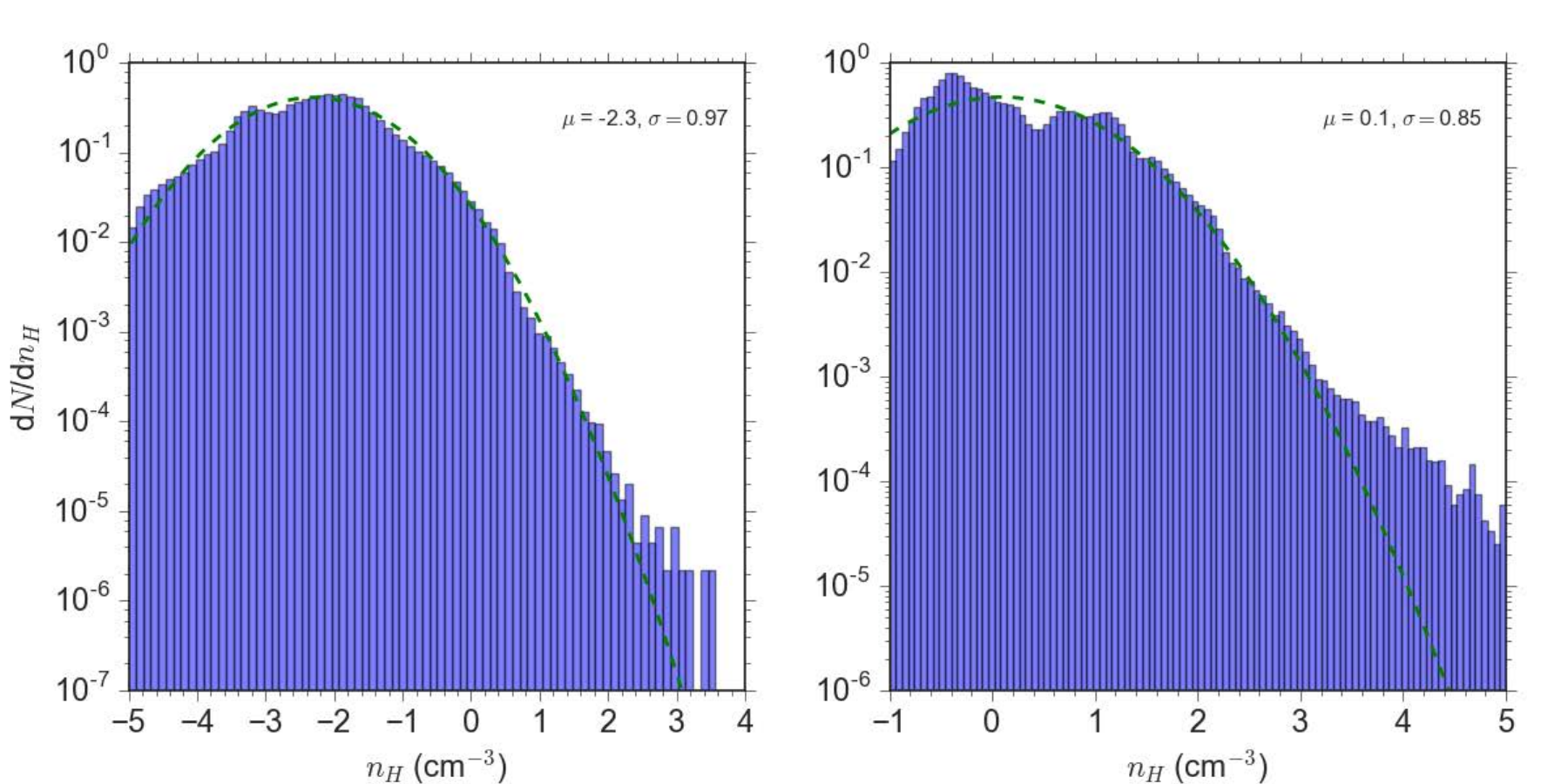}
\caption{Density PDF of the 3~kpc box (left) and the 200~pc box (right).
The green curves are best fit for the distribution function. }
\label{Fig:densityPDF}
\end{figure}

We show the density probability distribution function (PDF) 
for the 3~kpc region and the 200~pc disk region at $t$=6.2~Myr in 
Fig.~\ref{Fig:densityPDF}. The PDF at large scale can be fitted by 
a log normal distribution up to number densities of a few 100~cm$^{-3}$, 
where overdensed condensations along the filamentary structures start to 
develop. The lack of bins are related to the limited re-grid resolution 
at 3~kpc scale. As we enlarge the central disk GMC structure, its 
density PDF clearly shows a ``kink'' at number density of 
$\sim$10$^3$~cm$^{-3}$, turning away from the log-normal curve. 
The high density bins corresponds to the clumps formed at the center 
of the spiral and disk structures (see Fig.~\ref{Fig:inactiveComplex}). 
Therefore, in terms of density, the transition between the two scales 
appears to take place between a few 100 to $\sim$1000~cm$^{-3}$. 

\begin{figure*}
\includegraphics[width=0.5\columnwidth]{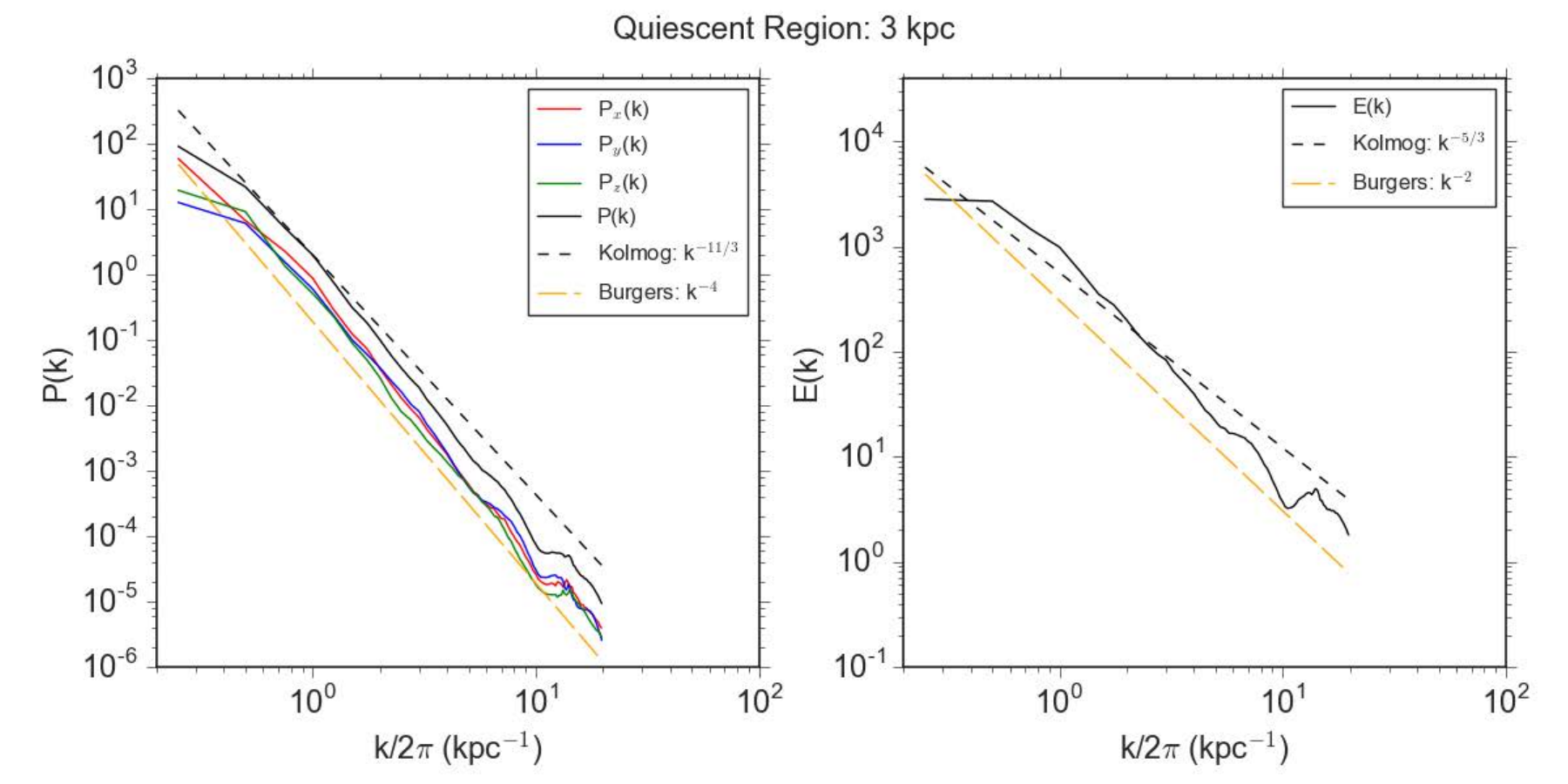}
\includegraphics[width=0.5\columnwidth]{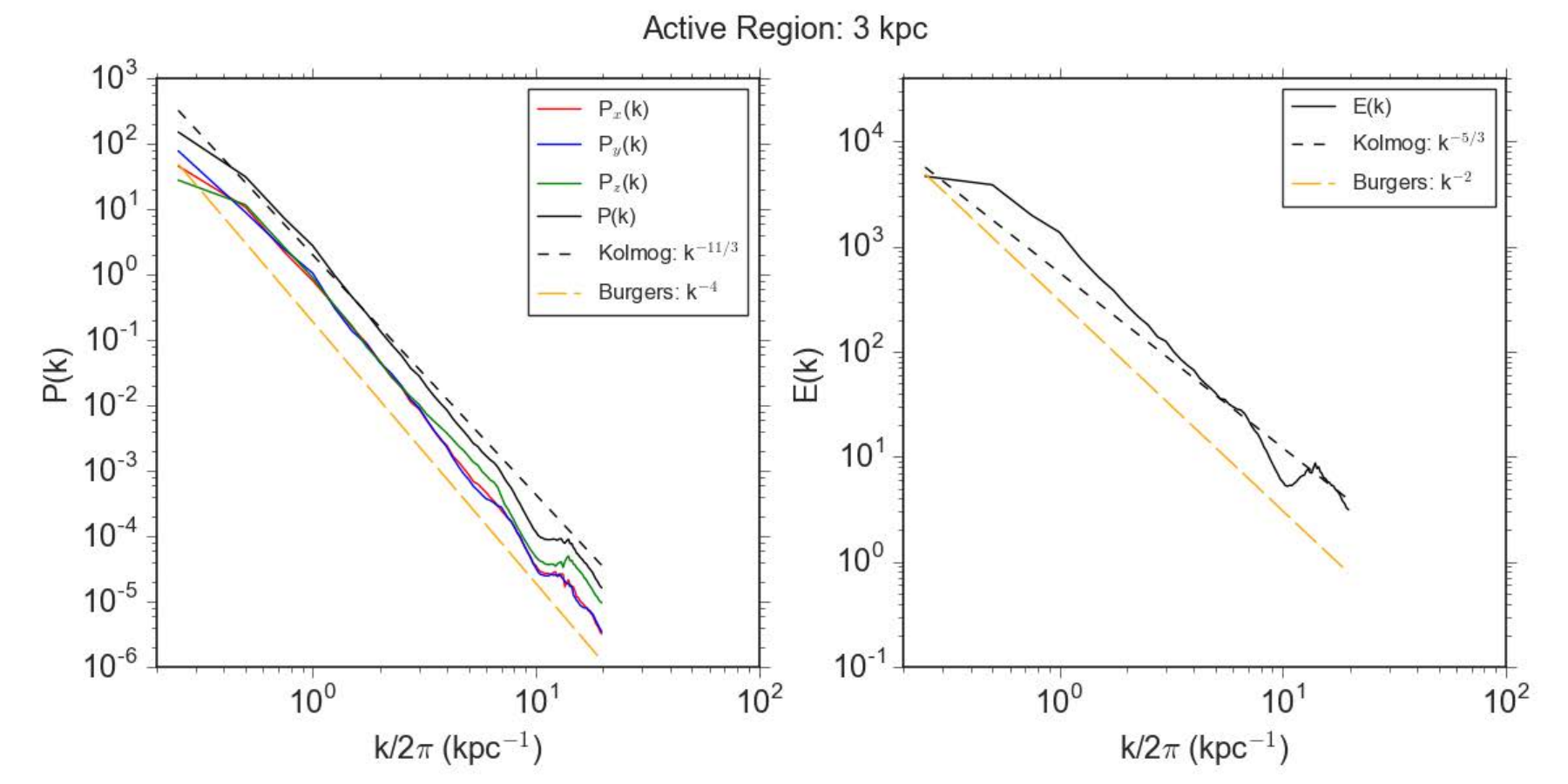}
\includegraphics[width=0.5\columnwidth]{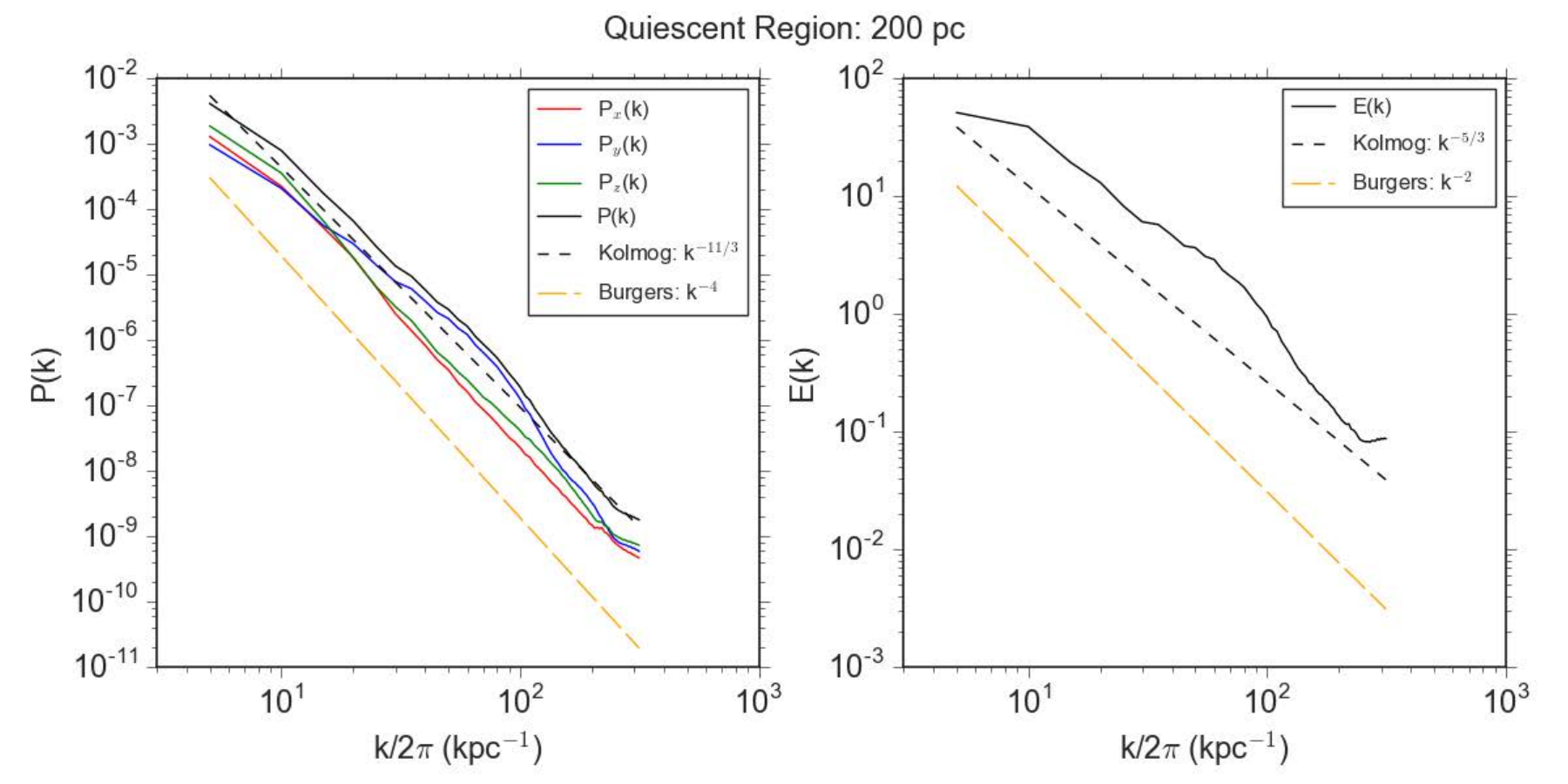}
\includegraphics[width=0.5\columnwidth]{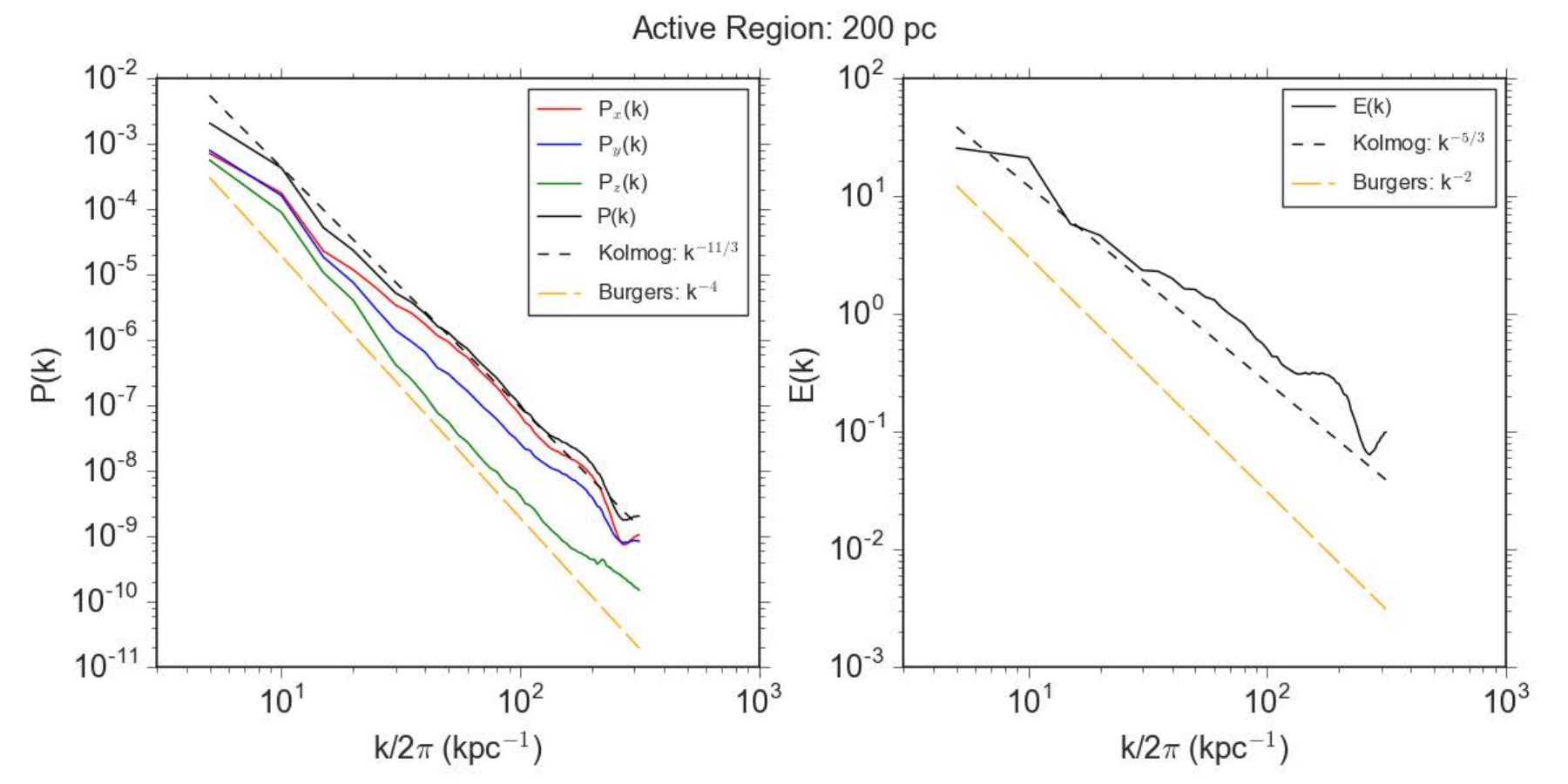}
\caption{Turbulence power spectrum of 
3~kpc box (top panels) and the 200~pc box (botom panels) of the quiescent 
region (left panels) and the active region (right panels) at a time frame 
of 6.4~Myr.}
\label{Fig:turbScale}
\end{figure*}

To further understand the spatial scale at which this transition 
occurs, we show the turbulence spectrum in Fig.~\ref{Fig:turbScale}. 
A consistent power law in turbulence spectrum indicates how kinetic 
energy is propagated and diffused from large to small scales, which 
is the case for the 3~kpc power spectrums (minimum re-grid resolution of 
$\sim$25~pc for the smallest $k$) in both the active and the quiescent 
regions. The spectrum follows closely 
the Burgers $k^{-4}$ law downto $k/2\pi$ of $\sim$10, which corresponds 
to $\sim$50~pc in spatial scale. As it is expensive to carry out 
Fourier transformation across many scales for the AMR grid, we show 
a separate power spectrum for the 200~pc box region instead (bottom panels). 
For both the power and energy spectrum, the Fourier components start 
to deviate at $k/2\pi$ $\sim$10--20, which also corresponds to 
a spatial scale of $\sim$50--100~pc. At scales smaller than this, 
turbulence cascade is likely dominated by the local processes in the 
GMCs, such as gravity and rotational motion. 

In the case of the disk-like GMC (first panel in 
Fig.~\ref{Fig:inactiveComplex}) formed in the quiescent region, 
the $y$ component spectrum is enhanced downto a few pc scale, 
whereas the $x$ and $z$ components are somewhat suppressed. 
Recall that the disk normal is more aligned with $y$-direction 
(see Fig.~\ref{Fig:helicalB}), it is likely that the turbulence 
power along $y$-component is more or less preserved from 100 to 10 pc 
scale, while the turbulence energy along $x$ and $z$ starts to transform 
into rotation motions. In comparison, the cluster-like complex in the 
active region (second panel in Fig.~\ref{Fig:activeComplex}) show 
different power law profiles along different spatial directions. 
It seems that vertical shear motions along the ($y$, $z$) plane is 
stronger in this region, so that the turbulence power in the $x$-direction 
experience a lesser cascade. Such an anisotropy in the turbulence property 
can also contribute to the cluster-like morphology in this region, 
in which multiple substructures evolve differently.


\label{lastpage}
\end{document}